\documentclass[fleqn,usenatbib]{mnras}

\usepackage{newtxtext,newtxmath}

\usepackage[T1]{fontenc}
\usepackage{ae,aecompl}

\usepackage{graphicx}	
\usepackage{amsmath}	
\usepackage{amssymb}	
\usepackage{ulem}
\def\xlinkspace#1 #2{%
 \ifx\relax#2%
 \xlinkdash#1-\relax
 \else
 \xlinkdash#1 -\relax
 \expandafter\xlinkspace\expandafter#2%
 \fi}

\def\xlinkdash#1-#2{%
 \ifx\relax#2%
 \tmp{#1}%
 \else
 \tmp{#1-}%
 \expandafter\xlinkdash\expandafter#2%
 \fi}
 
\newcommand{\ms}{M$_{\odot}$}
\newcommand{\zs}{Z$_{\odot}$}

\newcommand\nuk[2]{$\rm ^{\rm #2} #1$}
\newcommand{\ciso}{$\rm ^{12}C/^{13}C$}

\title[Evolution with rotating star yields]{Chemical evolution with rotating massive star yields \\
I. The solar neighborhood and the s-process elements }

\author[Prantzos et al.]{
N. Prantzos,$^{1}$\thanks{E-mail: prantzos@iap.fr}
C. Abia,$^{2}$
M. Limongi,$^{3,4}$
A. Chieffi,$^{5,6}$
S. Cristallo$^{7,8}$
\\
$^{1}$Institut d'Astrophysique de Paris, UMR7095 CNRS, Univ. P. \& M. Curie, 98bis Bd. Arago, 75104 Paris, France\\
$^{2}$Departmento de F\'\i sica Te\'orica y del Cosmos, Universidad de Granada, E-18071 Granada, Spain\\
$^{3}$
Istituto Nazionale di Astrofisica - Osservatorio Astronomico di Roma, Via Frascati 33, I-00040, Monteporzio Catone, Italy\\
$^{4}$
Kavli Institute for the Physics and Mathematics of the Universe, Todai Institutes for Advanced Study, the University of Tokyo, \\
  Kashiwa, Japan 277-8583 (Kavli IPMU, WPI)\\
  $^{5}$
Istituto di Astrofisica e Planetologia Spaziali, INAF, via Fosso del cavaliere 100, 00133 Roma - Italy \\
$^{6}$
Monash Centre for Astrophysics (MoCA), School of Mathematical Sciences, Monash University, Victoria 3800, Australia\\  
$^{7}$
Istituto Nazionale di Astrofisica - Osservatorio Astronomico d'Abruzzo, Via Maggini snc, I-64100, Teramo, Italy\\
$^{8}$
Istituto Nazionale di Fisica Nucleare - Sezione di Perugia, Via Pascoli, I-06123, Perugia, Italy
}

\date{Accepted XXX. Received YYY; in original form ZZZ}

\pubyear{2017}

\hypersetup{draft}

\begin{document}

\label{firstpage}
\pagerange{\pageref{firstpage}--\pageref{lastpage}}
\maketitle

\begin{abstract}
We present a comprehensive study of the abundance evolution
of the elements from H to U in the Milky Way halo and
local disk.  
We use  a consistent chemical evolution model, metallicity dependent isotopic yields from low and intermediate mass stars and yields from massive stars which include, for the first time, the combined effect of metallicity, mass loss and rotation for a large grid of stellar masses and for all stages of stellar evolution. The  yields of massive stars are weighted by a metallicity dependent function of the rotational velocities, constrained by observations as to obtain a primary-like $^{14}$N behavior at low metallicity and to avoid overproduction of s-elements at intermediate metallicities. We show that the solar system isotopic composition can be reproduced to better than a factor of two for isotopes up to the Fe-peak, and at the 10\% level for most pure s-isotopes, both light ones (resulting from the weak s-process in rotating massive stars) and the heavy ones (resulting from the main s-process in low and intermediate mass stars). We conclude that the light element primary process (LEPP), invoked to explain the apparent abundance deficiency of the s-elements with A$ < 100$, is not necessary. We also reproduce the evolution of the heavy to light s-elements abundance ratio ([hs/ls]) -  recently observed in unevolved thin disk stars - as a result of  the contribution of rotating massive stars at sub-solar metallicities. We find that those stars produce primary F and dominate its solar abundance and we confirm their role in the observed primary behavior of N. In contrast, we show that their action is insufficient to explain the small observed values of \ciso \ in halo red giants, which is rather due to internal processes in those stars.

\end{abstract}




\section{Introduction}
\label{sec:Intro}

In recent years, progress in our understanding of
the chemical evolution of the Milky Way came largely from
observations concerning the composition of stars in the halo and
the local disk. Several ongoing large spectroscopic surveys 
such as Gaia ESO, SEGUE, APOGEE, HARPS, RAVE or GALAH 
\citep{Gil12,Yan09,Wil10,Ste03,Hei12,Adi12,Bensby2014,Battistini2016,Del17}, are improving 
our understanding of the Galactic disk structure and its chemical evolution.
Probably one of the most significant results of these abundance surveys, when
combined with information on stellar kinematics and ages, 
is the existence of a different abundance pattern between thin and thick disk stars
regarding the alpha-elements (O, Si, Mg, Ca etc.),
i.e. the [$\alpha$/Fe] vs. [Fe/H]\footnote{In this paper the notation [X/Y] has the
usual meaning, [X/Y]$= \rm{log (X/Y) - log (X/Y)}_\odot$, where X (or Y) is the abundance by number of element X(Y).} ratios. In parallel, observations of halo stars with large scale surveys \citep{Cay04,Fre10,Yong2013,Roederer2014}
confirmed the constantly high [$\alpha$/Fe] ratio in low metallicities, and revealed a small dispersion for element ratios [X/Fe] up to the Fe-peak and the presence of a large dispersion in that ratio for heavier than Fe elements.

The interpretation of these data is not straightforward, however,since it has to be made in the framework of some appropriate model of galactic chemical  evolution (GCE). In general, GCE model predictions are hampered by our limited knowledge of the main ingredients: the initial mass function (IMF) and the star formation rate (SFR), the gaseous flows (infall and outflow), stellar migration and - last, but not least - the stellar yields.

Considerable progress in GCE studies has been made possible after
the publication of  yields from massive stars  (hereafter MS) for an extensive grid of isotopes (H to Zn), stellar masses and metallicities \citep{Woo95,LSC00,CL04,LC12,Chi13,UN02,NTUKM06,HW10,Nomoto2013}. 

The stellar models of the widely used yields of \cite{Woo95} and \cite{NTUKM06} do not include mass loss, but this ingredient was shown to affect in an important way the yields of several relatively light elements, like He, C, N, O, Ne and their isotopes (see e.g. \citealt{Maeder1983,Maeder1992} and references therein), as well as the isotopes produced by the so called "weak s-process" (\citealt{Prantzos1987}). The role of mass loss appeared to be important for stars of large masses ($>$25 \ms) and metallicities ($\sim$\zs), because radiation pressure is insufficient to efficiently remove the stellar envelope in lower stellar masses and/or metallicities. 

It was subsequently shown that rotation affects the yields of massive stars either directly and indirectly. Directly because a) the mixing induced by the combined effects of meridional circulation and secular shear brings in contact nuclear species that otherwise would remain well separated and b) it affects the size of the various convective regions (core and shells) changing therefore the physical evolution of a star. Indirectly because the inclusion of rotation alters significantly the surface properties of most of the stellar models, especially at subsolar metallicities, pushing them towards conditions where they lose an enormous amount of mass that would not be lost in absence of rotation.   For instance, \cite{Hirschi2007} finds that mixing of metals to the surface of a low metallicity (Z=10$^{-8}$) star of 85 \ms \ triggers mass loss of $\sim$65 \ms. Also, as the surface rotational velocity approaches the critical one,  the mass loss rate is largely enhanced (the "mechanical" wind discussed, e.g., by \citealt{Maeder2012} and references therein) but this phenomenon plays an important role only if the surface velocity gets very close to the critical one (more than 90\%). In the present set of models of massive stars such a phenomenon plays a minor role because the surface rotational velocity never exceeds 60\% or so of the critical one.

Several of the - potentially important - effects of rotating massive stars on GCE  are summarized  in \citet{Maeder2015}:

i) production of large amounts of N at low metallicity, from {\it both} rotating AGB and massive stars, explaining the observed primary behaviour of N in the Galactic halo \citep{Chiappini2006}; 

ii) production of quasi-primary \nuk{C}{13} at very low metallicities by massive stars, helping to understand the low \ciso \ ratio observed in halo stars {\citep{Chiappini2008}; 

iii) production of Galactic Cosmic Rays (GCR) mainly from the accelerated winds of massive stars, explaining the observed GCR excess of $^{22}$Ne \citep{Prantzos2012a} and helping to understand the observed primary behavior of spallogenic Be \citep{Prantzos2012b}; and 

iv) production of substantial amounts of "light s-nuclei" - resulting from the weak s-process in massive stars - which may help to understand the large dispersion of the "light/heavy" s-element ratio in halo stars \citep{Ces13}.

It should be emphasized that the aforementioned effects were not always studied with  the same set of yields: effects (i) to (iii) were studied with yields mainly from \cite{Hirschi2007}, while effect (iv) with yields from  Frischnecht (2011, PhD Thesis), \cite{Frischknecht2012}  and \cite{Fri16}, where stars of different masses, metallicities and rotational velocities were considered. Furthermore, none of the adopted stellar models was calculated to the final stage of stellar evolution and the subsequent explosion, neither were the yields of AGB stars properly considered. Finally, the issue of the overall validity of the adopted stellar yields to reproduce the key observable in GCE studies, namely the detailed elemental and isotopic composition of the proto-solar nebula, was never studied.

The latter point is of particular importance for
the study of the s-elements and their isotopes.
It is widely accepted that the main s-component, accounting for the s-process isotopic
distribution in the atomic mass range $90<$ A $<208$, occurs in low and intermediate mass stars (hereafter LIM) (M$\loa 8$ M$_\odot$) during their thermally pulsing asymptotic giant branch phase (TP-AGB; see \citealt{Bu99} and Section 2.3), where neutrons are mainly provided by the $^{13}$C($\alpha$,n)$^{16}$O reaction. The weak s-component, responsible for a major 
contribution to the s-process nuclides up to A$=90$, has been recognized as
the result of neutron-capture synthesis mainly during core He- and shell C-burning
phases of massive stars (hereafter MS, M$\goa 10$ M$_\odot$) \citep{Arn85,Pra90,Rai91,Pignatari2010} with 
the reaction $^{22}$Ne$(\alpha, n)^{25}$Mg as the major neutron source (see Section 2.2 below for details). Finally, the role of the strong s-component, introduced by \citet{Cla67} in order to reproduce more than $50\%$ of solar
$^{208}$Pb, has been demonstrated to be played by low metallicity and low mass 
($\leq 1.5$ M$_{\odot}$) AGB stars (\citealt{Ga98}, see also \citealt{Kap11} for  a recent review).

Previous GCE studies of the s-element evolution in the Milky Way were based
on grids of yields poorly sampled in stellar masses and metallicities, obtained
by post-processing nucleosynthesis calculations and/or just including either one of the possible stellar s-element sources (LIM stars or MS), and adopting an {\it ad hoc} contribution from the other source \citep{Tra04,Ces13}. Full (coupled) stellar evolutionary models and nucleosynthesis post-process calculations in LIM stars have shown the extreme sensitivity on the initial stellar metallicity of the s-process, namely to the ratio of the seed-nuclei (mainly Fe) to free neutrons (see e.g. \citealt{Kap11,Cr15}). Detailed calculations in massive stars also show this trend with metallicity, with an additional important contribution from rotationally induced mixing.

It goes without saying that the efficiency of the s-process is critically dependent also on the stellar mass  and, if rotation is taken into account, on the efficiency of the rotation-induced mixing (see Sections 2.2 and 2.3 below). Therefore, reliable GCE studies for these elements need the use of a grid of stellar yields as complete as possible in mass, metallicity and initial rotation velocity.

In this study we reassess the chemical evolution of "light" (up to the Fe-peak) and heavy (s-process) elements
in the Milky Way by using a new grid of stellar yields from LIM and
massive stars, covering a wide range of stellar masses and metallicities.
These yields also include the impact of stellar rotation in massive stars
for different rotation rates. We adopt an empirically constrained  metallicity-dependent weighted average for those yields, favoring faster rotation at low metallicities. We consider the ensemble of stable isotopes from H to U, in order to check the behavior of the adopted set of yields against all available observations. We use appropriate models for local Galaxy,  reproducing satisfactorily the main observational constraints. We put special emphasis on the comparison between predicted and observed abundances (isotopic and elemental) of the s-elements at solar system formation.  We specifically  assess the impact of our rotating MS yields on the evolution of nitrogen, \ciso \ and s-elements,  which are suggested to be "smoking guns" of rapidly rotating massive stars by \cite{Maeder2015}, as already mentioned.  We show that fluorine may be added to this group of observables, while the \ciso \ ratio should rather be dropped, as mostly affected by internal stellar processes.

The structure of the article is the following: in Section 2 we describe our GCE model and the  sets of yields from LIM stars and MS, as well as the adopted metallicity-dependence for the weighted (over rotational velocities) average yields. The results for  the isotopic and elemental abundances obtained at  solar system formation, as well as the predicted   abundance trends with metallicity are discussed in Section 3. Finally, we present our conclusions in Section 4.

\section{Model and stellar yields}
\label{sec:Model}

\subsection{The model}
\label{subsec:model}

Our simple, one-zone model, is based on \citet{Goswami2000}, as updated in \citet{Kubryk2015a} (hereafter KPA2015a). We assume  
that the local disk is built by infall of gas at an exponentially decreasing rate and a characteristic time-scale of 10 Gyr, where the star formation rate $\Psi$ is given by a Schmidt-Kennicutt law in both sub-systems:  
\begin{equation}
\Psi(t) \ = \alpha \ \Sigma_G(t)^{1.5}
\end{equation}
where $\Sigma_G$ is the  local gas surface density and the coefficient $\alpha$ is chosen as to obtain a gas fraction of $\sim$20\% at the end of the simulation (see Appendices A and B in KPA2015a for details on star formation and gas and star amounts in the Milky Way). We are fully aware that the adopted model reflects poorly the physical processes in both the halo and the disk (see our criticism in Sec. 2.1 and 3.4 concerning the hierarchical merging scenario for the halo and the thin/thick disk issue, respectively) but it is sufficient for our purpose, since our main concern  here is to test the implications of the new grid of stellar yields from rotating massive stars.

The chemical evolution code is described in detail in KPA2015a (see their Sec. 2.4 and Appendix C). Here we adopt the metallicity-dependent stellar life-times $\tau(M, Z)$ of \citet{Cr15} for stars in the mass range 1-7 M$_\odot$, 
and those from Limongi \& Chieffi (2018) (hereafter LC2018) for M $> 7$ M$_\odot$. We adopt the stellar initial mass function (IMF) of \cite{Kro02} in the mass range 0.1-120 \ms. 
Chemical evolution is calculated with the Single Particle Population (SSP) method.
We  use the metallicity-dependent  yields of \citet{Cr15} for LIM stars and of LC2018 for the massive ones,  which include  mass loss and rotation (see next sections). The latter include the yields of the final stellar explosion, but not those concerning the proton- and neutron- rich nuclei, produced by the p- and r-process, respectively. 

Since we are interested here on both isotopic and elemental evolution and since most heavy elements have a mixed origin, we adopt fiduciary yields for the r-isotopes.  Although core collapse supernovae (CCSN) have long been considered as the main site of the r-process, detailed nucleosynthesis studies in those objects have failed up to now to account satisfactorily for the production of the full range of  r-process elements  (see e.g. \citealt{wan13}). Alternative scenarios has been suggested, as  neutron stars mergers (NSM) and/or neutron-star-black-hole pairs \citep{Lat77,Fre99,Ros14,Metzger2010,Kasen2013,dro17}. The NSM scenario is given support by the recent joint detection of electromagnetic and gravitational signal from the $\gamma$-ray burst  GW170817/GRB170817A (\citealt{Pian2017} and references therein), but there is still no consensus on the role of that class of objects  in the production and evolution of r-elements (see e.g. \citealt{Ishimaru2015} and references therein).} For illustration purposes,
we assume here that they are produced in core collapse supernovae (CCSN) and their  yields for a star of mass $M$ and metallicity $Z$ are scaled to the yield of oxygen $Y_{^{16}\rm O}(M,Z)$:
\begin{equation}
Y_{r,i}(M,Z) \ =  Y_{^{16}\rm O}(M,Z) \ \frac{X_{r,\odot}}{X_{^{16}\rm O,\odot}} \ f_{r,i}
\label{eq:yields_r}
\end{equation}
where $f_{r,i}$ is the r-fraction of isotope $i$ in the proto-solar system \citep{Sne08}, and $X_{r,\odot}$ are the corresponding solar abundances. The choice of $^{16}$O,  produced exclusively by massive stars, as reference isotope ensures that if its solar abundance is well reproduced in the simulation, so will be the r-fractions of heavy elements. This will allow us to study the behavior of the other isotopes (of mixed origin), as well as the behavior of the elements and to constrain the adopted s-element yields.

We include all 285 stable isotopic species from H to U. For the few of them with lifetimes shorter than or comparable to the age of the Universe (\nuk{K}{40}, \nuk{Th}{232}, \nuk{U}{235,238}),  we take into account their radioactive decay within long lived stars and in the ISM. We calculate the evolution of those species and we sum up at each time step to obtain the corresponding evolution of their elemental abundances. We note that the use of the yields in GCE  calculations requires  interpolation in the mass range of the super-AGB stars and low mass CCSN (from $\sim$7 to 12 \ms, see Sec. 2.4), where no complete grids of yields are available; \cite[see, however,][for a recent study on super-AGB stars]{Doherty2014}. In contrast, no extrapolation in the high mass range is required, since the LC2018 yields go as high as 120 \ms.  We ensure that the sum of the ejected masses of all isotopes of a star to be equal to the original stellar mass minus the one of the compact residue (white dwarf, neutron star or black hole). This is important in order to ensure mass conservation in the system during the evolution. We include a detailed treatment for the production of the light nuclides Li, Be and B by spallation of CNO nuclei by cosmic rays as described in \citet{Prantzos2012b}.

For the rate of thermonuclear supernovae (SNIa) we adopt a semi-empirical approach: the observational data of recent surveys concerning the Delayed Time Distribution (DTD) are described well by a power-law in time, of the form $\propto t^{-1}$ (e.g. \citet{Mao12} and references therein). At the earliest times, the DTD is unknown/uncertain, but a cut-off must certainly exist before the formation of the first white dwarfs ($\sim$35-40 Myr after the birth of the stellar population). We adopt then the formulation of \citet{Gre05} for the single-degenerate (SD) scenario of SNIa. That formulation reproduces, in fact, the observations up to $\sim$4-5 Gyr quite well. For longer timescales, where the SD scenario fails, we simply adopt the t$^{-1}$ power law. Overall, our approach leads to 1.3 SNIa per 1000 \ms \ of stars formed; see Appendix C in \cite{Kubryk2015a}. As in \citet{Goswami2000} we adopt the SNIa yields of \citet{Iwa99} for Z=0 and Z=\zs , interpolating logarithmically in metallicity between those values.

\subsection{Massive stars}
\label{subsec:Masta}

\subsubsection{The stellar models}
\label{subsub:stellar_model}

The yields of massive stars used in this paper are based on a grid of models in the mass range $\rm 13-120~M_\odot$ and initial metallicities corresponding to [Fe/H]$=0, -1, -2, -3$. For each metallicity we computed models for three initial rotational velocities, namely v$_{\rm rot}=0,~150,~300$ kms$^{-1}$. These initial velocities were chosen in order to span the possible range of observed values \citep{duftonetal06,hunteretal08,ram17}. The adopted solar chemical composition is the one provided by \protect\citet{Asp09}, which corresponds to a total metallicity $\rm Z_\odot=0.01345$. At metallicities lower than solar we assume a scaled solar distribution for all the elements, except for C, O, Mg, Si, S, Ar, Ca, and Ti for which we adopt an overabundance with respect to Fe derived from the observations of unevolved low metallicity stars, i.e., [C/Fe]=0.18, [O/Fe]=0.47, [Mg/Fe]=0.27, [Si/Fe]=0.37, [S/Fe]=0.35, [Ar/Fe]=0.35, [Ca/Fe]=0.33, [Ti/Fe]=0.23 \protect\citep{Cay04,Spi05}. As a result of these enhancements, the total metallicities corresponding to [Fe/H]$=-1,-2,-3$ are $\rm Z=3.236\times 10^{-3},~3.236\times 10^{-4},~3.236\times 10^{-5}$, respectively.

All models used in this work have been computed with the same code and input physics (including all nuclear reaction rates) described in detail in our previous paper \citep{Chi13}. With respect to that code, however, two changes have been included, i.e. the mass loss triggered by the formation of dust and the mass loss triggered by the condition $\rm L>L_{\rm Eddington}$. 

The nuclear network chosen for this work includes 335 isotopes in total, from H to \nuk{Bi}{209} and is suited to properly follow all the stable and explosive nuclear burning stages of massive stars. The portion of the network including isotopes from H to \nuk{Mo}{98} takes into account all the possible links among the various nuclear species due to weak and strong interactions. For nuclei heavier than \nuk{Mo}{98} we consider only $(n,\gamma)$ and $\beta$-decays. Since in this paper we are mainly interested in following in detail the flux of neutrons through all the magic number bottlenecks and since in the neutron capture chain the slowest reactions are the ones involving magic nuclei, between \nuk{Mo}{98} and \nuk{Bi}{209} we explicitly follow, and include in the nuclear network, all the stable and unstable isotopes around the magic numbers corresponding to N$=82$ and N$=126$ and assume all the other intermediate isotopes at local equilibrium. In total $\sim 3000$ nuclear reactions are included in the various nuclear burning stages. 

 A crucial choice that must be made when computing models with rotation, regarding in particular the rotationally induced mixing, concerns the two free parameter $f_c$ and $f_{\mu}$ that control the efficiency of the stirring of matter in presence of rotation\footnote{We refer the interested reader to \citet{Chi13} for a comprehensive and detailed discussion of the two instabilities, meridional circulation and shear, and the two free parameter that are currently adopted in the FRANEC code.}. As already discussed extensively in \citet{Chi13} and references therein, the efficiency of the rotation driven mixing cannot be determined on the basis of first principles because, similarly to what happens for the thermal instabilities (convection), it is intrinsically a multidimensional physical phenomenon. In analogy to the mixing length parameter, that requires a calibration, the rotation driven mixing also requires a proper calibration of the two free parameters, $f_c$ and $f_{\mu}$. The solar metallicity models published in \cite{Chi13} were obtained by adopting $f_c=1$ and $f_{\mu}=0.03$. This choice was quite crude, in the sense that the value $f_c=1$ was (arbitrarily) chosen and $f_{\mu}$ was fixed then by requiring that a solar metallicity star of 20 \ms \ and initial rotation velocity of 300 km s$^{-1}$ increases its surface N abundance by roughly a factor of three. A similar approach was adopted also by, e.g., \citet{Heg00}. 

In the present case, a better calibration is adopted: the models  fit the main trend of N abundance versus initial rotational velocity observed in a sample of stars taken from the FLAMES survey of the LMC \citep[the so called Hunter diagram,][] {Hun09}, as originally done by \citet{Bro11}. The best values necessary to fit the FLAMES data with the new criterion are $f_c=1.5$ and $f_{\mu}=0.01$ and they are adopted in the present set of models. The reader should be aware that the efficiency of the rotation driven mixing heavily depends on these parameters: different choices may easily lead models computed with the same initial rotational velocity (and the same evolutionary code) to mix significantly more or significantly less than obtained in the present grid.

A paper  presenting the physical properties of these models, i.e. the evolution of the surface properties (HR diagram and surface abundances), the timescale of the various burning stages, the final fate of each model, the evolution of the angular momentum, along with the tables of the yields used in this paper is in preparation (Limongi and Chieffi 2018, hereafter LC2018)\footnote{The impatient reader may already download the yields from the repository: http://orfeo.iaps.inaf.it.}.

\subsubsection{Impact of rotation on the yields}
\label{subsub:yields_mas}

Since an important outcome of the present models is the synthesis of neutron rich nuclei up to Pb at low metallicities, we think to be useful to briefly remind the sequence of events that lead to Pb. In central He burning rotation driven mixing continuously brings matter from the He convective core up to the H burning shell and vice versa. Such an engine brings fresh carbon synthesized by the $3\alpha$ reactions up to the base of the H shell where it is quickly converted in \nuk{N}{14} that is then brought back towards the center where it is rapidly converted in \nuk{Ne}{22}, i.e. in a powerful primary neutron source. An easy way to quantitatively determine the amount of primary \nuk{Ne}{22} produced by the rotation driven mixing is to use the parameter $\chi({\rm N,Mg})=X({\rm ^{14}N})/14+X({\rm ^{18}F})/18+X({\rm ^{18}O})/18+X({\rm ^{22}Ne})/22+X({\rm ^{25}Mg})/25+X({\rm ^{26}Mg})/26$. In a non rotating massive star this quantity remains basically constant in central He burning because the amount of $\rm ^{14}N$ does not increases any more after the central H exhaustion and its burning goes only into $\rm ^{22}Ne$ first and $\rm ^{25}Mg$ and $\rm ^{26}Mg$ later. On the contrary, in a rotating massive star this number increases because of the continuous ingestion of fresh primary $^{14}$N produced in the tail of the H shell. Therefore its variation during core He burning provides a good quantitative estimate of the primary $^{22}$Ne produced. As an example, a 20 $\rm M_\odot$ star with initial rotation velocity v$_{\rm {ini}}=300$ kms$^{-1}$, the quantity $[\chi({\rm N,Mg})_{\rm end~He}-\chi({\rm N,Mg})_{\rm start~He}]/\chi({\rm N,Mg})_{\rm end~He}$ is equal to 0.422, 0.865, 0.975 and 0.995, for the four metallicities [Fe/H]$=0, -1, -2$ and $-3$, respectively. These values clearly show that the primary component of $\rm ^{22}Ne$ dominates the total \nuk{Ne}{22} abundance at metallicities $\rm [Fe/H] \leq -1$. Such a large increase of the neutron source in presence of rotation, is obviously not counterbalanced by a similar increase of the neutron seed (mainly Fe) so that in presence of rotation the neutron to seed ratio tends to increase. In particular it scales directly with the rotational velocity and inversely with the initial metallicity.

An environment with a large neutron to seed ratio favors a consistent production of heavy nuclei up to Pb. \cite{Ga98} were the first to demonstrate that the s-process nucleosynthesis in low mass stars may lead to a large production of Pb in presence of a primary neutron source at low metallicity (i.e. for high neutron to seed ratios). The results of LC2018 show that rotating massive stars produce similarly favorable conditions for the synthesis of heavy nuclei up to Pb at low metallicities. At solar metallicity the amount of "neutron poisons" (nuclei lighter than Fe) is too high to allow the synthesis of nuclides beyond the first neutron closure shell (the so called s-weak component). But as the initial metallicity drops, the neutron to seed ratio increases and the production of Pb raises. 

The continuous migration of matter from the He convective core to the base of the H burning shell and vice versa has other very interesting consequences other than an important primary production of \nuk{Ne}{22}. In fact, the \nuk{C}{12} brought in the H shell does not produce just \nuk{N}{14} but, obviously, all the nuclei involved in the CNO cycle: in particular \nuk{C}{13}, \nuk{N}{15} and \nuk{O}{17}. In addition, it is also obvious that, as time goes by, the radiative part of the He core (i.e. the region between the convective core and the H shell) progressively accumulates the local abundances of all these nuclei. Hence the He core progressively enriches in these three nuclei \nuk{C}{13}, \nuk{N}{15} and \nuk{O}{17} (plus obviously \nuk{N}{14} and \nuk{Ne}{22}). The simultaneous presence of large abundances of \nuk{N}{15}, \nuk{C}{13} and \nuk{N}{14} favors the synthesis of \nuk{F }{19} that, in fact, is significantly produced later on when the He convective shell forms.

\begin{figure}
\centering
\includegraphics[width=0.45\textwidth]{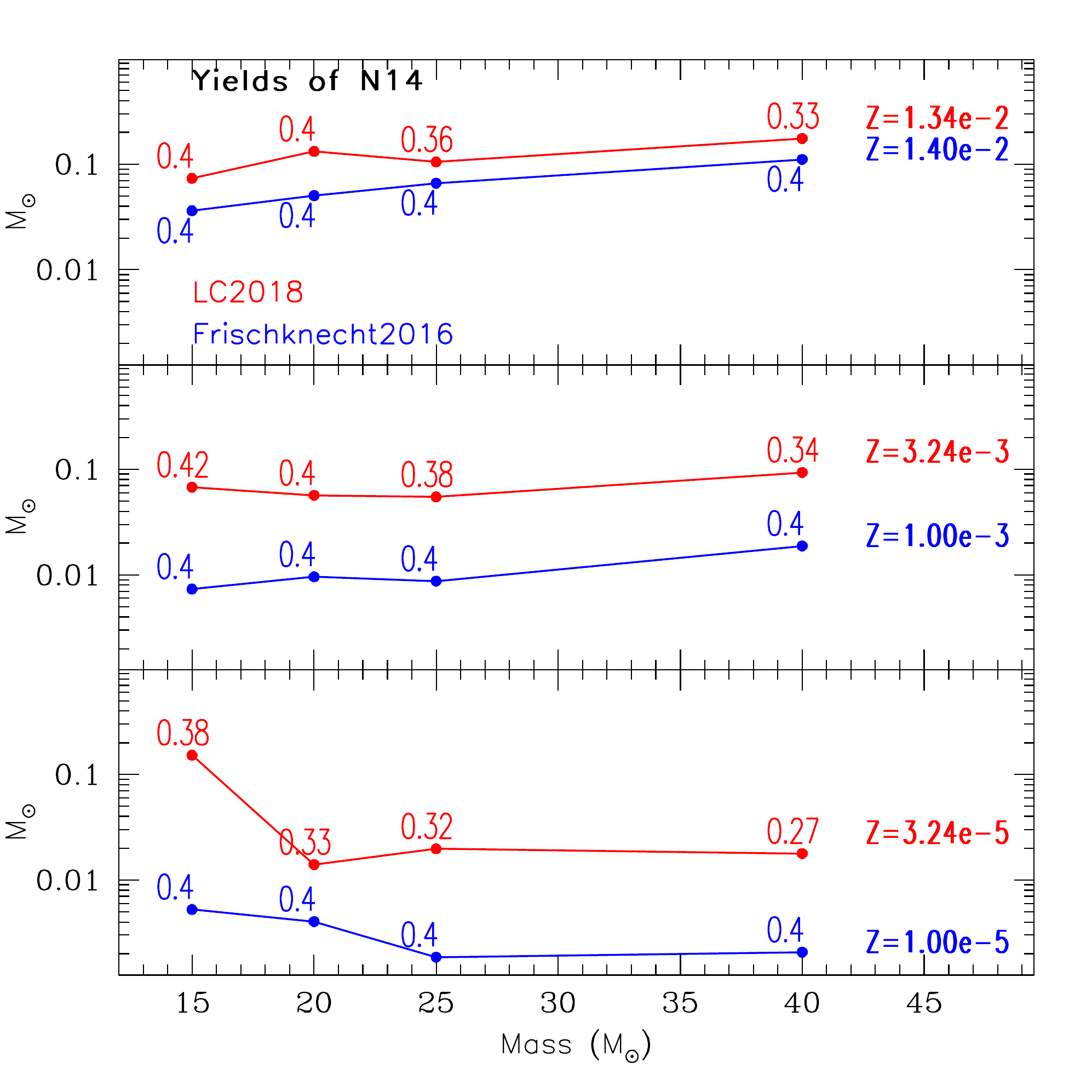} 
\caption{Ejected masses of $\rm ^{14}N$ as a function of the initial stellar mass.
Red lines and data refer to the LC2018 results and blue lines and data  to those obtained by \citet{Fri16} for the corresponding metallicity values indicated on the right of each panel. The numbers corresponding to each mass refer to the values of  v$_{\rm ini}/$v$_{\rm crit}$ ratio (0.4 for all the models of \citet{Fri16} selected for this comparison).
\label{fig:compn14}}
\end{figure}

As an example of the different results that may be obtained by different groups, we show in Figure \ref{fig:compn14} a comparison between the yields of \nuk{N}{14} as a function of the progenitor mass obtained by LC2018 and by \citet{Fri16}. The comparison shows that at solar metallicity the $\rm ^{14}N$ yields obtained by \citet{Fri16} are lower than those provided by our models by a factor of $\sim$ 2, on average. This difference increases to about an order of magnitude at lower metallicities. The reason of this difference is  most probably  due to to the differences in the general treatment of rotation and in particular to a different calibration. Note also that while our models have been computed for a fixed initial rotation velocity, \citet{Fri16} assume an initial rotation velocity corresponding to a constant v$_{\rm ini}/$v$_{\rm crit}$ ratio, which means that their initial rotation velocity is a function of the initial mass and metallicity and not flat. A more detailed comparison between the yields of the s-process elements obtained by the two groups goes beyond the scope of the present study but will be presented in LC2018.

Finally, we note that besides the issue of rotational mixing,  the cross sections of some key nuclear reactions constitute another source of uncertainty affecting the production of the s-process elements. For example, the recent study of \citet{Choplin2017} shows that the production of the s-process elements depends on the still largely uncertain cross section of the $\rm ^{17}O(\alpha,\gamma)^{21}Ne$ reaction, for which the LC2018 study  adopts the value provided by \citet{CF1988}.
It is beyond the scope of this study to perform a detailed investigation of that issue, but one should certainly keep it in mind.

\subsubsection{The explosive nucleosynthesis}
\label{subsub:expnuc}

The computation of an artificially induced explosion requires, a calibration of the amount of energy to inject in the deep interior of the star to trigger the explosion. The most adopted calibrations fix either the kinetic energy of the ejecta or the amount of \nuk{Ni}{56} that must be ejected. A few recent sets of explosions assume that the neutrino flux deposits a fraction of its energy in the star before escaping and this fraction is determined by requiring the reproduction of the key properties of the SN1987A.

In our previous sets of models the explosions were calibrated by requiring that each star, independently on the initial mass, ejects 0.1 \ms \ of \nuk{Ni}{56}. Such a choice, however, implies a steep increase of the kinetic energy of the ejecta with the initial mass because of the large increase of the binding energy of a star with the mass. In the last years there has been a quite general convergence towards the idea that stars more massive than 25 M$_\odot$ or so actually fail to explode and fully fall back in the remnant. The reasons for this are both observational and theoretical: on the observational side, \cite{pp15} found that the kinetic energy of the ejecta in a sample of Type IIP supernovae never exceeds 3 foes while on the theoretical one \cite{suck16}, \cite{oo11} and \cite{ertl16} find that stellar models more massive than $25-30$ \ms \ fail to explode (even if some massive stars -randomly distributed in mass- explode due to a specific overlap of the convective shells in the advanced burning phases). 

In the present set of models we adopt a similar calibration, i.e. we assume that all star more massive than 25 M$_\odot$ fully collapse in the remnant (failed supernovae) and therefore contribute to the chemical enrichment only through the wind. This procedure has been applied at all metallicities and initial rotational velocities. Star in the range 13 to 25 \ms \, vice versa are calibrated by taking into account the "mixing \& fall back" mechanism proposed by \citet{UN02} \citep[see also the discussion in][]{Nomoto2013} to explain the abundance pattern of some iron peak elements in extremely metal-poor stars.
In the present models the inner border of the mixed region is fixed by requiring that [Ni/Fe]$=0.2$ while the outer one is fixed at the base of the O burning shell. The final mass of the remnant is then determined by requiring the ejection of 0.07 $\rm M_\odot$ of $\rm ^{56}Ni$. This procedure has been adopted for each mass in the range 13-25 $\rm M_\odot$ and for all metallicities and initial rotational velocities. Different calibrations of the explosions are obviously possible and may be provided upon request.

\subsubsection{Rotation versus metallicity}
\label{subsub:MasStarYld+Rot}

\begin{figure}
\begin{centering}
\includegraphics[width=0.49\textwidth]{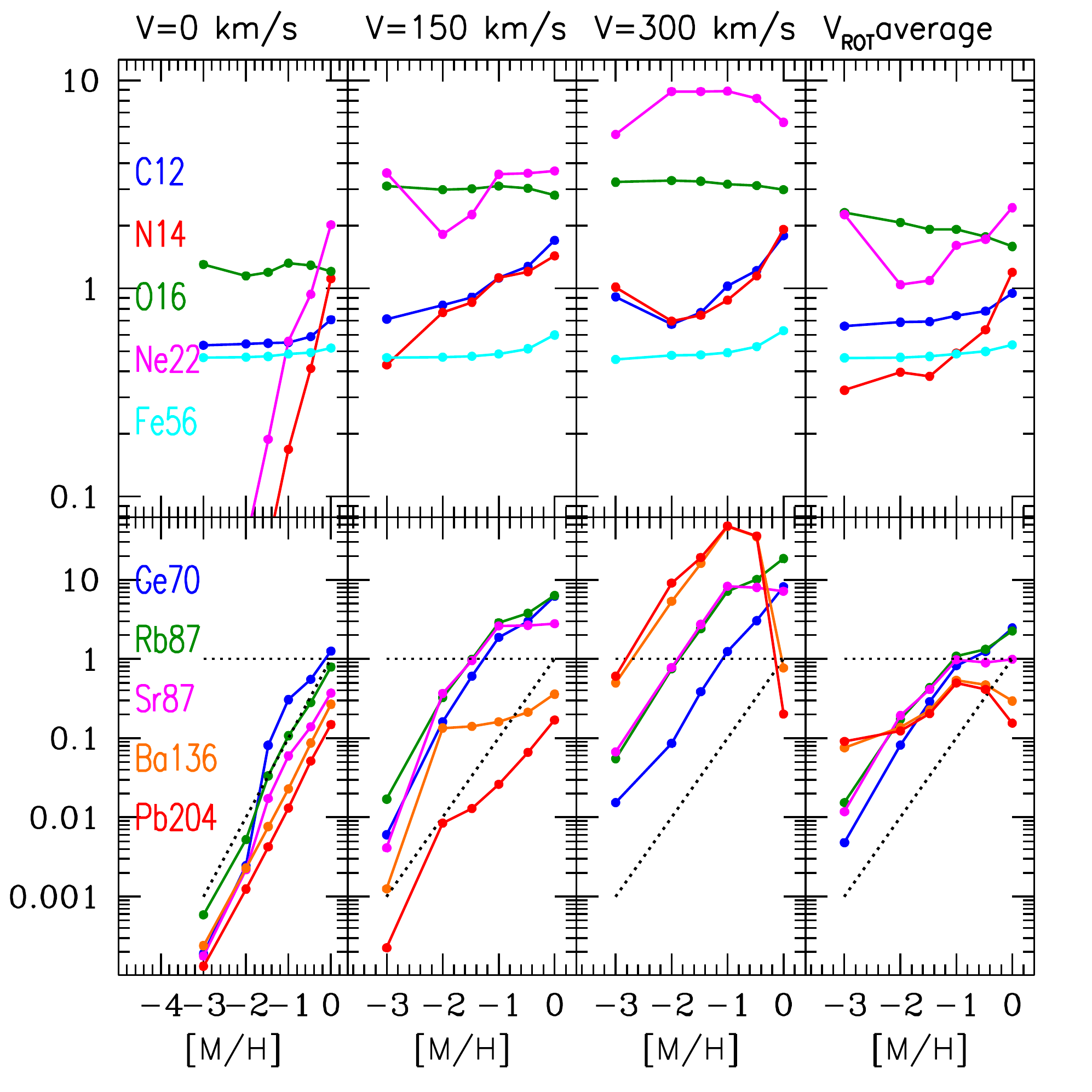} 
\end{centering}
\caption{Selected isotopic yields of massive stars as a function of stellar metallicity. They are total yields, integrated over the IMF of \protect\citet{Kro02} between 13 and 120 M$_\odot$ (Eq. \ref{eq:yldmas}) and normalized to the corresponding proto-solar abundances. Yields are displayed for three different initial rotational velocities (0, 150 and 300 kms$^{-1}$, from left to right). In the rightmost panel those yields are weighted by a metallicity dependent function of the rotational velocities (see text and Fig. \ref{fig:Vrot}); it is those weighted yields that are used in this work. In the bottom panels, the dotted lines indicate the behavior of a pure primary element ({\it horizontal, slope=0}) and a pure secondary one ({\it slope=1}).}
\label{fig:YldMassive}
\end{figure}

Selected yields of massive stars are displayed in Fig. \ref{fig:YldMassive} as a function of metallicity for the three aforementioned initial rotational velocities of stars. They are {\it total yields} including the part of the species originally introduced in the star at its formation and re-ejected at the end of its life; that contribution increases with stellar metallicity. They are integrated over the adopted IMF and  normalized to the corresponding pre-solar abundances $Y_{i,MS}$/X$_{i,\odot}$, where:
\begin{equation}
Y_{i_,MS} \ = \ \frac{1}{1-R} \ \int _{13} ^{120} y_i(M) \Phi(M) \ dM 
\label{eq:yldmas}
\end{equation}  
being $R$ the return mass fraction $\sim$0.42 for the adopted IMF, $y_i(M)$ the yields of individual stars of mass $M$, and $\Phi(M)$ the IMF.

Top panels display the trend with the metallicity of  $^{12}$C, $^{14}$N, $^{16}$O, $^{22}$Ne and $^{56}$Fe ($^{56}$Ni), while bottom panels display s-only nuclei from the first, second and third abundance peaks. The $\alpha$ elements show, as expected, the typical behavior of primary nuclei, i.e. a negligible dependence on the initial [Fe/H] at all rotational velocities. The main effect of rotation on these nuclei is that of increasing (on average) their yields because of the larger He core masses induced by the rotationally driven instabilities.  Note that the yield of $^{56}$Fe is constant by construction because, as mentioned above, all stars with mass $\rm M\leq25~M_\odot$ are assumed to eject the same amount of $^{56}$Ni. 

The trends of $^{14}$N and $^{22}$Ne with the initial [Fe/H], turn from a typical secondary behavior (in the non rotating case) to a typical primary trend (in the rotating case) because of the robust primary production of these two nuclei triggered by the rotation driven mixing (see the exhaustive discussion in Sections \ref{subsub:stellar_model} and \ref{subsub:yields_mas}). All these features are promptly visible in the top panels of Fig. \ref{fig:YldMassive}. The switch of both \nuk{N}{14} and \nuk{Ne}{22} from a secondary to a primary behavior may explain very naturally and without ad hoc assumptions, both the primary behavior of \nuk{N}{14} and a considerable production of s-process nuclei (up to Pb) at very low metallicity. 

The impact of rotation on the yields of s-only nuclei can be seen in the bottom panel of Fig.  \ref{fig:YldMassive}, where we plot a few selected yields as function of metallicity. The s-only isotopes\footnote{Our s-only list includes: $^{70}$Ge, $^{76}$Se, $^{80,82}$Kr, $^{86,87}$Sr, $^{96}$Mo, $^{100}$Ru, $^{104}$Pd, $^{110}$Cd, $^{116}$Sn, $^{122,123,124}$Te, $^{128,130}$Xe, $^{134,136}$Ba, $^{142}$Nd, $^{148,150}$Sm, $^{152,154}$Gd, $^{160}$Dy, $^{170}$Yb, $^{176}$Lu, $^{176}$Hf, $^{186}$Os, $^{192}$Pt, $^{198}$Hg, and $^{204}$Pb. We add  $^{152}$Gd in the list because it has an overwhelming s-process contribution, as discussed in \cite{Cr15b}. Note that $^{152}$Gd may have a contribution ($\sim 10\%$) from proton capture (p-process). } are produced only via the s-process or, at least, they have an overwhelming s-process contribution. Therefore, they can be used to efficiently constrain the evolution of the stars where the s-process is at work. The yields of light s-only nuclei ($^{70}$Ge, $^{87}$Sr) from stars of v$_{\rm rot} =150$ kms$^{-1}$ are increased by more than an order of magnitude with respect to non-rotating stars. In stars rotating at v$_{\rm rot} =300$ kms$^{-1}$, the yields of the heavy s-only nuclei ($^{136}$Ba, $^{204}$Pb) are enhanced by almost 2 orders of magnitude with respect to those of non-rotating stars, up to metallicities $\sim 0.1$ \zs.

\begin{figure}
\begin{centering}
\includegraphics[width=0.49\textwidth]{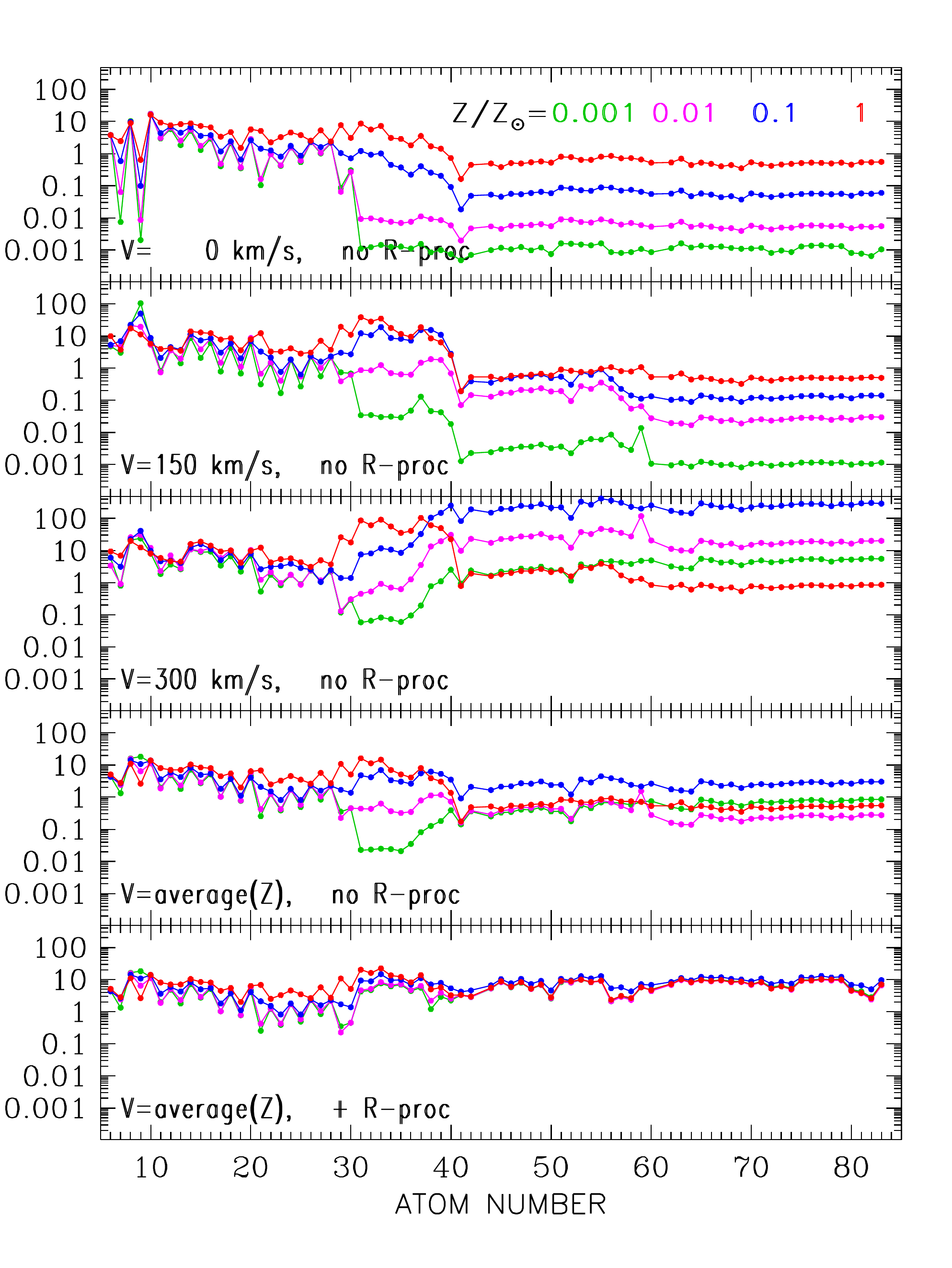} 
\end{centering}
\caption{ Overproduction factors for all elements in a 20 \ms \ star from Eq. \ref{eq:overprod}. They are displayed for (from top to bottom): the non-rotating case, the cases at v$_{\rm rot}=150$ and 300 kms$^{-1}$, and (last two panels) the adopted metallicity-dependent average $\left\langle \rm v_{\rm rot}(Z) \right\rangle$ (see Fig. \ref{fig:Vrot} and corresponding discussion in the text).
The first four panels display the yields of CL2018, which do not include the r-component, while the last one includes it (as assumed from Eq. \ref{eq:yields_r}).}
\label{fig:YldMasspR}
\end{figure}

 The effects of rotation described in the previous paragraph  are illustrated in Fig. \ref{fig:YldMasspR} for the elemental yields of a 20 \ms \ star. The overproduction factors 
\begin{equation}
f \ = \ \frac{Y_i(M,Z)}{X_{i,\odot} M_{ej}(M,Z)} 
\label{eq:overprod}
\end{equation}
are plotted for  four values of initial  metallicity Z ([Fe/H]$=-3, -2, -1$ and 0, respectively), $ M_{ej}(M,Z)$ being the ejected mass of the star in each case. The non-rotating models are displayed in the top panel, where it is seen that all heavy elements above the Fe-peak are produced as secondaries (in fact, the elements in the atomic number range 30-40 range start being overproduced above metallicity [Fe/H]$=-1$, the others are not overproduced at all but are just re-ejected). Models rotating at 150 and 300 km s$^{-1}$, respectively,  are displayed in the next two panels where the impact of the factor $\chi$ and the concomitant number of neutrons per seed - discussed in the previous paragraphs -  is clearly seen.  

In the case of v$_{\rm rot}=150$ km s$^{-1}$, elements in the atomic number range 30-40 are substantially affected, their yields increasing by one order of magnitude at [Fe/H]$=-1$ to two orders of magnitude for lower metallicities (in all cases, enhancements are with respect to the non-rotating case).  The impact is much more important for v$_{\rm rot}=300$ km s$^{-1}$. The number of neutrons per seed nucleus is so large that the neutron flow goes through the Z$=30-40$ region and enhances the heavier nuclei by three orders of magnitude. In contrast, the impact of rotation is negligible for the solar metallicity  star. 

\begin{figure}
\begin{centering}
\includegraphics[width=0.49\textwidth]{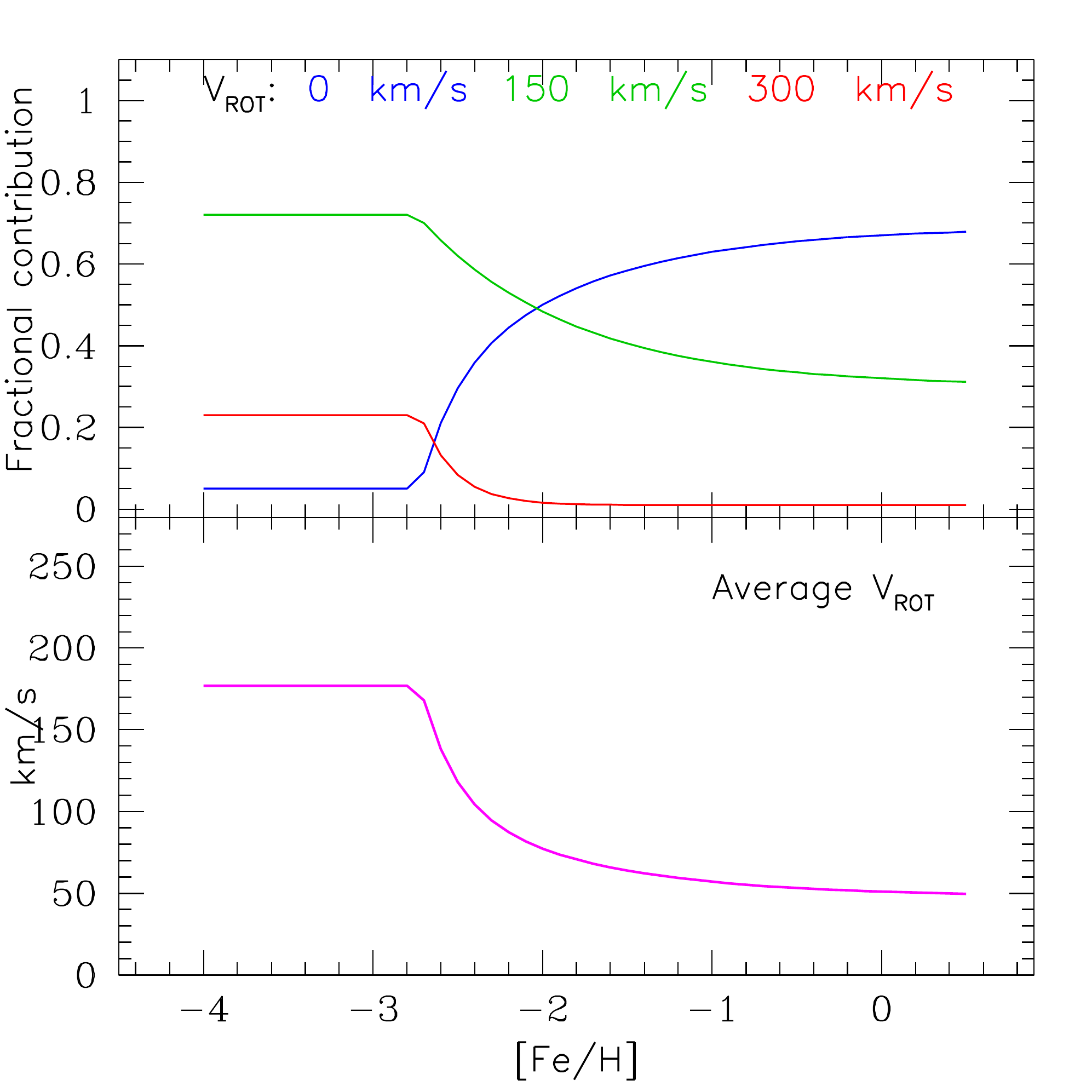} 
\end{centering}
\caption{\label{fig:Vrot}  
{\it Top:} Adopted fractional contribution with metallicity of the yields of rotating massive stars (see text). {\it Bottom:} 
Resulting average initial rotational velocity of massive stars as a function of metallicity.}
\end{figure}

The conclusion of the analysis of the previous two sections is that rotation in massive stars increases considerably the yields of almost all elements (except the Fe-peak nuclei whose abundances are in any case controlled by the choice of the mass cut  and the possible adoption of the mixing and fall back mechanism), by factors which depend strongly on the metallicity. As already discussed, rotation offers a natural solution to the problem of the primary $^{14}$N in the early Galaxy and the present rotating models do just that. However, at the same time, they largely overproduce the abundances of the s-only nuclei at metallicities in the range $\rm -2\leq [Fe/H] \leq 0$. 

The inclusion of rotating star yields in a galactic chemical evolution model requires, in analogy with the adoption of an IMF, also of an IDROV (Initial Distribution of Rotational Velocities) that in principle may depend in the initial metallicity. 
In this paper we tentatively fixed the relative contributions of the three available initial rotational velocities as a function of 
[Fe/H] guided by the observational requirements mentioned in the previous paragraph, namely: a) a primary behavior of $^{14}$N at the lowest metallicities (implying larger average rotational velocities at very low [Fe/H]) and b) the prevention of an overproduction of the s-nuclei at  metallicities [Fe/H]$\sim$ -2 to -1 (and hence low - but non-nul - average  rotational velocities for that metallicity range). 

\begin{figure}
\centering
\includegraphics[width=0.49\textwidth]{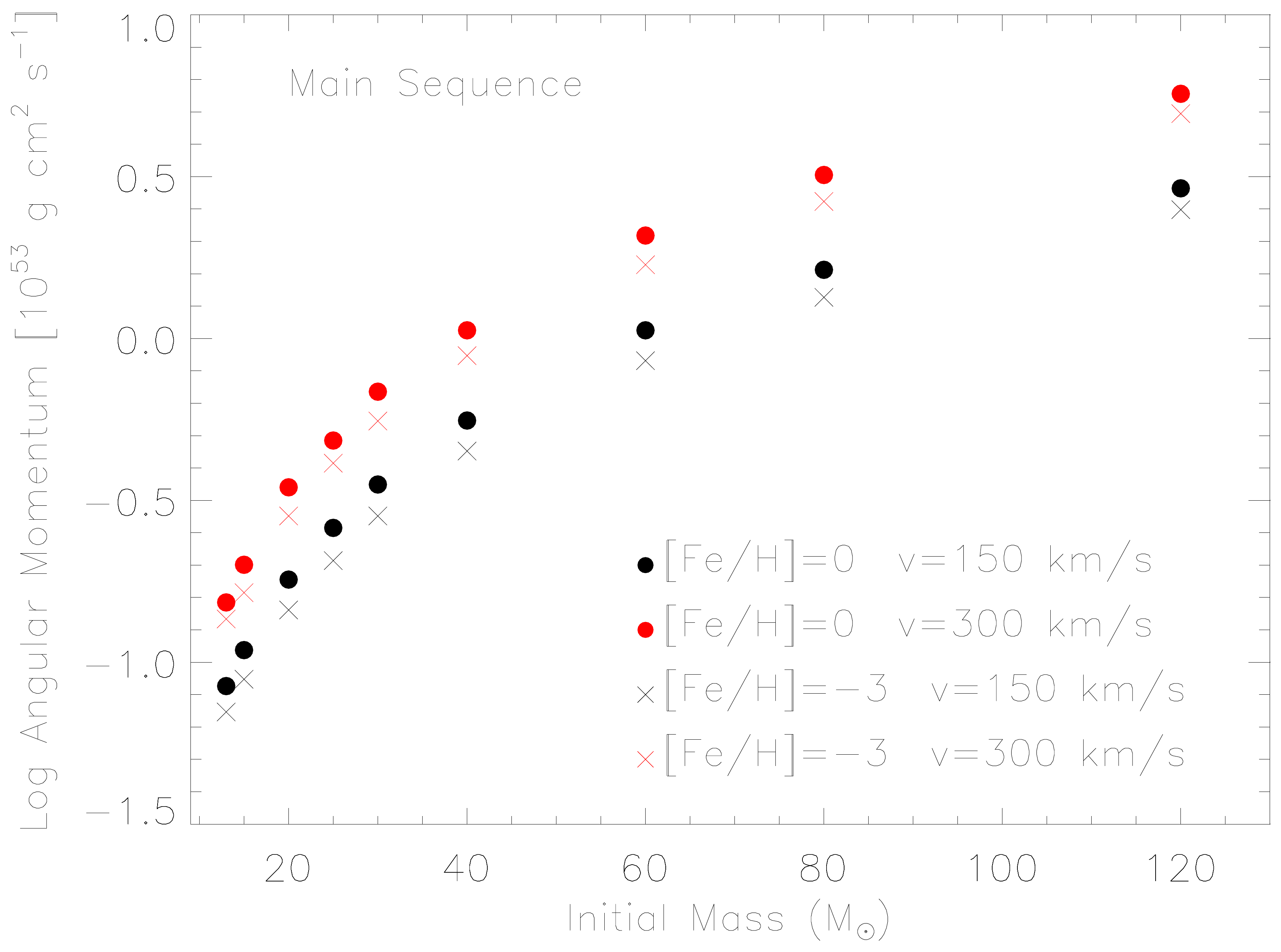} 
\caption{Angular momentum at the beginning of the Main Sequence stage for models with [Fe/H]$=0$ (filled dots) and [Fe/H]$=-3$ (crosses) and initial rotation velocities v$_{\rm rot}= 150$ kms$^{-1}$ (black) and 300 kms$^{-1}$ (red).}
\label{fig:compmomang}
\end{figure}

The adopted weighting factors are plotted in the upper panel of Fig. \ref{fig:Vrot} as a function of metallicity, while the bottom panel shows the resulting average rotational velocity of the massive star population as a function of [Fe/H]. We are well aware that this procedure is  questionable and introduces additional free parameters, but it finds some theoretical support by the argument put forward by the Geneva group 
\citep[see][]{Mey97}:  if the specific angular momentum is assumed to be conserved during the contraction of the proto-stellar nebula, lower metallicity stars should rotate more rapidly because the lower opacity leads to more compact structures. 
This is clearly demonstrated in Figure \ref{fig:compmomang}, where it is shown the angular momentum at the beginning of the main sequence stage as a function of the initial mass for two initial metallicities, i.e., [Fe/H]$=0$ (filled dots) and [Fe/H]$=-3$ (crosses), and two initial rotation velocities, i.e. v$_{\rm rot}= 150$ kms$^{-1}$ (black) and v$_{\rm rot} = 300$ kms$^{-1}$ (red). For the same initial mass and rotation velocity, lower metallicity models have a lower initial angular momentum compared to the high metallicity ones. Therefore, to have the same initial angular momentum, low metallicity models must rotate faster. 

Needless to say that other assumptions for the IDROV than the one adopted here may lead to equally good or even better results than found in this study.

Some of the resulting yields - obtained after applying the aforementioned weighting - appear on the right panel of Fig. \ref{fig:YldMassive}   as a function of metallicity. It can be seen that:

-  $^{12}$C, $^{16}$O, $^{28}$Si behave as primaries and their yields are  approximately at their respective solar values.

- the yields of $^{14}$N are slowly increasing with metallicity (by a factor of $\sim$10 for three orders of magnitude in metallicity), indicating that $^{14}$N is behaving almost  as primary.

- all the s-nuclei behave more or less as secondaries, but only  the light ones (like $^{70}$Ge and $^{87}$Sr) have their yields at approximately the corresponding solar values at Z$\sim$\zs; the yields of heavier s-nuclei are sub-solar by large factors  at Z$\sim$\zs.

   In a similar vein, the two bottom panels of Fig. \ref{fig:YldMasspR} display the  yields of the 20 \ms \ star  as a function of metallicity  {\it after adopting the metallicity-weighted IDROV of Fig. \ref{fig:Vrot}}. The fourth panel shows that overproduction factors do not exceed those of the light elements (lighter than Fe) for all metallicities. More specifically, elements in the Z$=30-40$ range display a secondary-like behavior while heavier ones a primary behavior. But only in the former case the overproduction factors at near solar metallicities are comparable to the one of oxygen (i.e. around 10), while they are considerably smaller in the latter. This implies that the massive star contribution to the light s-elements is expected to be small at low metallicities and dominant at quasi-solar metallicities. In contrast, their contribution to the main s-elements (Z$>40$) is always sub-dominant: at high metallicities it is overwhelmed by the LIM stars and at low metallicities by the r-process, as shown in the bottom panel of Fig. \ref{fig:YldMasspR}, where
 we include a fictitious primary  r-component for each isotope, as described by eq. \ref{eq:yields_r}: now all heavy elements are dominated by the r-component at low Z and behave essentially as primaries in the whole metallicity range. We shall analyze their behavior  with respect to observations in Sec. \ref{subsub:Elm_heavies} and we shall discuss the contribution of the weak and main s-processes to the elemental abundances as a function of metallicity.

\subsection{Yields of low and intermediate mass stars}
\label{subsec:YldLIM}

AGB stars are major chemical polluters of the interstellar medium, in particular regarding He, C, N, F, Na and s-elements (see e.g. \citealt{Cr11}).
During the AGB phase, stars suffer for thermonuclear runaway events (Thermal Pulses, TPs) in the He-intershell, triggered by the sudden activation of 3$\alpha$ reactions. Due to the
large energy released in these events the layers above the He-intershell expand and cool. If expansion is powerful
enough, the H-shell switches off and the convective envelope can penetrate the H-exhausted He-intershell: this phenomenon is known as Third Dredge-Up (TDU) episode. As a consequence of a TDU,
the products of the internal nucleosynthesis can appear on the stellar surface (for reviews see \citealt{IR83, He05, St06, KL14}).


One important product of the 3$\alpha$ reactions is carbon, whose surface abundance is normally less abundant than oxygen, but can overtake it in case of efficient TDUs. This has important consequences on the spectrum of AGB stars, depending on which molecules (C-bearing or O-bearing) are dominant. AGB stars
are responsible for the synthesis of about 50\% of the heavy elements (A$>56$), via slow neutron captures during the so-called "main" s-process \citep{Ga98, Bu99}. 
The requested neutrons are mainly produced by the $^{13}$C($\alpha$,n)$^{16}$O reaction, with a marginal contribution from the $^{22}$Ne($\alpha$,n)$^{25}$Mg reaction. The former reaction works in radiative conditions \citep{St95} between two TPs (T$\sim 10^8$ K), while the latter releases neutrons in the convective shell triggered by a TP at higher temperatures (T$\sim 3\times 10^8$ K).
The $^{13}$C left in the ashes of the H-burning shell is definitely too low to account for the observed AGB surface s-element distributions. Therefore, an extra source of
$^{13}$C is needed at the base of the convective envelope. 

The physical mechanism triggering the formation of such a $^{13}$C-pocket is still matter of debate, with different proposed solutions, as magnetic fields \citep{Tr16}, gravity waves \citep{DT03} or opacity-induced overshoot \citep{Cr09}. The latter occurs when a H-rich (opaque) layer approaches a He-rich (transparent) region, as it happens during a TDU episode. A detailed description of this
situation and how, as a by product, the formation of a large enough ($\Delta $M$\sim 10^{-3}$ M$_\odot$) $^{13}$C-pocket is obtained, can be see in \cite{St06} and \citet{Cr09}.
\citet{Cr09,Cr11,Cr15} have constructed evolutionary stellar models coupled to a full nuclear network including
all the relevant isotopes, up to the termination point of the s-process path (Pb-Bi). Those models are available on the web pages of the FRUITY database\footnote{fruity.oa-abruzzo.inaf.it} and have been used in this study.

Indeed, other AGB yield sets are available in the literature, even if most of them are not covering the full range of metallicities and stellar masses needed to properly calculate a GCE model
(e.g. \citealt{St04,WF09,Ve13,Ba16}). Nevertheless, apart from FRUITY, the only extensive AGB yields set available is that of Mt Stromlo group (hereafter MST; see \cite{KL16} and references therein for a recent up date). \citet{Cr11} already presented a detailed comparison between the FRUITY and the MST sets available at that time; here we briefly remind major differences. Contrary to FRUITY models, in which a time-dependent mixing scheme is adopted (see, e.g., \citealt{St06}), in MST models instantaneous mixing is
assumed within the convective zone and no extra-mixing beyond the convective boundaries is applied.
Thus, in order to obtain a $^{13}$C pocket, a proton profile is added by hand after each TDU episode. Then, the s-process nucleosynthesis is calculated with a post-process technique. A third difference arises from the adopted mass-loss law. AGB envelopes are eroded by radiative stellar
winds at very high rates ($10^{-7}-10^{-4}$ M$_\odot$yr$^{-1}$). Up to date, theoretical AGB models use empirical period-mass loss relations determined by observations of galactic giant
stars. Depending on the adopted sample, different relations have been proposed: MST models include the mass-loss formula from \citet{vw93}, while FRUITY models use the relation published
by \citet{St06}.

We notice that these features play an important role on the resulting stellar yields and on the output of the GCE models (see also discussion in Sec. \ref{subsub:MasStarYld+Rot}  concerning the yields from massive non-rotating/rotating stars). For instance, in the framework of the opacity-induced overshoot mechanism for the formation of the $^{13}$C-pocket described in \citet{Cr09}, a different choice of the maximum allowed penetration of the convective envelope during a TDU episode may lead to a variation of the $^{13}$C-pocket mass extension, with sizable consequences on the s-process production (see \citealt{Cr15b,Cr16}). Moreover, the inclusion of rotation in the computation of low mass AGB stars may lead to significant changes in the surface abundance distribution of these objects, as a function of the initial rotation velocity and of the initial metallicity (see \citealt{Pi13}). \\

\subsection{LIM vs massive star yields}
\label{subsec:Lim_Mas}

\begin{figure}
\begin{centering}
\includegraphics[width=0.49\textwidth]{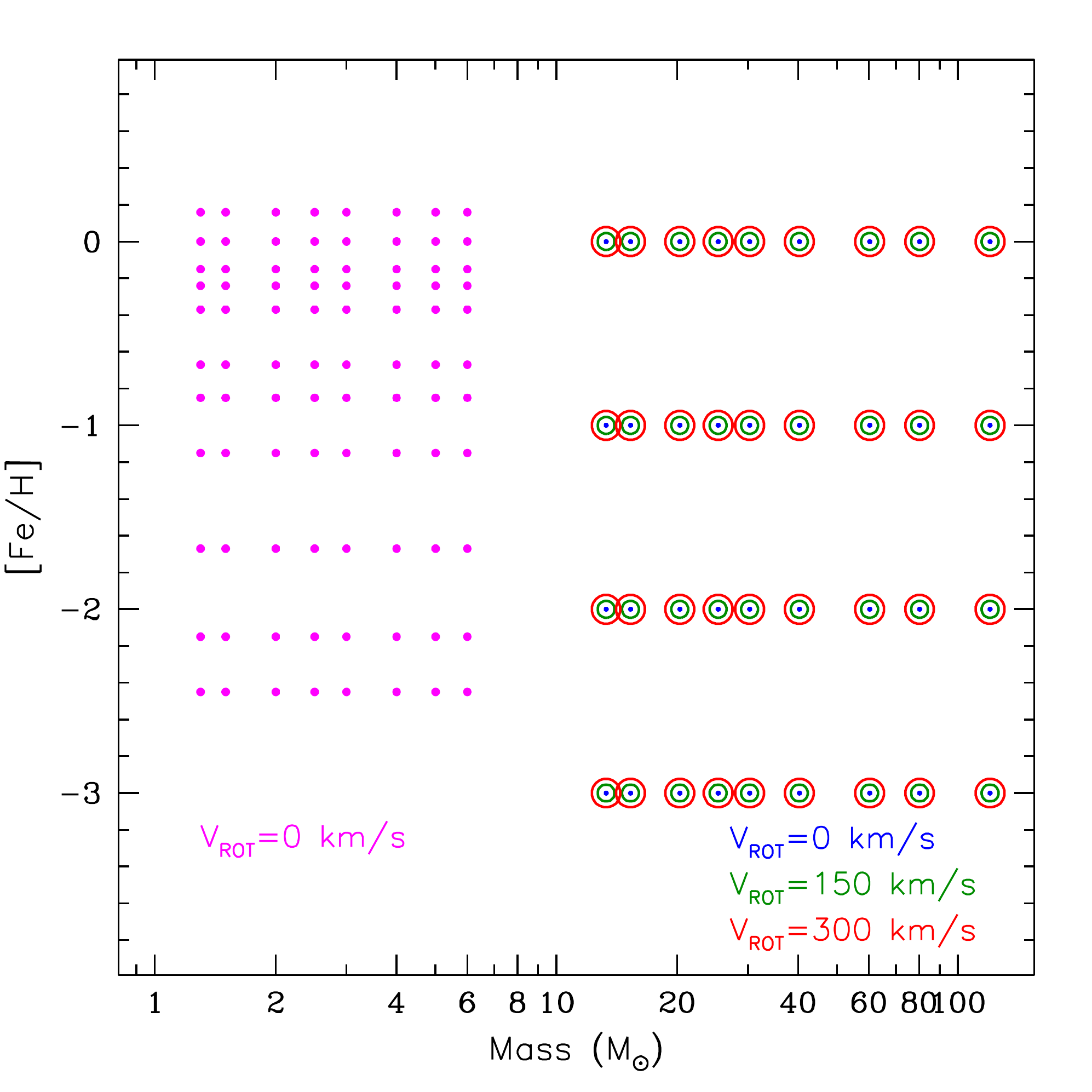} 
\end{centering}

\caption{\label{fig:YieldGrid}  Grid of stellar yields used in this work.  For LIM stars, non-rotating models are considered, while for massive stars we consider 3 rotational velocities, as indicated in the figure, empirically weighted as a function of metallicity (see Fig. \ref{fig:Vrot}).}

\end{figure}

 Figure \ref{fig:YieldGrid} displays the grid of stellar yields adopted in this work, in the stellar mass vs. [Fe/H] plane. The [$\alpha$/Fe] enhancement at metallicities [Fe/H] $\leq$ -1 is appropriately taken into account in the stellar models:    
   while  the LC2018 MS models have a different enhancement for the various $\alpha$ elements, as presented in Sec. \ref{subsub:stellar_model},  the LIM stars models have a uniform enhancement of 0.5 dex, i.e. the same as the one for O adopted for MS models. For all models, all the stable isotopes between H and Bi are considered. 

An inspection of the figure shows that the grid covers satisfactorily the whole metallicity range. Massive star models at metallicities lower than [Fe/H]$=-3$ and higher than 0 would be required for the study of the most metal poor halo stars and the inner disk, respectively, but this is not the subject of this work. Also, the grid covers reasonably well the whole stellar 
mass range, from the lowest to the highest masses, with no need for extrapolation to either direction. 

In order to obtain a continuous sampling of the IMF between the most massive LIM star  of our set (6 \ms) and the lightest massive star (13 \ms), some kind of interpolation has to be made
(a point rarely discussed in the literature, despite the fact that the 6-13 \ms \ mass range contains $\sim$15\% of the ejecta of a stellar generation). Instead of a simple log-log interpolation, we adopt here the following scheme. We assume that for stars up to M$_{\bf *}$, here taken to be M$_{\bf *}$=10 \ms, stars evolve as AGBs, and their yields can be obtained by extrapolation of the LIM yields of M$<6$ \ms, weighted by the corresponding ejecta mass  $E(M)=M-m_R(M)$ where $m_R(M)$ is the mass of the stellar remnant:
\begin{equation}
y_i(M) \ = \ \frac{y_i(6)}{E(6)} \ E(M)
\end{equation}
For stars above M$_{\bf*}$, we assume that they evolve up to Fe-core collapse and their yields are obtained than by log-log interpolation between the yields of M$_{\bf*}$ and those of the 13 M$_\odot$ star. We find that this procedure helps avoiding the overproduction of the massive-star products by the massive AGBs, which is introduced artificially by a simple log-log interpolation.

\begin{figure}
\begin{centering}
\includegraphics[width=0.49\textwidth]{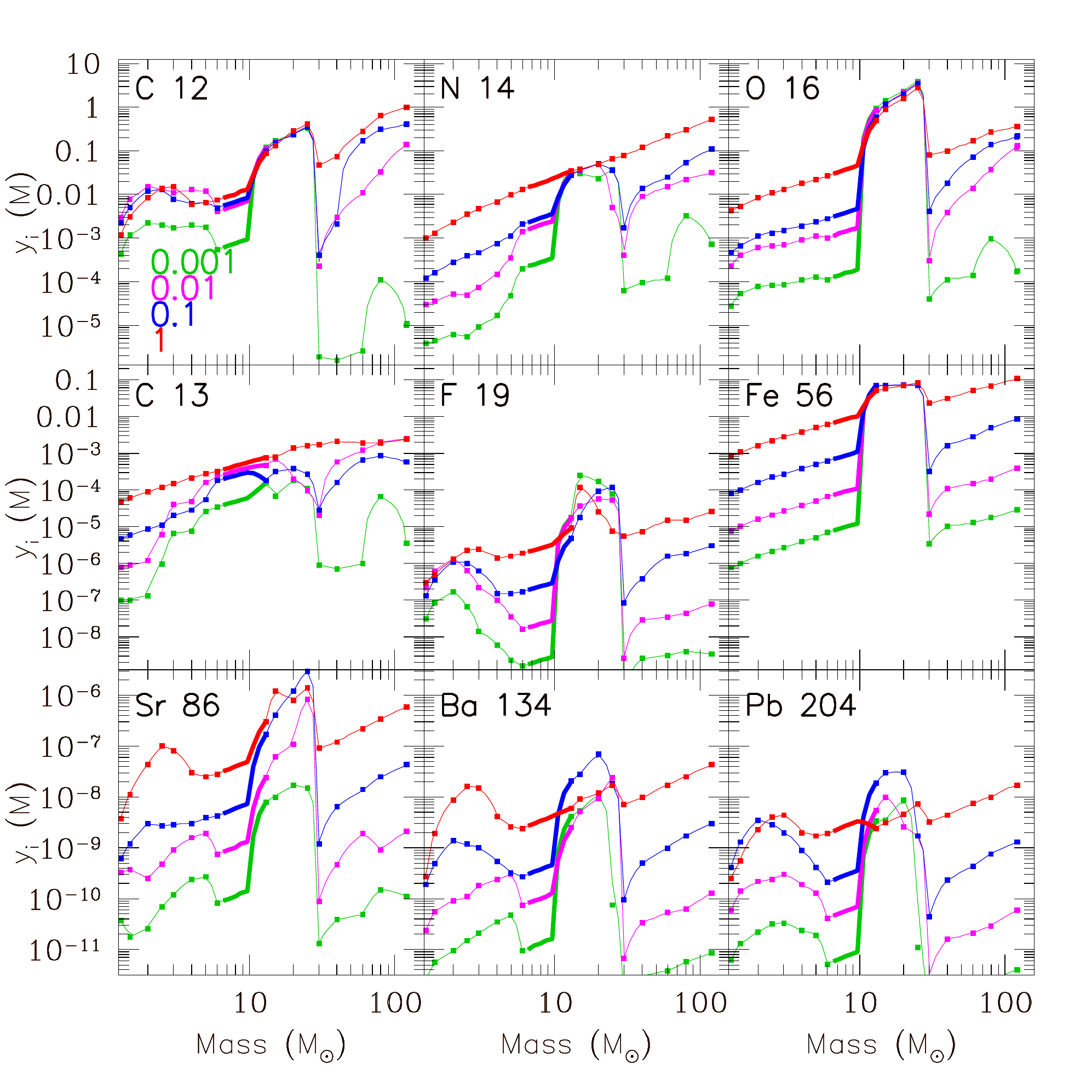} 
\caption{\label{fig:YieldVsMass}  Total yields $y_i(M)$ (in \ms) of selected nuclei as function of the initial stellar mass for four initial metallicities Z/\zs \ (color coded in top left panel), weight by the adopted distribution of V$_{\rm rot}$ vs Z from Fig. \ref{fig:Vrot}. Points indicate the actual grid of used yields (Fig. \ref{fig:YieldGrid})  and are connected by solid curves through interpolation. The thick portion of the curves indicates the adopted interpolation in the "desert" between 6 and 13 \ms \ (see text).}
\end{centering}
\end{figure}

\begin{figure*}
\begin{centering}
\includegraphics[angle=-90,width=1.0\textwidth]{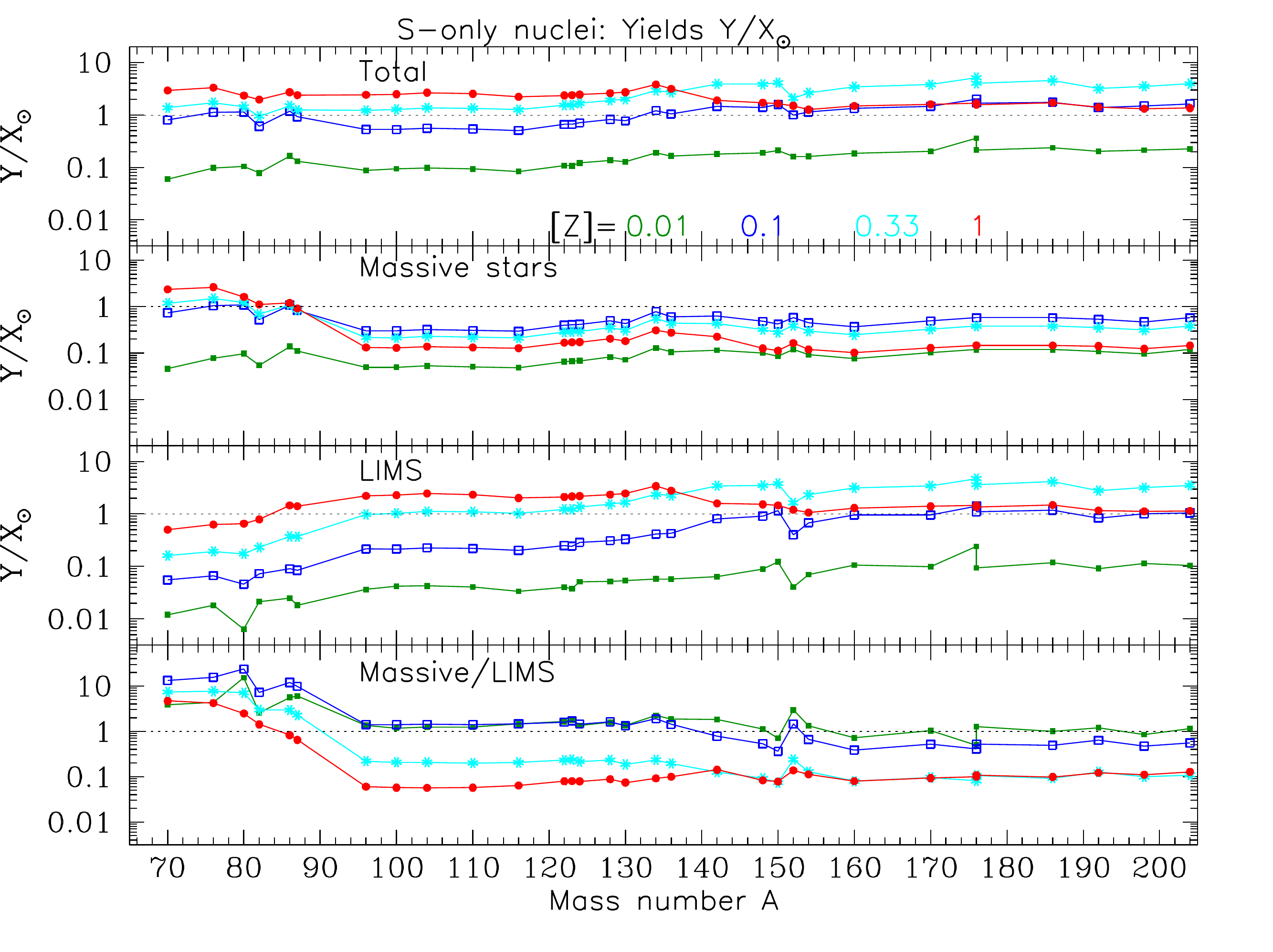} 
\caption{\label{fig:YldSonly}  
Yields of s-only nuclei, averaged over the adopted IMF for four metallicities (green filled squares: Z=0.01 \zs, 
blue open squares: Z$=0.1$ \zs; cyan asterisks: Z$=1/3$ \zs; red dots: Z=\zs). From top to bottom: total yields, yields of massive stars (13-120 \ms), yields of LIM stars (1-6 \ms). Bottom panel: Corresponding ratios of the MS  yields to those of LIM stars.
}
\end{centering}
\end{figure*}

Some of the key features discussed in the previous sections are illustrated in Fig. \ref{fig:YieldVsMass}
which  displays the yields $y_i(M)$ for individual stellar models of several selected isotopes as function of the initial stellar mass $M$ and for four different metallicities. The yields of massive stars are the metallicity-dependent weighted average over the rotational velocity, as discussed in Sec. \ref{subsub:MasStarYld+Rot}.

- Stars more massive than 25 \ms \ contribute to the total yields only with the mass ejected through the wind because these stars are assumed to collapse in the remnant.

- $^{16}$O is produced as primary by massive stars. Lower mass stars eject only the initial $^{16}$O
of their envelope. $^{40}$Ca and $^{56}$Fe display exactly the same behavior. Note that, as mentioned above, the $^{56}$Fe produced by massive stars depends on the choice of the mass cut. In this case $^{56}$Fe is essentially independent on both the initial mass and initial metallicity simply by construction.

- $^{12}$C is produced as primary by massive stars but, to some extent, also by LIM stars for metallicities $\rm Z\geq 0.01~Z_\odot$.

- $^{14}$N is produced by massive stars, as a quasi-primary (its  yield increases weakly with metallicity). $^{23}$Na has a similar behavior. According to our yields, there is no primary $^{14}$N production from hot-bottom burning in LIM stars.

- The light s-only isotope $^{86}$Sr is produced essentially by massive stars at all metallicities as secondary. LIM stars have a small contribution at solar metallicity.

- The heavy s-only isotope $^{134}$Ba is produced mostly as secondary by massive stars at low $Z$; at \zs \ it is clearly produced by LIM stars. The same behavior is qualitatively displayed by $^{204}$Pb.

Figure \ref{fig:YldSonly} illustrates the behavior with metallicity and the role of LIM vs. massive stars for the yields of all the s-only nuclei. 

In the upper panel, the total normalized yields Y$_{\rm total}$=Y$_{\rm MS}$+Y$_{\rm LIM}$ are displayed, where Y$_{\rm MS}$ are the massive star yields from Eq. \ref{eq:yldmas} and Y$_{\rm LIM}$ are the corresponding yields from LIM stars (integrated in the 1-6 \ms \ range). In the second and third panels, we report the yields of massive and LIM stars, respectively. 
As expected, in LIM stars lead is mainly synthesized at low metallicities. Then, for larger metallicities, elements belonging to the second s-process peak (Ba-La-Ce-Nd) start being efficiently produced. At solar-like metallicities, the production of elements belonging to the first s-process peak (Sr-Y-Zr) reaches its maximum and dominates the overall heavy element nucleosynthesis. The key quantity regulating this nucleosynthesis is the {\it neutron-to-seed} ratio, i.e. the ratio between the neutron number density and the seed (mainly $^{56}$Fe) number density. While seeds scale with the metallicity, the $^{13}$C abundance in the $^{13}$C-pocket (the main neutron source) does not depend on the initial CNO abundance. Therefore, in LIM stars the  number of available neutrons is roughly the same at all metallicities and produces the reported yields.

In the bottom panel of Figure \ref{fig:YldSonly} we plot the ratios Y$_{\rm MAS}$/Y$_{\rm LIMS}$ as function of metallicity. For nuclei up to Sr, massive stars clearly dominate the production at all metallicities, especially at Z$<0.33$ \zs. In fact, massive stars produce also significant amounts of the heavy s-nuclei at low metallicities, competing with LIM stars. Above Z$=0.33$ \zs, however, LIM stars clearly dominate the production of heavy s-nuclei.

\section{Results}

\subsection{Evolution of solar neighborhood}
\label{subsec:Local_Evol}

\begin{figure} \label{fig:f_GCE_model}
\begin{centering}
\includegraphics[width=0.49\textwidth]{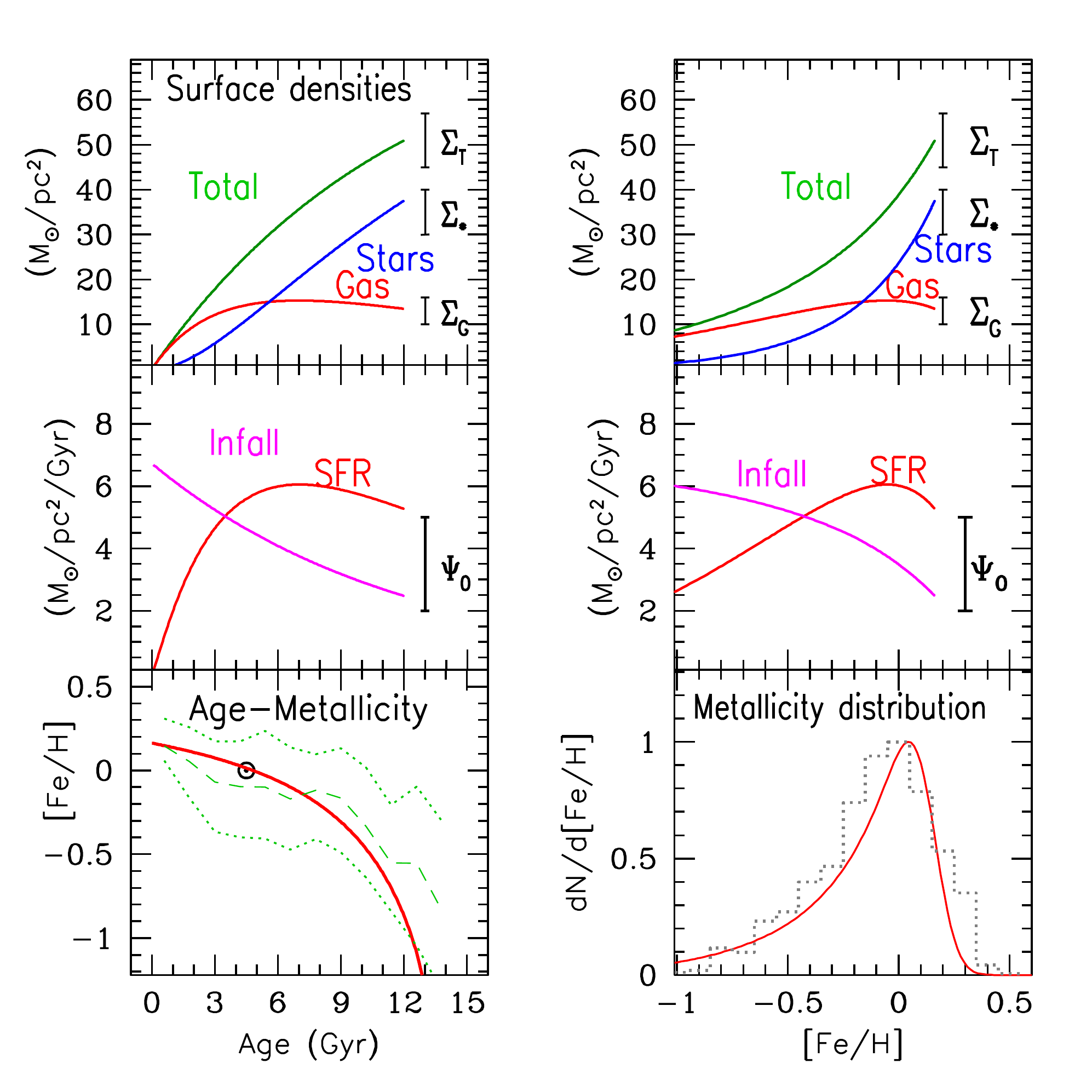} 
\end{centering}
\caption{Results of the chemical evolution model for the solar neighborhood (solid curves in all panels represent model results). {\it Top left}: Evolution of surface densities of stars and gas; vertical bars at 12 Gyr represent corresponding present day values of those quantities. {\it Top right}: Same as on the left, as a function of metallicity [Fe/H]. {\it Middle left}: Evolution of star formation and infall rates; vertical bar indicates present days estimates of local SFR. {\it Middle right}: Same as on the left as a function of [Fe/H]. {\it Bottom left}: Age-metallicity relation (age running opposite to time of the previous panels); dashed and dotted green curves indicate the average and $\pm$1$\sigma$ values, respectively,  of the local age-metallicity relation as derived by observations of \protect\citet{Cas11}. {\it Bottom right}: Local metallicity distribution  compared to data (dotted histogram) from \protect\citet{Adi12}.}
\label{fig:GCE_model}
\end{figure}

The evolution of some key quantities as function of time or metallicity are plotted in Fig. \ref{fig:GCE_model}. The final results are compared satisfactorily to present day observables in the solar neighborhood, namely surface densities of gas, stars and star formation rate (top and middle panels).

The two main observables of the solar neighborhood, namely the age-metallicity relation and the metallicity distribution are also well reproduced. Metallicity  increases substantially with age at early times and flattens considerably after reaching the solar value $\sim$4.5 Gyr ago. The resulting curve is well within the error bars of recent surveys, like the one of \citet{Cas11} within the Geneva-Copenhagen survey based in the analysis
of more than 16000 FGK stars. Also the computed metallicity distribution in the solar
neighborhood is consistent with the observed one. 

We notice, however, that the adopted simplified one-zone model is known, for sometime now, to be far from satisfactory regarding several observables in the solar neighborhood. In particular, despite the difficulties in estimating stellar ages,  the early age-metallicity relation is flatter than the theoretical one obtained here. Moreover, there is considerable dispersion of metallicity at any age, much larger than the one in the local gas \citep{Cartledge2006}; this is  impossible to reproduce with 1-zone models where gas is instantaneously and completely mixed. As for the metallicity distribution, one-zone models cannot simultaneously reproduce the local gas metallicity ($\sim$\zs \ today) and the most-metal rich stars locally (with metallicities 
$\sim$2 \zs \ or more). Neither can they explain the observed presence of both old and young stars at all metallicities 
\citep{Cas11} in the solar neighborhood.

An elegant solution to the aforementioned problems is provided by radial migration \citep[see e.g.][]{sel02,sch09,Kubryk2015a}.
In this work we shall content ourselves to the exploration of the impact of the adopted yields on the simple one-zone model of local GCE. The study of those yields in the framework of a more "realistic" model for the Galactic disk, including radial migration, will be made in a forthcoming study.

\subsection{Isotopic  abundances at  Solar System formation}
\label{subsec:Isotopes}

\begin{figure*} 
\begin{centering}
\includegraphics[angle=-90,width=1.0\textwidth]{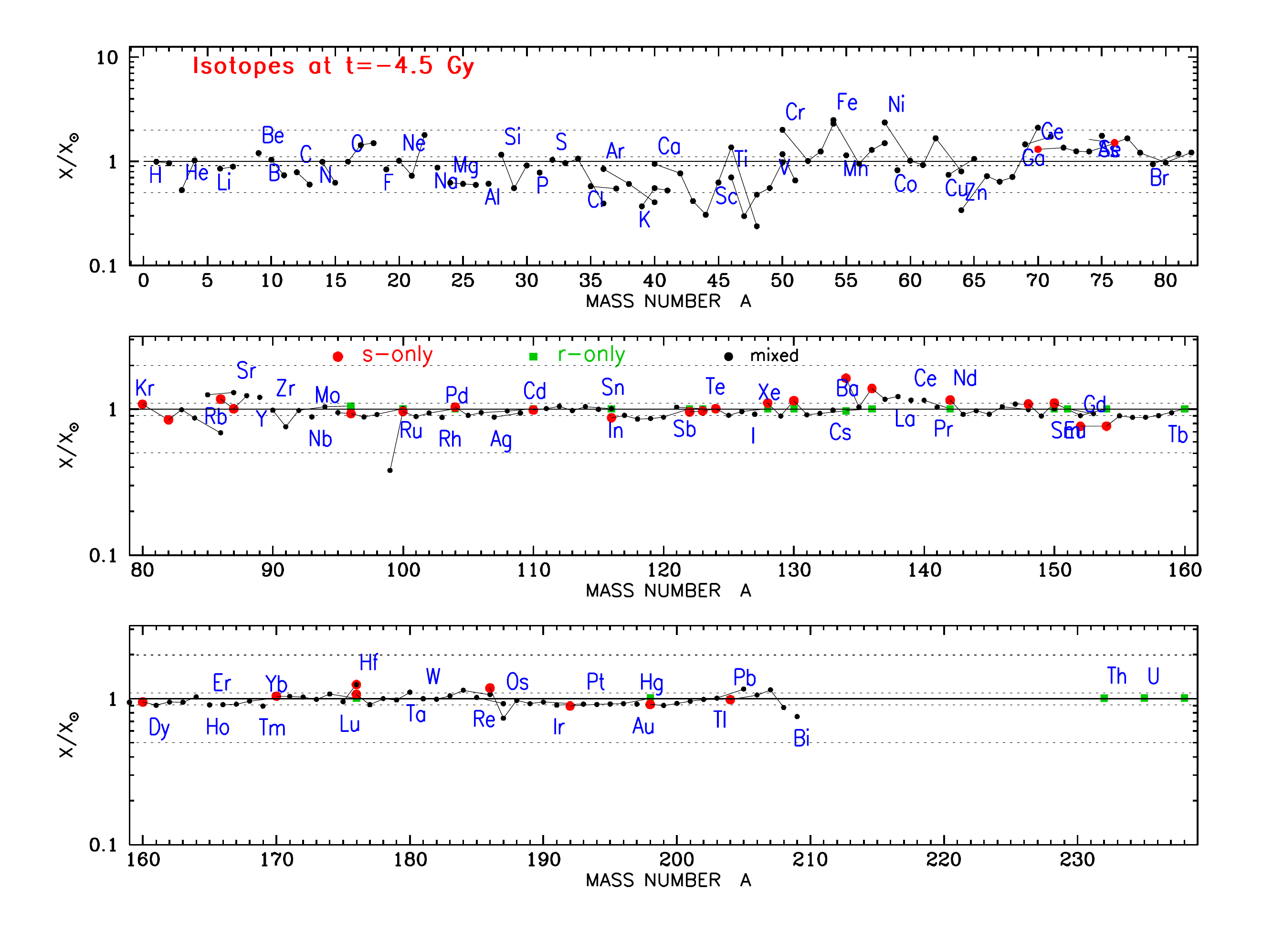} 
\par\end{centering}

\caption{Model distribution of isotopic abundances (plotted as X/X$_{\odot}$) obtained at the time of the formation of the solar system and compared to solar system data (\citealt{Lod09}). Yields for isotopes lighter than the Fe-peak and those on the s-process path are from stellar nucleosynthesis models. r- isotopes are  assumed to originate in massive stars and their yields are fiduciary (see Sec. \ref{subsec:model}). Dotted horizontal lines bound the regions where over/underproduction factors of 2 and of 10\%, respectively, are obtained. Red dots denote s-only nuclei, green squares r-only nuclei and black dots those of mixed origin. p-isotopes do not appear on the figure. Element symbols appear close to the lightest isotope of a given element.}
\label{fig:all_isotopes}
\end{figure*}

\subsubsection{The global picture}
\label{subsub:Global_isotopes}

In Figure 7 we present our results for the abundances of all the isotopes obtained at the time of solar system formation, i.e. 4.5 Gyr ago. This is an important "sanity check" of both the chemical evolution model and the adopted stellar yields. Such a comparison of isotopic abundances to the corresponding solar system values has been first made in \citet{Timmes1995} and subsequently in other studies \citep{Goswami2000,Kubryk2015a} for isotopes up to Zn or Ge. The result of that comparison has not varied by much over the time: the solar system isotopic composition is globally reproduced within a factor of two, a relative deficiency is found for the A$\sim 40-50$ mass number region, while  some Fe-peak isotopes, like $^{54}$Fe or $^{58}$Ni are overproduced by a factor of two. The latter feature results from a well known problem of the "standard" W7 model of SNIa nucleosynthesis, both in its original \citep{Thielemann1986} and more recent \citep{Iwa99} version, the latter being adopted here.

An inspection of the upper panel of Fig. \ref{fig:all_isotopes} shows that the aforementioned features also appear in our results. A closer inspection reveals some interesting points:

a. All the major isotopes of the multi-isotopic elements up to Fe ($^{12}$C, $^{14}$N,     $^{16}$O,     $^{20}$Ne,  $^{28}$Si,     $^{32}$S,     $^{36}$Ar,     $^{40}$Ca,     $^{54}$Cr,     $^{56}$Fe)  are reproduced to better than 15\% and,  in most cases, to better than 10\%. Exceptions to that "success story" are $^{24}$Mg and $^{48}$Ti, which are under-produced by $\sim$40\%,  $^{39}$K which is under-produced by more than a factor of 2, and $^{58}$Ni, which is overproduced by a factor of $\sim$2, as previously discussed.     These results are reflected in the corresponding elemental composition, to be discussed in Sec. \ref{subsec:Elements}.

b. The fact that $^{16}$O and $^{56}$Fe are fairly well reproduced is a guarantee that the adopted combination of SFR, IMF, massive star yields and SNIa rate is successful regarding its main nucleosynthesis implications and validates the model. Whether other isotopes are well reproduced depends then exclusively on the adopted stellar yields.

c. The isotopes of  mono-isotopic elements are, in general, less well reproduced, being deficient by 20-30\% ($^{19}$F, $^{23}$Na,     $^{31}$P) or more ($^{27}$Al, $^{45}$Sc), except those of the Fe peak ($^{55}$Mn, $^{59}$Co) which are fairly well reproduced, at the 10\% level.

d. The case of fluorine is of particular interest. Fluorine is not made in conventional (non-rotating) massive star models. \cite{Woosley1990} suggested that F could be produced by neutrino spallation on $^{20}$Ne nuclei, the energetic neutrinos being released from the collapse of the Fe-core.
Since we do not take into account such neutrino-induced nucleosynthesis, in our model F is mainly produced by rotating massive stars by the sequence \nuk{N}{14}($\alpha$,$\gamma$)\nuk{F}{18} ($\beta ^+$)\nuk{O}{18}(p,$\alpha$)\nuk{N}{15}($\alpha$,$\gamma$)\nuk{F}{19} \citep{GORI90}. The protons necessary to the activation of this sequence come from the \nuk{N}{14}(n,p)\nuk{C}{14} nuclear reaction that it is in turn activated by the \nuk{C}{13}($\alpha$,n)\nuk{O}{16}. This means that the synthesis of F requires the simultaneous presence of both \nuk{N}{14} and \nuk{C}{13}. Rotating models may produce large amounts of \nuk{F}{19} because the stirring of matter between the central He burning and the H-burning shell creates a buffer of either \nuk{N}{14} {\it and} \nuk{C}{13} in the radiative part of the He core. At the end of the central He burning, the growth of the He convective shell leads to the quick ingestion of both \nuk{N}{14} and \nuk{C}{13} at temperatures high enough that all the sequence described above may activate efficiently. About 2/3 of the F abundance at solar system formation comes from that source in our model, the remaining 1/3 resulting from LIM stars. In contrast, when the yields of non-rotating massive stars are adopted, we find that their contribution is negligible w.r.t. the one of LIM stars.  Overall, in our baseline model, $\sim$85 \% \ of proto-solar F is produced, a quite satisfactory achievement in view of the uncertainties in the physics of rotating stars and chemical evolution modeling.

e. The minor isotope of N, $^{15}$N, is  rather well produced in our model, since we get 60\% of its solar value. Notice that we adopt here the protosolar isotopic ratio $^{14}$N/$^{15}$N=441 of\cite{mar11}, based on solar wind measurements, and not the value $^{14}$N/$^{15}$N=272 of \cite{Lod09}.  Without rotation, we obtain a  severe underproduction, by a factor of $\sim$6. This isotope is, in general, not found to be produced in non-rotating massive stars, so  \cite{Woo95} and \cite{Timmes1995} invoke  neutrino-induced nucleosynthesis to explain its abundance. In our case, the production of $^{15}$N is again due to the role of rotating massive stars and it is produced by the same sequence that leads to the synthesis of \nuk{F}{19}.  Our proto-solar $^{14}$N/$^{15}$N ratio is 740, larger than found by \cite{mar11} but certainly within acceptable limits. We notice here that novae are considered as possible sources of $^{15}$N \citep{jos16} and that some AGB carbon stars show inexplicable low $^{14}$N/$^{15}$N ratios ($<100$), which  points out to a possible $^{15}$N contribution from these stars \citep{hed13}.

f. Regarding the minor isotopes of elements up to the Fe peak, one sees a relative underproduction (by 40\% or more) of the Mg isotopes, of $^{29}$Si and of most of the isotopes between A$=35$ and 50. We notice that the problem of Mg underproduction (including $^{24}$Mg) is also present in the \citet{Woo95} yields, whereas the underproduction of most intermediate mass isotopes  also characterizes the \citet{Woo95} (for A$=40$ to 50) and \citet{Nomoto2013} (for A$=35$ to 50). It is beyond the scope of this study to analyze the reasons for these discrepancies, but they certainly point out to interesting physical phenomena in advanced phases of stellar nucleosynthesis, which are poorly modeled at present.

g.  It is worth reminding that the two first nuclei beyond the Fe peak, Cu and Zn, have a composite production. \nuk{Cu}{63} is the most abundant of the two Cu isotopes and it is made by both  central He burning and Si burning (as \nuk{Ge}{63}). At low metallicity \nuk{Cu}{63} is mainly produced by the explosive nucleosynthesis. At solar metallicity, on the contrary, it is mainly produced by the explosion in the two lowest masses (13 and 15 \ms) and by core He burning in the more massive ones \citep{LC03,CL04}. Since the integration of the yields over a Salpeter IMF favors stars in the range 20-25 \ms, \nuk{Cu}{63} and hence Cu, may be basically considered a product of the He burning. \nuk{Zn}{64} is the most abundant isotope of Zn and has a production very similar to that of Cu, in the sense that both the central He burning and the Si burning (as \nuk{Ge}{64}) contribute to its yield. It is difficult to understand why Cu is well reproduced while Zn is not, since both integrated yields (over a Salpeter IMF) are dominated by the hydrostatic production. Note that the \nuk{Cu}{63}(n,$\gamma$) nuclear reaction rate produces the unstable nucleus \nuk{Cu}{64} that has a terrestrial half life of the order of 12.7 h, and may decay either in \nuk{Zn}{64} or in \nuk{Ni}{64}. Hence, a possible solution to the Zn underproduction could simply be a wrong branching ratio between the two possible decays. Another possibility could be that the explosive component of the \nuk{Zn}{64} is underestimated. 
 
h. Isotopes above Zn are produced by neutron captures, either in the s- or the r- process, or as a mixture of the two. As stated in Sec. \ref{subsub:MasStarYld+Rot}, in the absence of reliable r-yields we adopted here  fiduciary yields, assuming the r-isotopes are produced in massive stars with yields proportional to those of $^{16}$O and to their solar system fraction (Eq. 2). Thus, it is not a surprise that at solar system formation, the abundances of the pure r-isotopes match well their solar values, since this is also the case for $^{16}$O.

We turn now to one of the main themes of our study, namely the production and evolution of the s-isotopes from both massive and LIM stars in the framework of a successful chemical evolution model (i.e. satisfying all the main observational constraints), as the one described in the previous two sections.

\subsubsection{The s-only isotopic distribution: no need for a solar LEPP}
\label{subsub:sonly_isotopes}

\begin{figure*}
\begin{centering}
\includegraphics[angle=-90,width=1.0\textwidth]{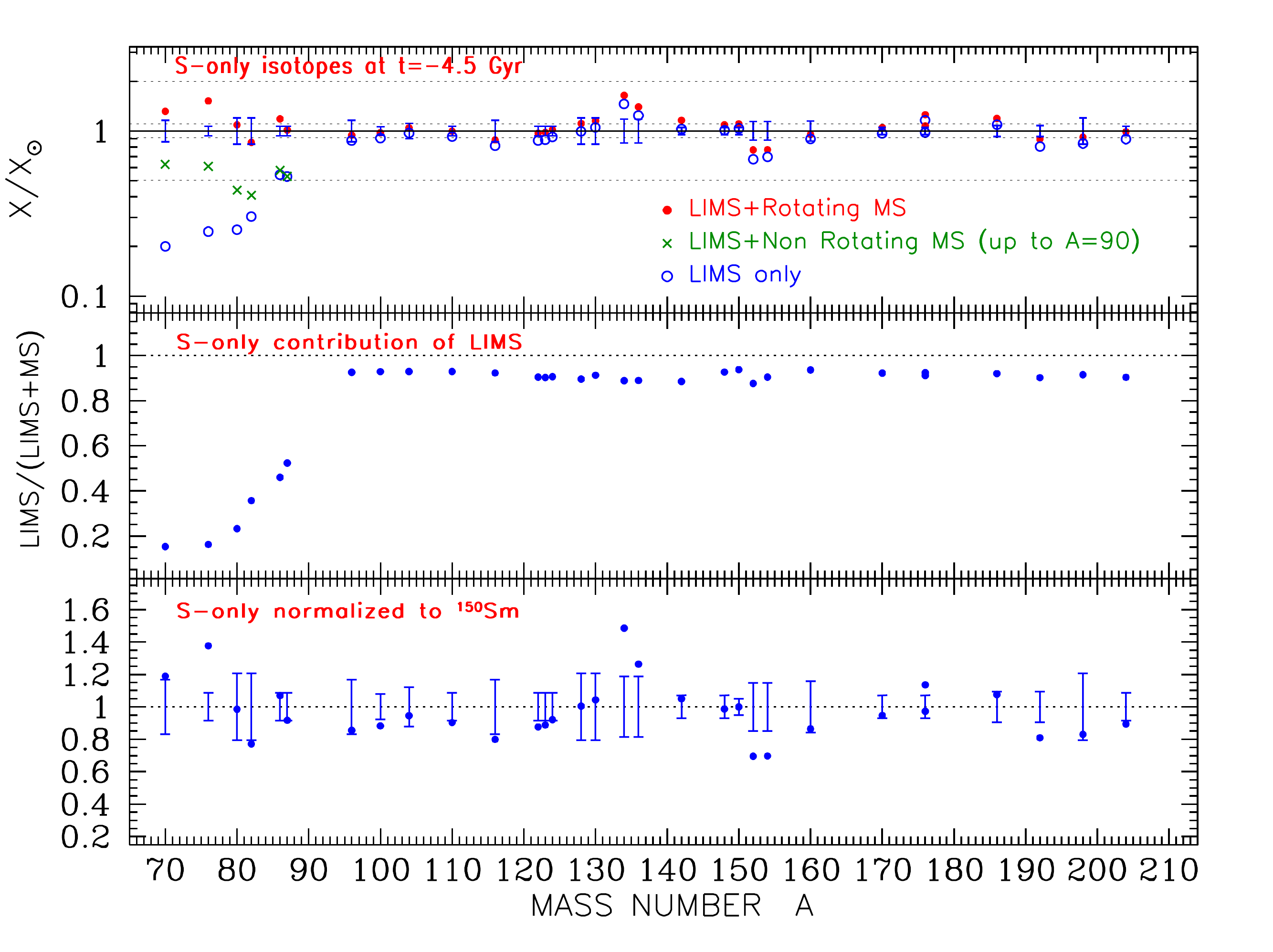} 
\par\end{centering}

\caption{\label{fig:f_Sonly} Results for s-only isotopes. {\it Top}: Comparison to proto-solar abundances  with the contribution of a) LIM and rotating massive stars (our baseline model as in Fig. \ref{fig:all_isotopes}, red filled symbols), with LIM stars and non-rotating massive stars (green crosses, only up to A$=96$ to avoid confusion at higher A)  and with the contribution of LIM stars only (open symbols). Abundance uncertainties (1 $\sigma$ from  \citealt{Lod09}) for each isotope are indicated. The dotted horizontal lines show deviations of 10$\%$ and 50$\%$, respectively,  from the proto-solar values. {\it Middle}: Contribution of LIM stars to the total production of s-only isotopes in the baseline model.  {\it Bottom}: Production factors normalized to $^{150}$Sm (see text for details). }
\end{figure*}

Figure \ref{fig:f_Sonly} shows the s-only isotopic abundance distribution obtained at the time of the solar system formation (4.5 Gyr ago), compared to the measured protosolar values.\footnote{Note that those abundances differ from the current ones observed in the solar photosphere due to the impact of gravitational settling. We adopt as the isotopic protosolar distribution that from \citet{Lod09}.} The upper panel shows our baseline model (LIM and rotating massive stars, filled red symbols), and two models run for comparison: one with LIM and non-rotating massive stars
(green crosses)\footnote{For the case of "LIM + non-rotating massive stars", results are displayed only for A$<90$, to avoid overlapping of the data at larger A values with those of the "LIM stars only" model, which are quasi-identical.} and one with LIM stars only (open blue symbols). It can be seen  that:

a.  The distribution of s-only nuclei in our baseline model is essentially flat. In fact, most of the computed X/X$_{\odot}$ ratios are within $10\%$ (or less) of the corresponding proto-solar values. For some isotopes, however ($^{76}$Se, $^{134,136}$Ba,$^{152}$Gd) differences are beyond  this limit, at the 30-40\% level. Nevertheless, their deviation form the mean value may be connected to significant uncertainties of their nuclear inputs, i.e. neutron capture cross sections and weak $\beta$-decay rates \citep{Cr15b}. While a great effort has been made by the nTOF collaboration to derive experimental neutron capture cross sections \citep{Gu13}, weak rates are still frozen to the compilation by \citet{TY87}. A systematic theoretical (and/or experimental) study of those rates might significantly improve our knowledge of many s-process branchings.

b. The impact of {\it rotating massive stars} on the production of light s-nuclei, through the so-called "weak s-process",  is quite significant. LIM stars alone produce only 20$\%$ to 50$\%$ of the s-only nuclei in the mass region A$<90$, as can be seen in the middle panel of Fig. \ref{fig:f_Sonly}. Yields from non-rotating massive stars (upper panel) improves the situation considerably for the lightest isotopes, especially $^{70}$Ge and $^{76}$Se, but it is insufficient to bring a satisfactory agreement
(here meaning at the 10\% level) with the proto-solar composition for the isotopes with A$<90$.

c. In the bottom panel of Figure \ref{fig:f_Sonly} we display  the s-only distribution of the baseline model normalized to the value of $^{150}$Sm, which has been chosen as reference due to its unbranched origin \citep{Ar99}. Isotopes belonging to the first s-process peak (Sr-Y-Zr) are fully reproduced thanks to the contribution of the weak s-process in rotating massive stars.  This is a remarkable improvement with respect to previous studies \citep{Tra04,Cr15b,Bi17}. The computed distribution of s-only isotopes with $95<$A$< 125$ is also improved but still show a mild underproduction with respect to $^{150}$Sm, of $\sim$10\% on average. This basically confirms the results by \cite{Cr15b}, who already identified such a trend. In the same panel, we also display the corresponding uncertainties on the measurements of those isotopic abundances \citep{Lod09}. It can be seen that most isotopes are  within (or better) than one $\sigma$ of their proto-solar abundance. Exceptions are $^{76}$Se, $^{134}$Ba and $^{152,154}$Gd, found at $\sim 2\sigma$. Taking into account the theoretical and observational uncertainties, we believe that the production of the pure s-nuclei over the whole mass range by the combined action of rotating massive stars and LIM stars is utterly successful (and as close to observations as it can be). 

Other GCE models that combine the s-process contribution from AGB stars (main and strong components) and massive stars (weak-s and r-processes) exist. \citet{Tra04} and, more recently, \citet{Bi14,Bi17}, reported a deficit of the predicted solar system abundances of the s-only isotopes in the Sr-Te region, connecting this deficit to the existence of a missing contribution: the so-called solar {\it light element primary process} (LEPP)\footnote{Note that a different LEPP has also been proposed to explain the abundances of a large group of light elements with an important contribution from the r-process. For instance, \citet{Mo07} and \citet{Am11} distinguished between 'solar' and 'stellar' LEPP, the latter being linked to r-enhanced  low-metallicity halo stars. In this study, we only focus on s-only isotopes in the protosolar nebula.}. In contrast, \citet{Tr16},  ruled out the existence of the LEPP, on the basis of their analysis of single AGB models. 

On the other hand, \citet{Cr15b}, basing on FUNS code calculations \citep{St06} and a simple GCE model for the solar neighborhood,  investigated the effects on the solar system s-only distribution (and yields) induced by the inclusion of phenomena normally ignored in the evolution of AGB stars (as rotation), or by the variation of physical processes (convective overshoot and mass-loss rate) and micro-physics inputs (strong and weak reaction rates). These authors also concluded that a LEPP is not  necessarily required to understand the solar system s-only abundances in the range 96$\leq$ A$\leq$ 124  - due to the uncertainties still affecting both stellar and galactic chemical evolution models - but they did not rule it out definitely. 

\begin{figure*}
\begin{centering}
\includegraphics[angle=-90, width=1.0\textwidth, viewport=183 10 593 760,clip=true]{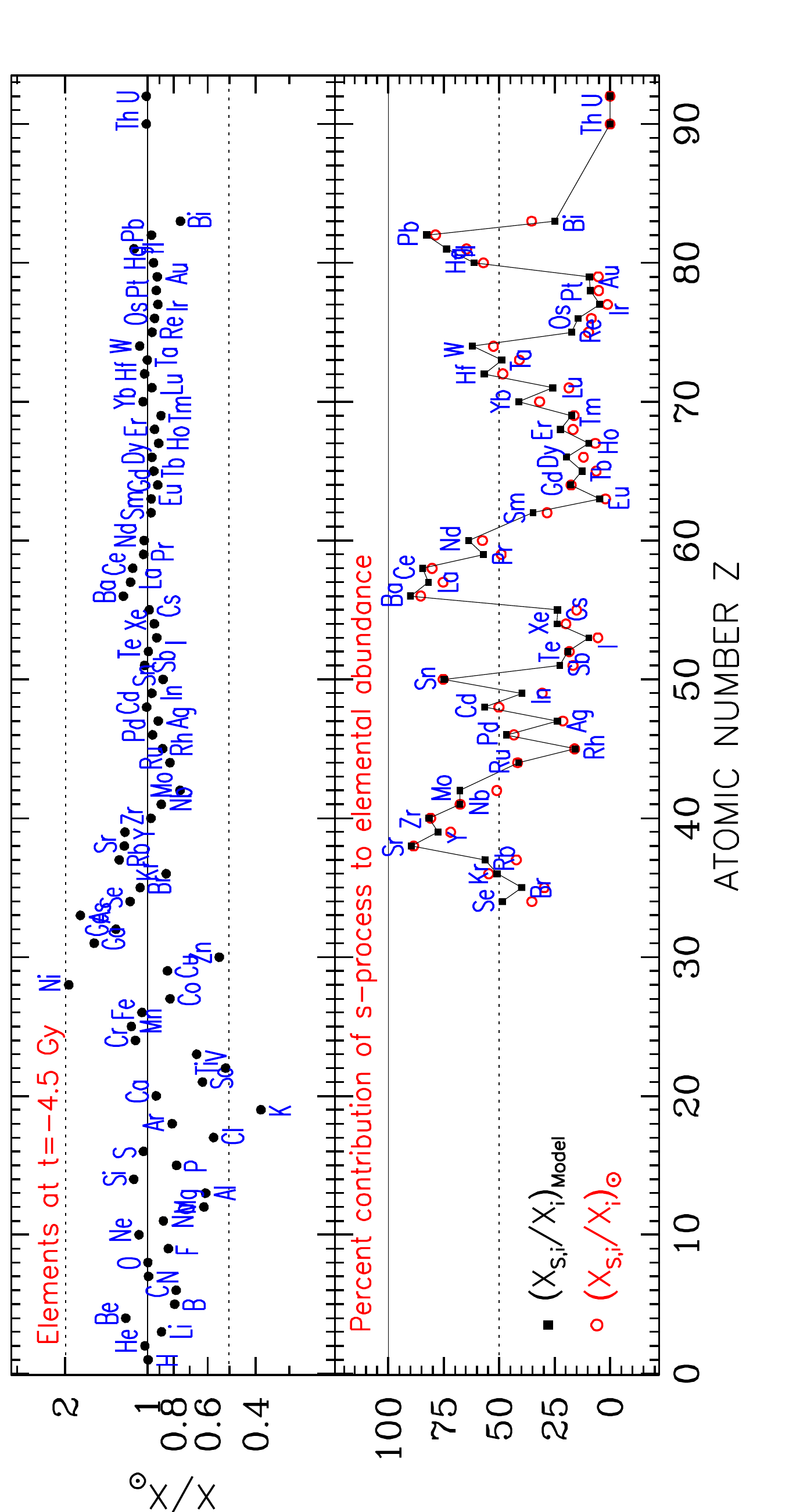}

\par\end{centering}

\caption{\label{fig:f_elements} {\it Top}: Model distribution of elemental abundances obtained at the time of the formation of the solar system compared to the observed solar system data
\citep{Lod09}. Dotted horizontal lines indicate a factor of two with respect to the solar system value. {\it Bottom}: Percentage contribution of the s-process to the elemental abundances at solar system formation. Model results are in black squares and measured solar system data in red open circles. The dotted horizontal line at 50 \% defines elements produced mostly by the s- or the r- processes (above or below it), respectively.}
\end{figure*}

Here we have addressed again the solar LEPP issue making use of our s-element yields in massive stars computed by the first time in a large grid of masses and metallicities including rotation. Furthermore, we have used a much more consistent GCE model than that in \citet{Cr15b}. Our GCE model reasonably fits all the observational constrains in the solar neighborhood (see Fig. \ref{fig:GCE_model}), most of the observed [X/Fe] vs. [Fe/H] relationships (see Section \S \ref{xsufe}) and, as already mentioned, the absolute abundance distribution observed in the solar system (to better than a factor of two, see Fig. \ref{fig:all_isotopes}  ) for almost all the isotopes; we consider this as a notable achievement, taking into account the many uncertainties still affecting stellar and GCE models. As already tested and discussed by \citet{Cr15b}, different stellar assumptions and/or GCE model recipes may lead to flatter distributions than that shown in Figure \ref{fig:f_Sonly}. We note that in the aforementioned analyses, the contribution from massive stars (mainly the weak-s component)  to the s-only distribution was considered in the "classic" way, i.e. starting from the derived solar weak s-contribution and assuming a secondary-like behavior for lower metallicities. Here we include a "realistic" weak s-process contribution from rotating massive stars, based on metallicity-dependent model yields.

We conclude that, considering the large uncertainties at play (both theoretical and observational), our results show clearly that  a solar LEPP mechanism is not required. 

\subsection{Elemental proto-solar composition}
\label{subsec:Elements}

In Fig. \ref{fig:f_elements}  (upper panel) we show the distribution of elemental abundances  of our model  at the time of solar system formation. They are obtained by summing the corresponding isotopic abundances (see Sec. \ref{subsub:Global_isotopes}). These results can be understood in terms of the isotopic results presented in Sec. \ref{subsec:Isotopes} and Fig. \ref{fig:all_isotopes}. 

Most of the intermediate mass elements (C, N, O, F, Ne, Na, Si, S, Ar, Ca, Cr, Mn, Fe, Co and Cu) are co-produced within better than $\sim$20\% of the corresponding solar system values. In fact, the $\alpha$-elements of that list (with  the exception of Mg) are co-produced to better than 10\%. 

A second class of elements, mostly (but not always) of odd charge number Z, are systematically underproduced, ranging from 75\% 
to 40\% of their solar system values. These are Mg, Al, P, Cl, K, Sc, Ti, V, and Zn. On the other hand, Ni is the only element significantly overproduced (by a factor of 2); as discussed in Sec. \ref{subsub:Global_isotopes} this overproduction stems from  the W7 model of SNIa
adopted here; that model is, however, extremely successful regarding its other nucleosynthesis predictions for Fe-peak isotopes, as already discussed in Sec. \ref{subsub:Global_isotopes} (with the exception of $^{54}$Fe).

These results constitute a success of the rotating massive star yields, weighted here with the rotational velocity as described in Sec. \ref{subsec:model}, at least for the majority of the products of their hydrostatic nucleosynthesis. However, the production of the odd charge elements and of those produced through explosive nucleosynthesis  (Sc, Ti, V) as well as Zn,  requires further improvements in stellar nucleosynthesis models (see also Sec. \ref{subsub:Global_isotopes}).

Beyond the Fe peak, we note a small overproduction of the elements produced by the weak s-process, because of the important contribution of rotating massive stars in that range of atomic masses (including Ga, Ge and As). Taking into account not only the uncertainties in the stellar nucleosynthesis models but also the approximate and difficult to calibrate weighting over the rotational velocities, we think that this small excess is well within acceptable limits.

Finally, for all elements above As, the fitting to the solar system abundances is more than satisfactory. Of course, this is obtained here by construction for the pure r-elements (Th and U) and for the pure r-component of all the others. But it is important to check what happens with the s-component, depending on the adopted yields of rotating massive stars and LIM stars. As it turns out, the s-component of each element is  fairly well reproduced. This can be seen in the bottom panel of Fig. \ref{fig:f_elements}, where we display the solar system s-fraction of the heavy elements (red circles; \citealt{Sne08}) and the corresponding values of our model (black squares). 
Note that the r-fractions estimated by Sneden et al. 2008 are determined by considering the average r-process distribution in a handful of very metal-poor stars, all showing a similar (pristine) r-process pattern. {\it The agreement is better than 10\%}, with the exception of Se, Mo and Bi. In the former case, the reason is the overproduction of Se by rotating massive stars, as reported in the previous paragraph. In the case of Mo, the underproduction of its total abundance that we obtain (upper panel) is due to p-isotopes of that element, which are missing from our analysis: $^{92,94}$Mo  make up $\sim$20\% of solar Mo, and for that reason the contribution of the s-process to Mo appears overestimated in the bottom panel of Fig. 
\ref{fig:f_elements}. Finally, the  mono-isotopic Bi is under-produced  in our model (see also Fig. \ref{fig:all_isotopes}), because of insufficient LIM yields, and this is also reflected in the s-fraction of that element, in the bottom panel of Fig. \ref{fig:f_elements}.

Summarizing the content of this section, we wish to emphasize that we manage to reproduce fairly satisfactorily the proto-solar composition of all the heavy elements  but also the s-component of each element, in a model satisfying all the key observational constraints in the solar vicinity. This is not a trivial enterprise, because it involves several factors: star formation rate, LIM stars yields, appropriate weighting of the rotating massive star yields, and a stellar IMF correctly balancing the heavy vs light s-nuclei, produced by the LIM and the massive stars, respectively. 

In addition to that, we also adopted a relation between IDROV (Initial distribution of rotational velocities) and metallicity ($\rm v_{\rm rot}^{\rm ini}$ vs Z),  calibrated in order to have primary \nuk{N}{14} at low metallicities but not overproduce s-elements at intermediate metallicities. Such a relation follows qualitatively the suggestion of the Geneva Group that low metallicity stars should rotate faster than their respective more metal rich counterparts. Such a large number of choices, necessary  to predict the distribution of abundances of all the nuclear species at the time of the solar system formation, implies that our solution is not necessarily unique. Other combinations may produce acceptable results  as well. However, it is important to have such a satisfactory solution if one wishes to extent the investigation to the evolution of the heavy elements and to the study of the role of the s- vs. r- components during that evolution. This is done in the next section.

\subsection{Evolution of [X/Fe] vs metallicity} \label{xsufe}
\label{subsec:Element_evolution}

\begin{figure*}
\begin{centering}
\includegraphics[angle=-90,width=1.0\textwidth]{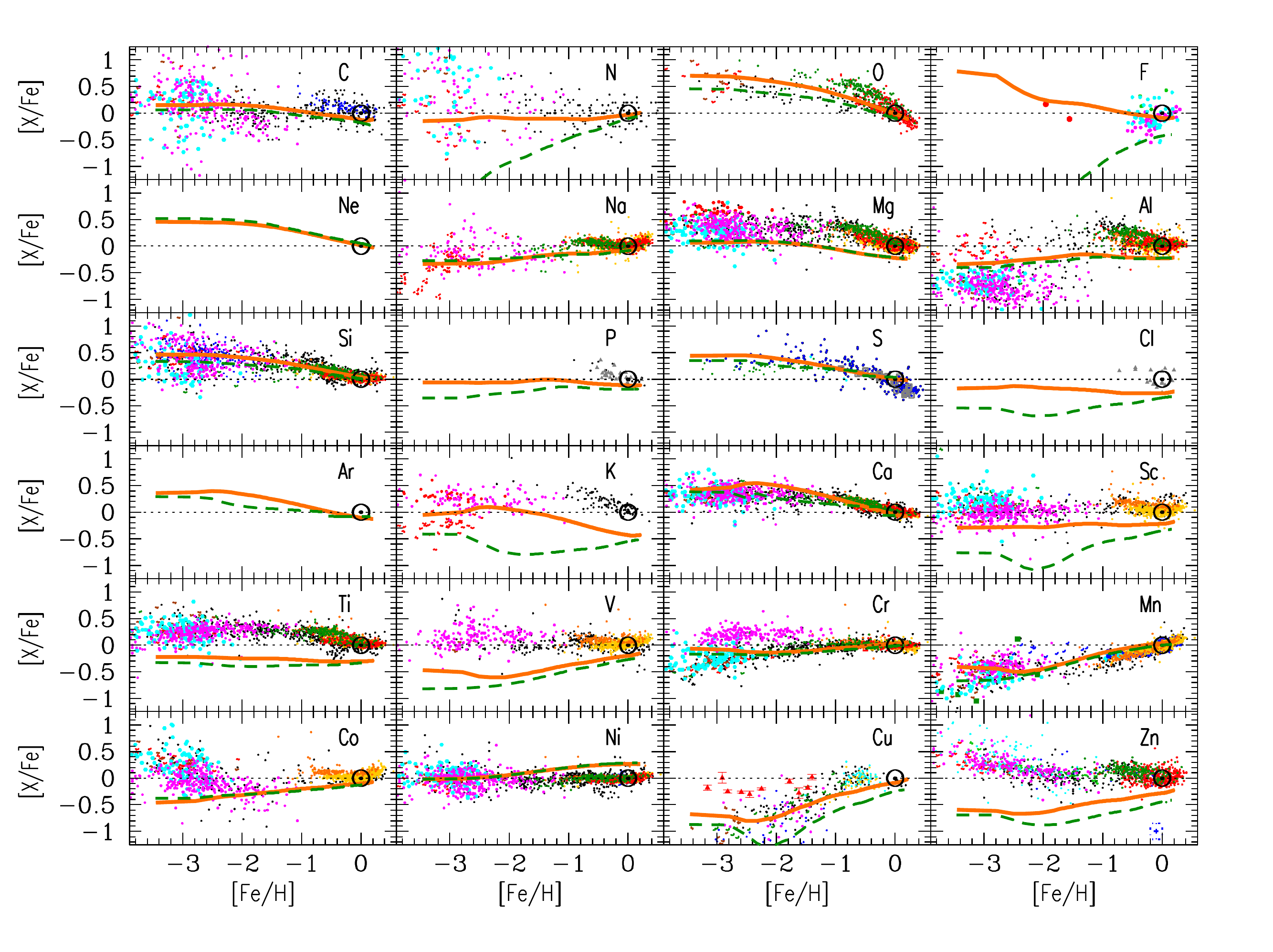} 
\par\end{centering}

\caption{\label{fig:f_elm_evolFepeak} Evolution of abundance ratios [X/Fe] as a function of [Fe/H] for elements up to the Fe-peak and comparison to observational data. 
Our baseline model with rotating massive star yields is in  solid orange curves; the same model but with non-rotating massive star yields is in dashed green curves. Observational data have been
taken from references in Table 1. }
\end{figure*}

In this section we compare the results of our model to a large body of observational data, concerning [X/Fe] abundance ratios in halo and the disk (thick and thin) stars of the Milky Way. We adopt data from a few recent surveys, listed in Table 1, as to keep the data set as homogeneous as possible. However, the dispersion in the data (Figs. \ref{fig:f_elm_evolFepeak} and \ref{fig:f_evol_heavies}) is due, at least partially, to systematic differences in the analysis between different 
data sets. We note also that part of the observed dispersion
at very low metallicities ([Fe/H]$<-2.0$) probably reflects chemical inhomogeneities in the interstellar medium at very early epochs in the evolution of the Galaxy. For heavy elements though, dispersion may result both from inhomogeneities in the ISM \citep[e.g.][]{Ces13}  and from  production in sub-haloes evolving at different rates \citep[see discussion in][]{Prantzos2006,Ishimaru2015}.

\begin{table}
\begin{center}
\caption{ Observational data of [X/Fe] ratios in halo and disk stars used for Figs. \ref{fig:f_elm_evolFepeak} and \ref{fig:f_evol_heavies}. 
}

\begin{tabular}{c c c c }
\hline
Element & Z  & Mostly S or R & References\\
\hline
\\
 C   &    6  &      & 1,3,17     \\
 N   &    7  &      & 1,3,17    \\
 O   &    8  &      & 2,3,7,17    \\
 F   &    9  &      & 9,10,11,12,19,20,21,22,23    \\
 Na  &   11  &      & 1,2,3,5,7     \\
 Mg  &   12  &      & 1,2,3,5,7    \\
 Al  &   13  &      & 1,2,3,5,7     \\
 Si  &   14  &      & 1,2,3,5,7     \\
 P   &   15  &      & 13,14        \\
 S   &   16  &      & 8,13,15     \\
 Cl  &   17  &      & 15     \\
 K   &   19  &      & 3     \\
 Ca  &   20  &      & 1,2,3,5,7   \\
 Sc  &   21  &      & 1,3,5,17    \\
 Ti  &   22  &      & 1,2,3,5,7,17     \\
 V   &   23  &      & 3,5,17     \\
 Cr  &   24  &      & 1,2,3,5,17     \\
 Mn  &   25  &      & 1,3,5,17     \\
 Co  &   27  &      & 1,3,5,17     \\
 Ni  &   28  &      & 1,2,3,5,7     \\
 Cu  &   29  &      & 3,16,17,18     \\
 Zn  &   30  &      & 2,3,17     \\
 Sr  &   38  &  S   & 3,4,17     \\
 Y   &   39  &  S   & 2,3,6     \\
 Zr  &   40  &  S   & 3,4,6     \\
 Mo  &   42  &  S   & 3     \\
 Ba  &   56  &  S   & 2,3,6     \\
 La  &   57  &  S   & 3,4,6    \\
 Ce  &   58  &  S   & 3,4,6     \\
 Pr  &   59  &  S   & 3     \\
 Nd  &   60  &  S   & 3,5    \\
 Sm  &   62  &  R   & 3,4,6    \\
 Eu  &   63  &  R   & 3,4,6     \\
 Gd  &   64  &  R   & 3     \\
 Dy  &   66  &  R   & 3     \\
 Er  &   68  &  R   & 3    \\
 Yb  &   70  &  R   & 3     \\
 Pb  &   82  &  S   & 3     \\   
\hline
\end{tabular}
\end{center}
{\it References}: 1. \cite{Yong2013}; 2. \cite{Bensby2014}; 3. \cite{Roederer2014}; 4. \cite{Battistini2016}; 5. \cite{Adi12}; 6. \cite{Mishenina2013};  7. \citet{Chen2000}; 8. \cite{Caffau2005}; 9. \citet{Ale12};  10. \citet{Jon14}; 11. \citet{Pil16}; 12. \citet{Jon17}; 13. \citet{Caffau2011}; 14. \citet{Maas2016}; 15. \citet{Maas2017}; 16.
\citet{Yan2015}; 17. \citet{Lai2008} ; 18. \citet{Andrievsky2017}; 19.
\citet{li13}.; 20. \citet{mai14}; 21. \citet{nau13}; 22. \citet{rec12};
23. \citet{cun08}.

\label{tab:Obs_data}

\end{table}

Before starting, we note that  a comparison of observations to one-zone models like this one is a standard practice in the field: the vast majority of studies of the chemical evolution of the local halo and disk have been made in such a framework 
\citep{Prantzos1993,Timmes1995,Matteucci1996,Chiappini1999,Goswami2000,Romano2010,Kobayashi2011,Nomoto2013}. However, we stress here that such a model is an oversimplification of the real situation.

Indeed, it is well established now that the Galactic halo did not evolve in a "monolithic collapse", as suggested by \cite{Eggen1962}, but rather by hierarchical merging of smaller sub-haloes, according to the current cosmological paradigm of galaxy formation \citep[e.g.][]{Bell2008}. In that case, there is no unique relation between metallicity and time, since the different sub-haloes evolved at a different pace \citep{Prantzos2006}.	 Models of different degrees of sophistication have been developed along those lines, from a simple semi-analytic models summing up  the different sub-halo chemical histories \citep{Salvadori2007,Komiya2014}  up to full cosmological simulation \citep[e.g.][]{she15}. In particular, it has been shown that such models may help to explain the fact that the observed dispersion of heavy elemental ratios in the halo is much larger than that observed for the intermediate mass elements \citep{Ishimaru2015}. Furthermore, the possibility of inhomogeneous mixing of the stellar ejecta may also alter the results of simple one-zone models and produce dispersion in abundance ratios, e.g. \citet{Ces13,ces14}.

In a similar way, it is now obvious that the history of the Milky Way disk(s) is more complex than described by 1-zone models, as we discussed  in the end of Sec. \ref{subsec:model}: several local observables cannot be interpreted in the framework of such models, and require some mixing of stellar populations with different histories, bringing in the solar vicinity mostly stars from the inner galactic regions. 

Despite those shortcomings, simple one-zone models still play an important role in studies of galactic chemical evolution, since they allow one to probe some key features, like e.g. the dependence of yields on metallicity, the relative importance of various metal sources evolving on different timescales (e.g. massive stars vs. SNIa or LIM stars), the local star formation history (through the G-dwarf distribution), etc. This is why we shall still use such a simplified model here, and compare our model predictions only with the average observed [X/Fe] vs. [Fe/H] trends, leaving a more detailed work for the future in the framework of the radial migration model of the galactic disk by \cite{Kubryk2015a}. 	

A word of caution is also required regarding the validity of one-zone models in the case of the halo. In the absence of reliable stellar ages, [Fe/H] is used as a proxy for time.  In the local disk the observed age-metallicity relation serves as a useful constraint to models (Fig. 6), establishing a one-to-one relation between age and metallicity. In the halo, however,  no such  relation is observed and, consequently, no constraint may exist on the timescale 
in which metallicity reached a given value, say [Fe/H]$=-3$ or $-2$.
As a result, it is difficult  to establish at what metallicity  a long-lived source, like AGBs, enriched the Galactic gas with its nucleosynthetic products. These considerations are important regarding the  predictions of one-zone model for  the earliest stages of the Galaxy, as we shall discuss in the following subsections.

\subsubsection{Up to the Fe-peak}
\label{subsub:Elm_upto_Fe}

Our results for the evolution of intermediate mass elements, up to the Fe-peak, are displayed in Fig. \ref{fig:f_elm_evolFepeak}. The solid orange curve is our baseline model with the averaged yields of rotating massive stars. For comparison purposes, we display also the results with non-rotating massive star yields (dashed green curve), everything else (SFR, IMF, etc.) being kept the same.

Comparison of the two model curves shows that rotation
affects the evolution of only a handful of  intermediate mass elements. The most significant difference is obtained for N and F, their behavior turning from a secondary one (without rotation) to a primary one (with rotation), as we discuss in more detail in the next sub-section. 

\noindent
{\it \underline{Carbon}}: Although carbon is clearly an $\alpha$ element from the theoretical point of view, observationally it behaves as Fe.
The most straightforward way to interpret this trend is by assuming that C and Fe are exclusive products of massive stars at halo metallicities, while the Fe increase through SNIa at higher metallicities is balanced by C production from long-lived LIM stars. Alternatively,
metallicity-dependent C yields from massive star winds may enhance C production at high metallicities, with no need for C production by LIM stars \citep{Prantzos1994}. The actual situation is much more complicated by uncertainties in C yields due to stellar evolution and nucleosynthesis (e.g. the $^{12}$C($\alpha$,$\gamma$)$^{16}$O rate and the treatment of convection during the late stages of core He burning), to the fact that C may be produced by both massive and LIM stars and to the uncertainties in the chronology of the halo (making it unclear at what metallicity a LIM star of a given mass ejected its C in the halo ISM). These uncertainties are reflected in the detailed investigation of \cite{Romano2010} adopting different sets of yields. Here we simply display in Fig. \ref{fig:f_elm_evolFepeak}  our own results (top left), showing a slight overproduction of C ($\sim$0.2 dex) at low metallicities, which is shared by both non-rotating and rotating models. This is in general not the case with yields of \cite{Woo95} \citep[e.g.][]{Timmes1995,Goswami2000} or of \cite{Nomoto2013} which produce [C/Fe]$\sim 0$. It is possible that the use of the $^{12}$C($\alpha,\gamma$)$^{16}$O rate from \citet{kun02} in the case of LC2018 models, leads to somewhat larger production of  C; the amount of mixing assumed in those models may also have a similar effect. Alternatively, an IMF somewhat steeper than adopted here, that is with a slope $\alpha<$--1.35, should also reduce the C/Fe ratio of massive stars. In any case, we emphasize that the [C/O] ratio of our models at low metallicity is clearly sub-solar, in agreement with observations of halo stars and low-metallicity HII regions \citep[e.g.][]{est09,est14,nis14,nak16}.

On the other hand, in our baseline model about 1/3 of the solar carbon is produced by LIM stars while $\sim 2/3$ comes from massive stars. We notice that the contribution by LIM stars should be considered as an upper limit since we have not included net yields from stars in the mass range 7-10 M$_\odot$. These stars may undergone hot hydrogen burning at the base of the convective envelope (HBB) at the end of their evolutionary phase (the super-AGB phase) and, in consequence, they may deplete $^{12}$C and produce some $^{14}$N. However, the actual stellar mass range where HBB may occur and its dependence on the metallicity is basically unknown. In particular it depends dramatically on the treatment of the coupling between burning and mixing, mass-loss rate etc therefore,  theoretical yields in this mass range are very uncertain \citep[see e.g.][and references therein]{KL14}. Nevertheless, the contribution to the chemical evolution of the Galaxy of the stars in this stellar mass range, when weighted with the IMF, should be limited except for a few nuclei as $^7$Li, $^{17}$O, $^{26}$Mg and $^{26}$Al. Therefore, considering the above discussion and the observational errors we believe that our predicted [C/Fe] vs. [Fe/H] trend agrees with the average observed one.

\noindent
{\it \underline{$\alpha$-elements}}: The observed evolution of most $\alpha$ elements like O, Si, S, Ca, is fairly well reproduced by our model. The "plateau-like" behavior of [$\alpha$/Fe] at low metallicities (halo stars) is followed by a slow decline at higher metallicities (disk stars) and is attributed to the delayed action of SNIa, producing about half of solar Fe. The noble gases Ne and Ar, for which there are no observations in stars, display a similar behavior. Rotation changes very little or not at all the results. The evolution of $\alpha$ elements is well reproduced, in general, by most (if not all) GCE models and for all sets of stellar yields (see Table 2). This is one of the well established results in the field of stellar nucleosynthesis and GCE studies. In fact, it is reassuring that rotation does not affect that result (even if it increases slightly the amount of C and O at low metallicity).

\noindent
{\it \underline{Magnesium}}: The evolution of Mg is not well reproduced by our models. As we discussed in Sec. \ref{subsub:Global_isotopes}, the  Mg isotopic abundances obtained at solar system formation are under-produced with the yields adopted here, and this feature characterizes the whole evolution of Mg. We notice that similar results are obtained with the yields of \cite{Woo95} \citep[see e.g][]{Timmes1995,Goswami2000,Francois2004}, while those of \cite{Nomoto2013} reproduce correctly the behavior of Mg as an $\alpha$ element \citep[see also][]{Romano2010}.

\noindent
{\it \underline{Odd elements}}: The odd mono-isotopic elements Na and Al display the theoretically expected odd-even effect, due to a lower production at lower metallicities. That behavior is in qualitative agreement with observations, but their abundance determinations are known to be affected by non-LTE effects, thus precluding any strong conclusions. The yields of both \cite{Woo95} and \cite{Nomoto2013} also show that behavior.

Few observations exist for the next odd-Z elements, namely P (which is also mono-isotopic) and Cl (which has two isotopes $^{35}$Cl and  $^{37}$Cl).  These observations concern disk, not halo, stars and in the case of Cl they concern its major isotope $^{35}$Cl \citep{Maas2017}. We obtain a primary, Fe-like, behavior for those two elements with our rotating massive star yields, while the non-rotating ones display a small metallicity effect. Such an effect is also obtained in \citet{Timmes1995} but not in \citet{Nomoto2013}.

\noindent
{\it \underline{Before the Fe-peak }}: The four elements before the Fe-peak, K, Sc, Ti and V, are persistently found to be deficient  in all sets of massive star yields and GCE studies, when compared to both solar abundances \citep{Goswami2000,Kubryk2015a}  and observations of halo stars \citep{Timmes1995,Goswami2000,Francois2004,Romano2010,Nomoto2013}. The case of Ti is particularly intriguing: its main isotope is $^{48}$Ti and comes from the decay of $^{48}$Cr, nucleus synthesized by the incomplete explosive Si burning. It correctly shows the typical trend of a primary nucleus but its ratio Ti/Fe is systematically shifted downwards with respect to the observed value. A proper analysis of the physical conditions in which $^{48}$Cr is produced and the possible importance of some net production out of the equilibrium (some kind of $\alpha$-rich freeze out) should be investigated to shed light on the presence of this puzzling offset. Sc and V are odd and mono-isotopic elements, while K is odd but has three isotopes and Ti is even and has six isotopes. An inspection of Fig. \ref{fig:f_elm_evolFepeak} shows that rotation improves slightly the results for all those elements -  particularly in the cases of K and Sc at low metallicities - however,  without solving the problem of their overall underproduction. 

\noindent
{\it \underline{Fe-peak elements}}: A large fraction of the solar abundance of the Fe-peak elements Cr, Mn, Co and Ni (more than 50\%) is produced by SNIa; since no metallicity-dependence in the yields of those elements is considered here from SNIa, the increase of [Mn/Fe] with metallicity obtained in our results is solely due to the metallicity-dependent yields of Mn from massive stars. This feature, consistent with observations, characterizes also the 
yields of \cite{Woo95} and \cite{Nomoto2013}. However, X-ray observations of Mn/Fe and Ni/Fe in the SNIa remnant 3C 397 with {\it Suzaku}, lead
\cite{yam15} to model SNIa explosions at various metallicities,  finding a substantial metallicity dependence of those ratios. Thus, the roles of CCSN and SNIa in the late production of Mn is an issue not settled yet.

In the case of Cr, all sets of yields, including ours, produce a $\sim$constant [Cr/Fe] ratio at all metallicities \citep{Timmes1995,Goswami2000,Romano2010,Nomoto2013}. This trend is confirmed by observations down to [Fe/H]$\sim-1.5$ but at lower metallicities the observational situation is unclear, probably being affected by systematic errors and/or inhomogeneities in the ISM.

The case of Co remains puzzling. Models predict either a small decline of [Co/Fe] as [Fe/H] decreases
\citep[this work][]{Timmes1995,Goswami2000} or a $\sim$constant value \citep{Nomoto2013}. This is, however, in stark contrast with observations who show a rise of [Co/Fe] below [Fe/H]$\sim-2$. Several ideas have been put forward to interpret this trend, like energetic supernovae explosions or the "mixing and fallback" occurring in the inner supernovae regions \citep[see the discussion in][]{Nomoto2013}, but the situation is unclear yet. Note that the yields of massive stars have been computed here with "typical", and not high, explosion energies and take into account the mixing-fallback mechanism, as described in section \ref{subsub:expnuc}.

Finally, the rise of [Ni/Fe] with metallicity in our results is due to the well known overproduction of that element in the "standard" models W7 (and W70 for white dwarfs of initially metallicity Z$=0$) of SNIa. The above-mentioned results of \cite{yam15}, finding a Ni/Fe ratio increasing with progenitor metallicity for SNIa, would certainly enhance that trend, making the discrepancy with the observed flat behavior of Ni/Fe even worse.

\begin{table}
\begin{center}
\caption{ Assessment of massive star yields, compared to observations of halo and local disk stars. based on analysis of \citet{Timmes1995,Goswami2000,Francois2004,Romano2010,Kobayashi2011,Nomoto2013}
}

\begin{tabular}{c c c c c c}
\hline
Elm. & Z & WW95  & NKT13 & LC2018 &  Comments\\
\hline
 C   &    6  &  + & + &   +  &      \\
 N   &    7  & --  & -- &  +   & Primary from RMS$^a$     \\
 O   &    8  & +  & + &   +  &  $\alpha$ - OK  \\
 F   &    9  & --  & -- &  +   &  2/3 of X$_{\odot}$  from RMS    \\
 Ne  &   10  &   &  &     &      \\
 Na  &   11  & +  & + & +   &  Odd  - OK  \\
 Mg  &   12  & --  & + &  --   & $\alpha$     \\
 Al  &   13  & +  & + &  +   &   Odd - OK  \\
 Si  &   14  & +  & + &  +   &  $\alpha$ - OK     \\
 P   &   15  &   &  &     &    Odd  \\
 S   &   16  & +  & + &  +   &  $\alpha$ - OK   \\
 Cl  &   17  &   &  &     &   Odd   \\
 Ar  &   18  &   &  &     &      \\
 K   &   19  & --  & -- & --    &  RMS improve at low Z    \\
 Ca  &   20  & +  & + &  +   &  $\alpha$ - OK \\
 Sc  &   21  & --  & -- &  --   & RMS improve at all Z   \\
 Ti  &   22  & --  & -- &  --   &  $\alpha$ ?  - Problematic  \\
 V   &   23  & --  & -- &  --   &  Problematic    \\
 Cr  &   24  &  + & + &  +   &      \\
 Mn  &   25  & +  & + &  +   &      \\
 Co  &   27  & --  & -- & --  & HESN$^b$ at low Z?    \\
 Ni  &   28  & +  & + &  +   &  SNIa overproduction    \\
 Cu  &   29  & +  & + &  +   &   OK  \\
 Zn  &   30  & --  & -- & --   & HESN at low Z?\\

\hline
\end{tabular}
\end{center}
WW95: \cite{Woo95}; NKT13: \cite{Nomoto2013}; LC2018: \cite{LC18}, with metallicity-dependent weighted rotational velocities (this work). \\
{\it Assessment:} +: Broad agreement with observations over the whole metallicity range; -- : disagreement with observed evolution in some metallicity range and/or solar abundance.
A blank entry means lack of enough observational data or partially in agreement with observations. (a) RMS: rotating massive stars. (b) HESN: high energy supernovae.

\label{tab:Yield_comparison}
\end{table}

\noindent
{\it \underline{Cu and Zn}}: The observed decline of [Cu/Fe] with decreasing metallicity is well reproduced by our baseline model; rotation increases the [Cu/Fe] ratio by $\sim$0.2 dex over the whole metallicity range. This decline is due to the secondary-like nature of the dominant isotope $^{64}$Cu, produced mainly by neutron captures in the He-core for metallicities higher than [Fe/H]$\sim -2$ \citep[see discussion in Sec. \ref{subsub:Global_isotopes} and in][]{Romano2007}. We note, however, that recent NLTE analysis of high resolution observations of a few halo stars \citep{Andrievsky2017} suggests quasi-solar [Cu/Fe] at those low metallicities (red points with error bars in the corresponding panel of Fig. \ref{fig:f_elm_evolFepeak} ; if confirmed, those observations may put in question our current understanding of Cu nucleosynthesis.

In contrast, our model fails to reproduce the observed evolution of Zn, which is clearly underproduced, a feature shared by the yields of \cite{Woo95} and \cite{Nomoto2013}. To explain the behavior of both Cu and Zn at low metallicities, specific models have been suggested, invoking simultaneously: inhomogeneous early chemical evolution, a large fraction of hypernova at low metallicities ($\sim50\%$ of stars with M$>2$ \ms) and a peculiar mechanism of "mixing and fallback" of the inner supernova layers \citep[see discussion in][]{Nomoto2013}. Let us again point out that the yields for massive stars adopted in this paper have been computed for typical explosion energies and with the mixing-fallback as described in \ref{subsub:expnuc}. The analysis of how the GCE results change by changing one or both these two parameters is beyond the scope of this study, the aim of which is to test average yield trends in the framework of a homogeneous model. It is clear, however, that the early evolution of our Galaxy may have been much more complex than assumed here; in particular, the proportion of high-energy supernova may have been larger than at the low metallicities of the halo than at the higher metallicities of the disk (perhaps in line with the higher fraction of rotating massive stars), thus explaining the early evolution of [Zn/Fe].

In Table 2 we asses briefly the status of GCE models vs. observations in the halo and local disk, considering the three main sets of stellar yields currently available,
namely \cite{Woo95}, \cite{Nomoto2013} and LC2018. Obviously, such a comparison cannot be quantitative, but only qualitative.
Indeed, the yields of the former two studies
concern stars with no mass loss or rotation, have different prescriptions for the explosive nucleosynthesis  and do not extend above 40 \ms \ in the case of \cite{Woo95} or 
70 M$_\odot$  in the case of \cite{Nomoto2013}, thus requiring an extrapolation to higher masses. For those reasons, our comparison will be only indicative, trying to present generic features on the basis of published GCE studies with those sets of yields.

An inspection of Table 2 shows  the few cases where rotating massive stars bring improvement over non-rotating ones: primary N, F, K and Sc (although in the last two cases disagreement with observations still remains). It also reveals common problems in all sets of yields for K, Sc, Ti, V, Co and Zn (the latter two possibly solved by invoking early hypernovae). For the remaining elements, agreement with observational data is rather (or very) satisfactory.

In the next sub-sections we discuss in more detail the specific cases of N and F, as well as the one of the $^{12}$C/$^{13}$C
isotopic ratio and of the s-element evolution, since the yields of the involved species are found to be critically affected by rotation.

\subsubsection{Nitrogen and fluorine}
\label{subsub:NandF}

Nitrogen is produced in the CNO cycle by conversion of almost all the initial C and O; it should then behave as a secondary element. However,
the observed [N/Fe] vs [Fe/H] trend in galactic stars (Fig. \ref{fig:f_elm_evolFepeak}) and 
the [N/O] vs [O/H] one derived in metal-poor extragalactic HII regions rather show a primary-like behavior (for sub-solar metallicities) \citep[see e.g.][]{pil10}.

The origin of early primary N remained elusive for a long time. Intermediate mass stars can make, indeed, primary N through HBB, but they are expected to enrich the ISM not much earlier than SNIa, i.e. at metallicities $-2<$[Fe/H]$<-1$.
Stars of intermediate mass, rotating at 300 kms$^ {-1}$ and mixing primary C from the inner He-layers into the outer proton-rich zones, have been 
suggested as the main sources of primary N in  
\cite{MM02,MM2002b}. However, 
\cite{Prantzos2003}, using full scale GCE models, showed that the evolutionary timescale of such stars leads to sufficient amounts of primary N only above [Fe/H]$>-2$.

Subsequent calculations of the Geneva group for much faster rotating stars (v$^{\rm ini}_{\rm rot}=800$ kms$^{-1}$) suggested the production of large amounts of primary N  from massive  stars at the lowest  observable metallicities. Using GCE models \cite{Chiappini2006} showed then that such stars may explain the appearance of primary N in the early Galaxy. Our own results (Fig. \ref{fig:f_elm_evolFepeak}) confirm this conclusion. However, our model includes a mixture of stars with different rotational velocities, the fastest of which rotate at 300 kms$^{-1}$, i.e. much less than the models put forward by the Geneva group. This difference shows clearly that the outcome of rotating star models depends not only on the values of the rotational velocity but also (and in a critical way) on the adopted mixing prescription and remains an unsettled issue. It is also important to mention that
in our GCE model LIM stars contribute by $\sim 45\%$ to the solar nitrogen. We notice again that this might be considered a lower limit since we have not included the possible contribution from LIM stars through the HBB.

The evolution of fluorine attracted considerable attention recently, regarding the contributors to this element in the Galaxy \citep{ren04,Kobayashi2011b,Abi15,Pil16,Jon17}.
As discussed in Sec. \ref{subsub:Global_isotopes}, we obtain $\sim 85\%$ of the proto-solar F abundance with $\sim 1/3$  coming from LIM stars and $\sim 2/3$ from rotating massive stars. Non-rotating massive stars produce essentially no F through nuclear fusion reactions. For that reason, \cite{Woosley1990} favoured the role of $\nu$-process in CCSN, producing \nuk{F}{19}, along with some \nuk{Li}{7}, as well as part of \nuk{B}{11}.
However, the many uncertainties still affecting CCSN explosions, and in  particular the $\nu$-spectrum, make the resulting yields highly uncertain.   
Taking rotation into account allows to explain at one stroke both the primary behaviour of N and the proto-solar F, with no need for $\nu$-induced nucleosynthesis.

\begin{figure}
\begin{centering}
\includegraphics[angle=-90,
width=0.48\textwidth]{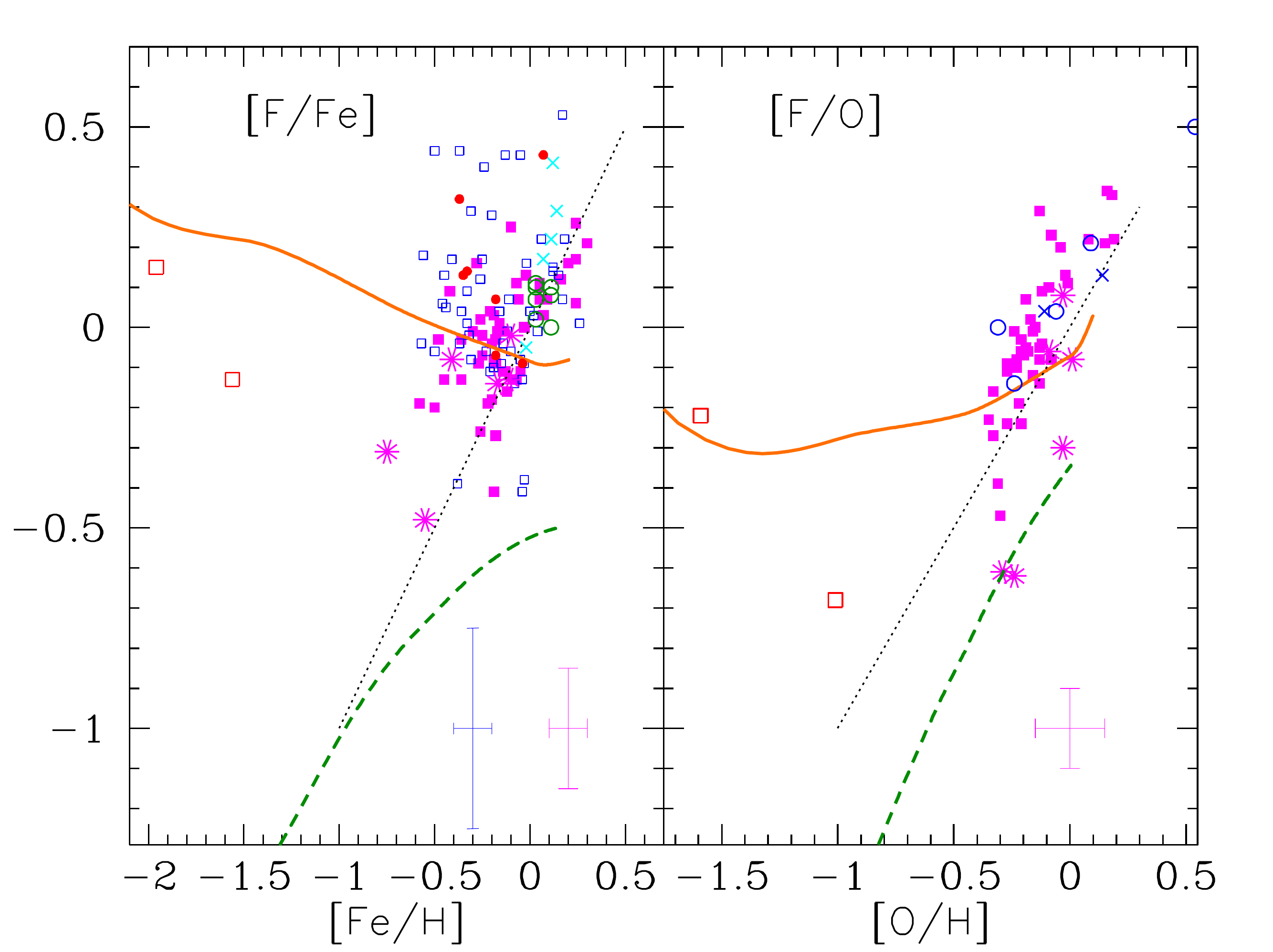} 
\par\end{centering}

\caption{\label{fig:f_fluorine} 
Evolution of abundance ratios of [F/Fe,O] vs. [Fe/H] ({\it left}) and vs. [O/H] ({\it right}). Model curves are as in Fig. \ref{fig:f_elm_evolFepeak} ({\it orange solid}: baseline model; {\it green dashed}: non-rotating stars). In both panels, {\it dotted} lines indicated a slope of 1 (purely secondary behaviour). Observational data for disk stars are from \citet{Pil16} (blue open squares), \citet{mai14} (green open circles), \citet{nau13} (cyan crosses), \citet{rec12} (red filled squares)
and \citet{Jon17} (magenta filled squares), while for halo stars are from \citet{li13} (red open squares) and for bulge stars from 
\citet{Jon14} (magenta asterisks) and \citet{cun08} (blue open circles).
}
\end{figure}

Regarding the F evolution, data of [F/Fe] for disk stars and for metallicities in the range $-0.5 <$[Fe/H]$<0.3$ display important scatter (Fig. \ref{fig:f_elm_evolFepeak} and Fig. \ref{fig:f_fluorine}, left panel). The scarce available data at lower metallicities \citep[the two red points in Figs. \ref{fig:f_elm_evolFepeak} and Fig. \ref{fig:f_fluorine} from][]{li13} is rather uncertain since they correspond to a CH-star (probably polluted by the companion star when it was in the AGB phase) and a star whose kinematics might indicate that it has been accreted from a satellite galaxy. Thus, these data points do not allow to constrain the models.

The observed [F/O] ratios have been obtained by \cite{Jon17} for disk stars; data for bulge stars are from \cite{cun08} and \cite{Jon14} (right panel of Fig. \ref{fig:f_fluorine}). They suggest a secondary behavior of F against O for  stars with [O/H]$>-0.5$, which may be also exist against Fe (see purple circles in the left panel of Fig. \ref{fig:f_fluorine}). This secondary-like behavior at metallicities close to solar, if confirmed, would imply a more important role of LIM stars in late F production than found here. The small upwards trend of [F/O] vs. [O/H] that we obtain around  [O/H]$\sim 0.0$ is indeed due to the secondary F production in LIM stars. We checked that an upwards revision of their yields by a factor of two would bring a better agreement of our model with the data in that metallicity range, bringing the LIM stars contribution to proto-solar F at the same level as the one of rotating massive stars. On the other hand, if a secondary-like behaviour of [F/O] vs [O/H] (or [F/Fe] vs [Fe/H]) is found to exist at lower metallicities, it would disagree with our finding of primary-like F evolution. It would be an indication that 
either rotating massive stars do not contribute significantly to \nuk{F}{19} or that the calibration of the rotationally induced mixing must be revised.

In that respect, we note that the steep rise of the computed [F/Fe] ratio for [Fe/H]$<-2.0$ in our models (Fig. \ref{fig:f_elm_evolFepeak}) is entirely due to the strong increase of the F production by rotating massive stars with [Fe/H]$=-3$. In particular it is due to a specific phenomenon that occurs in the 15 and 20 \ms \ stellar models rotating initially at 150 kms$^{-1}$. In both these models, during the central Ne burning a small He convective shell forms in the radiative He tail below the main He convective shell. As the evolution proceeds, this small convective shell eventually merges with the main He convective shell and the net result is the formation of a new He convective shell whose base is located more internally than before and hence it is exposed to a higher temperature. Since the He convective shell is still very rich in $^{15}$N, the higher temperature at its base  leads to a burst of F production. It is very hard to know if this sequence of events is "realistic" or not, the more so, since it affects only two stellar models. Our understanding of the growth of the instabilities that lead to  mixing of stellar material is still too poor. Therefore we need to  be guided by  observations to evaluate the extent of various mixing processes. This is the reason why we are testing the new very extended grid of stellar models of LC2018 with a detailed GCE model considering all available observational data.

Other choices of the fractional contribution  of rotating massive stars than the one we made (Fig. \ref{fig:Vrot}), could  give a milder increase, or perhaps a flat trend at low [Fe/H]. Therefore, measurements of [F/Fe] ratios in halo stars (a difficult task),  will be critical  to {\it calibrate} the rotational yields in low metallicity massive stars.  On top of all these arguments it must always be reminded that the fluorine nucleosynthesis is also strongly affected by uncertainties affecting some key nuclear cross sections \citep{cr14}, as the $^{19}$F(p,$\alpha)^{16}$O and the $^{19}$F($\alpha$,p)$^{22}$Ne reactions. Their rates have been recently studied by \cite{ind17} and \cite{piz17}, who propose increased values by a factor 1.5 and 4, respectively, at the temperatures of interest. 

Finally, we note that even if the primary production of both N and F is due to rotation, the two elements are made in different locations and times in the stars, so their behaviour with metallicity should not be expected to be necessarily the same. N is made in the H burning shell during central He burning, from C mixed  in that shell from the He rich zone.
Only later when central He is exhausted, the He convective shell forms and part of the previously formed N is used to produce F.  As a result, the  production of N is, in principle, not tightly correlated to the one of F. For instance, a larger convective shell would reduce the amount of N and increase the one F, while a smaller one would work in the opposite direction.

\subsubsection{The evolution of the $\rm {^{12}C/^{13}C}$ ratio}
\label{subsub:Cratio}

\begin{figure}
\begin{centering}
\includegraphics[width=0.48\textwidth]{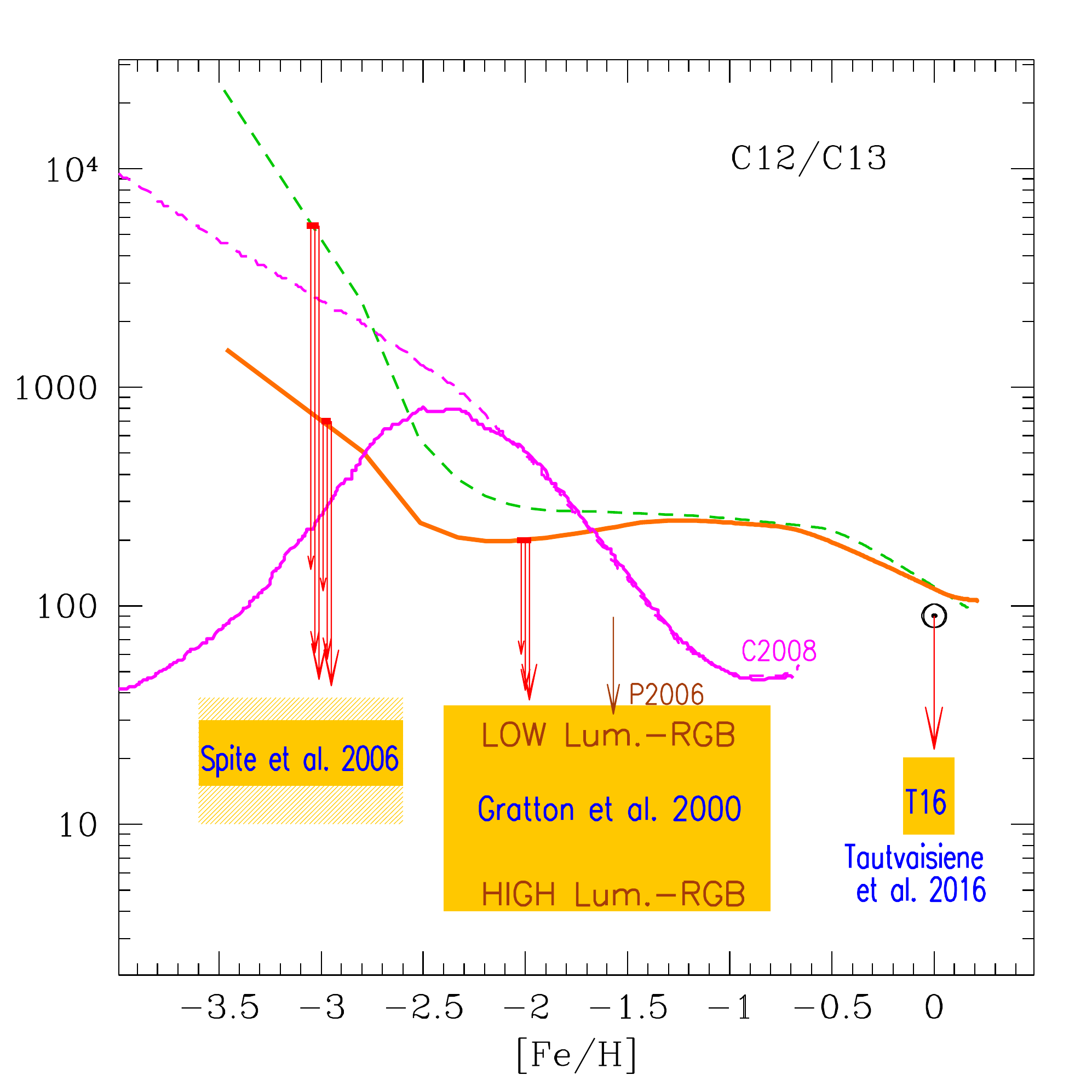} 
\par\end{centering}
\caption{\label{fig:f_Cisotopes} 
Evolution of the $^{12}$C/$^{13}$C isotopic ratio in our baseline model (orange solid) and with non-rotating massive stars (green dashed). The magenta curves represent the results of \citet{Chiappini2008}, for non-rotating stars (dashed) and for rotating ones (solid). Shaded regions indicate the range of observations for red giants at various metallicities, from \citet{Spite2006,Gratton2000,Taut2016}; the latter correspond to open clusters of age $\sim$1 Gyr, i.e. turn-off mass of $\sim$2 \ms. Arrows (in triplets) indicate internal depletion in red-bump stars of 0.8, 0.85 and 0.9 \ms (from left to right), starting  from initial \ciso \ ratios obtained in our GCE model: $\sim$6500 for non-rotating and 750 for the baseline model at [Fe/H]$=-3$, and 200 for the baseline model at [Fe/H]$=-2$. 
P2006 indicates the result by Palacios et al. (2006) for a 0.85 \ms \ model with solar initial \ciso. At solar metallicity we show a 2 \ms \ model, as apropriate for clusters of age $\sim$1 Gyr.
}
\end{figure}

The evolution of  the \ciso \ ratio at low metallicities has been suggested  as another indication of the role played by massive fast-rotating stars in the early halo phase \citep{Chiappini2008}. GCE models with non-rotating stars predict a secondary production for \nuk{C}{13}, similar to the one of \nuk{N}{14}. This results in a very high \ciso \ ratio at low metallicity, of the order of several $10^3$ \citep{Prantzos1996,Kobayashi2011b,Chiappini2008}, much higher than the solar value of $\sim 90$. Figure \ref{fig:f_Cisotopes} shows the trend predicted by \cite{Chiappini2008} (magenta dashed line) together to our current predictions (green dashed line) for the non rotating case. The flattening shown by our predictions for [Fe/H]$>-2.5$ is due to the quasi primary production of \nuk{C}{13} by massive AGB stars.

The inclusion of rotation in the modeling of the stars changes considerably this scenario because of the primary production of \nuk{C}{13} by the rotationally driven mixing. By using yields of massive stars rotating at 800 kms$^{-1}$ at Z$=10^{-8}$ and interpolating to models rotating at 300 kms$^{-1}$ at disk metallicities, \citet{Chiappini2008} found a dramatically different evolution of \ciso:   at the lowest metallicities that ratio is found to be lower than its solar value (i.e. a factor of $\sim 100$ decrease with respect to the non-rotating case at [Fe/H]$=-3.5$), then it increases constantly up to $\sim 10^3$ at [Fe/H]$\sim-2.5$ and then it follows the steeply decreasing trend of the non-rotating case at larger metallicities (solid magenta curve in Fig. \ref{fig:f_Cisotopes}).

Our GCE model with rotating star yields (orange solid curve) also predicts a significant reduction of the \ciso \ ratio at low metallicities, but not as extreme as in the \citet{Chiappini2008} models. The reason for such a difference lies on both the different yields and on the different trend of the initial rotational velocity with the initial stellar metallicity, that are adopted in the two studies. 

The comparison between the predicted trends and the observational data could, in principle, help to understand the role played by rotation in determining the \ciso \ ratio at  various metallicities. Unfortunately, this ratio is observed only in red giants where it is well known that various phenomena alter the initial \ciso \ ratio. The first one is the first dredge-up (FDU), that lowers this ratio by a factor depending on the initial chemical composition and stellar mass.  In order to quantify the effect of the FDU at various metallicities,  when the initial ratio is the one predicted by our GCE models, we have thus computed the evolution at  [Fe/H]$=-3$ and  [Fe/H]$=-2$ of three stars with masses 0.8, 0.85 and 0.9 \ms \ with corresponding evolutionary timescales of $\sim$12.5, 10 and 8.5 Gyr, respectively. The  initial \ciso \ was adopted from our GCE baseline model and was equal to 750 for [Fe/H]$=-3$ and 200 for [Fe/H]$=-2$, respectively. The red arrows in Fig. \ref{fig:f_Cisotopes} show the effect of the FDU: in all cases the \ciso \ ratio drops by large factors, the larger masses showing larger decrease: the 0.85 and 0.9 \ms \ stars reach values close to the upper limit of available observations, which concern red giants at the red bump \citep[i.e. after the FDU,][]{Spite2006,Gratton2000}.   For completeness, we show in Fig. \ref{fig:f_Cisotopes} the effect of the FDU  for  higher initial metallicities, and a solar initial \ciso : for [Fe/H]$=-1.5$ and a 0.85 \ms \ star calculated by \cite{Palacios2006} and for [Fe/H]$=0$ a 2 \ms \ star calculated by us. The latter case is for comparison with the data of \cite{Taut2016} concerning the open clusters NGC 2324, 2477 and 3960, estimated to be $\sim$1 Gyr old. 

On the basis of the obtained results, we emphasize that the FDU alone is able to deplete the initial stellar \ciso \ by  large factors, those factors increasing with decreasing stellar metallicity and increasing stellar mass. At [Fe/H]$=-3$, we find that the decrease may reach two orders of magnitude for stars of 0.85-0.9 \ms \ reaching the red bump. This result is in good agreement - within error bars - with the highest \ciso \ ratios observed in red bump stars of low metallicity, as shown in Fig. \ref{fig:f_Cisotopes}. 

Furthermore,
 observations \citep[see e.g.][and references therein]{Gratton2000} show that this ratio drops by  another order of magnitude as the stars clump from the red bump to the tip of the Red Giant Branch (RGB). Different physical mechanisms have been invoked to explain such an unexpected behavior, like rotation itself through shear effects and meridional circulation, gravity waves, magnetic buoyancy and/or molecular weight inversions (thermohaline mixing) \citep[see e.g.][and references therein]{Charbonnel1995,den10}. Although none of these mechanisms has been definitely identified as  responsible for such a continuous mixing, it is well established that such mixing occurs in nature and contributes to alter the surface chemical composition of the stars climbing along the RGB. Those additional mechanisms, which are not considered in the stellar models that we show in Fig. \ref{fig:f_Cisotopes}, would
 extend the displayed arrows downward by an order of magnitude or so, i.e down to the values found in high luminosity RGBs by \cite{Gratton2000} and \cite{Spite2006}.

 We note here that there are some unevolved very metal-poor stars which show low \ciso \ (<15) ratios \citep[e.g.][]{luc03,coh06,mas12}. However, the overwhelming majority
of these stars are known to be carbon enhanced metal-poor stars with s-element
enhancements (CEMP-s). These objects belong to binary systems, in which they accreted mass from the primary star (now a WD) when it
was on the AGB phase. We remind that the non-standard mixing processes mentioned above may occur not only during the RGB but also during the AGB phase decreasing the \ciso \ ratio to unexpected low values as observed in many AGB stars \citep{bus10,abi11}. Therefore, the
observed chemical peculiarities (carbon and s-element enhancements, and low \ciso \ ratios) in these stars can be easily explained in this scenario.  We have also to mention that the warm temperatures of these unevolved stars make it extremely difficult to derive their \ciso \ ratio, even for large $^{13}$C enhancements, because the
spectrocopic lines used (mainly $^{13}$CH lines) are very weak and blended features in a very crowded spectral region (see references above). These features are probably undetectable if the \ciso \ ratio was initially much larger than the solar one. 

Our conclusion is
then that the \ciso \ ratio observed in red giants descending from single stars  is affected mainly by stellar internal processes and can hardly be used to infer their initial \ciso \ ratio. GCE model predictions should rather be compared to \ciso \  ratios of turn-off and/or sub-giant stars of different metallicities, that have preserved, in principle,  the initial ratio in their envelopes.
An alternative would be the derivation of that ratio in metal-poor interstellar gas, such as observed in some Damped Lyman-alpha systems at high redshift. However observations of this type are still very scarce \citep[see e.g.][]{les06}.

\begin{figure*}
\begin{centering}
\includegraphics[angle=-90,width=1.0\textwidth]{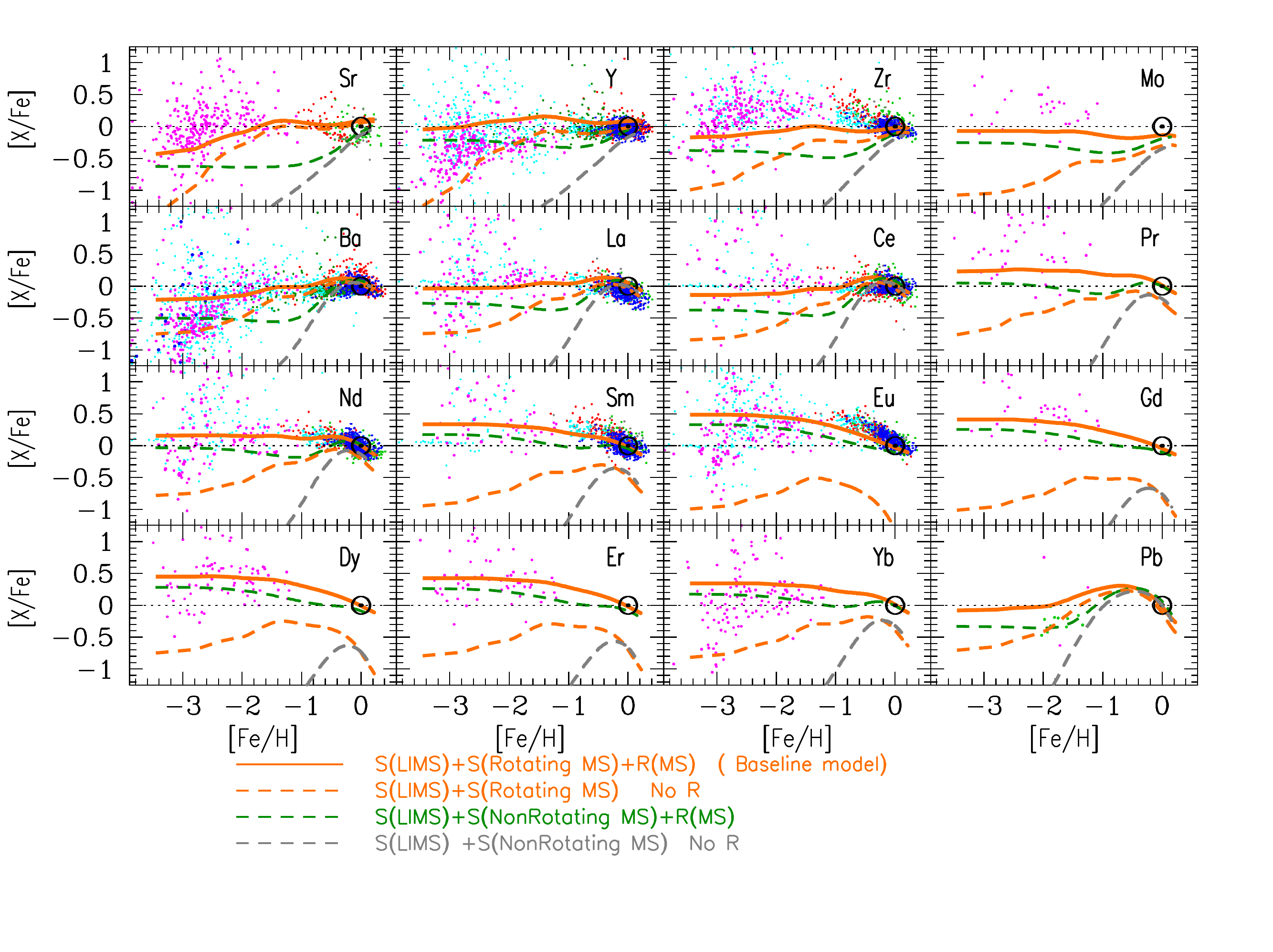} 

\par\end{centering}

\caption{\label{fig:f_evol_heavies} Same as in Fig. \ref{fig:f_elm_evolFepeak}, but for elements heavier than the Fe-peak. Two models have been added, both without the r-component, one for the rotating massive stars ({\it orange dashed)} and one for the non-rotating massive stars ({\it gray dashed}). }
\end{figure*}

\subsubsection{The heavy elements}
\label{subsub:Elm_heavies}

Figure \ref{fig:f_evol_heavies} shows the predicted [X/Fe] vs [Fe/H] evolution for some representative elements beyond the Fe-peak. This includes elements
with a significant s-process contribution belonging to the first
(Sr, Y, Zr), second (Ba, La, Ce, Nd, Sm) and third (Pb) s-element peaks, and elements with a main r-process contribution (Eu, Gd, Dy, Er and Yb) (see Fig. 8).
For clarity, we show the predicted evolution for the cases in which different contributing sources are considered: 

i) LIM stars, rotating massive stars plus our fiduciary r-process (the baseline model, orange continuous curve); 

ii) LIM stars, non-rotating massive stars and r-process (green dashed curve);

iii) LIM stars and non-rotating massive stars without r-process contribution (gray dashed curve); and

iv) LIM stars plus rotating massive stars without the r-process contribution (orange dashed curve). 

The first two cases are also considered in Fig. \ref{fig:f_elm_evolFepeak}, while the last two ones are introduced here to evaluate the respective roles of LIM  and massive stars to the production of heavy elements, as well as the role of the r-process. Case (iii), for instance, illustrates the role of LIM stars only (with their time-delayed contribution) to the s-component, since the non-rotating massive stars have a negligible contribution to that component. In a similar vein, case (ii) shows the s-component alone (from both LIM and massive stars) of our baseline scenario. 

Several interesting features can be extracted from Fig. \ref{fig:f_evol_heavies}:

- LIM stars begin to contribute significantly at metallicities [Fe/H]$\ga-1.$ ($t\geq 1$ Gyr), but somewhat  earlier for Pb, which has a strong s-component of  $\sim 80\%$. Their contribution (gray-dashed curve) has a secondary-like behavior with [Fe/H], showing a maximum in the [X/Fe] ratio at a different metallicity depending on the corresponding s-element, namely: for the first s-element abundance-peak (Sr, Y, Zr) at [Fe/H]$\approx 0.0$, for the second peak (Ba, La, Ce, Nd, Sm) at [Fe/H]$\approx-0.4$, and for the third peak (Pb) at slightly lower metallicity ($\sim-0.6$). This is a direct consequence of the dependence on metallicity of the s-element yields from LIM stars (see section 2.3). The contribution of LIM stars is, therefore, important to account for the observed [X/Fe] trends of s-elements in stars with (both thin and thick) disk metallicities, in particular at the epoch preceeding solar system formation.

- Pb is  clearly the sole element for which LIM stars
are the dominant source for [Fe/H]$\geq -1.5$. 
In our model, its abundance at  solar system formation is provided  almost exclusively by these stars.
Unfortunately, there are very few Pb abundance determinations to constrain the actual contribution of the other stellar sources in the production of this element at sub-solar metallicities.

- Including the r-component, as discussed in Sec. \ref{subsec:model} and Equ. \ref{eq:yldmas}, allows us to reproduce correctly the observed evolution of the elements with a known predominant r-process origin (Eu, Gd, Dy, Er, Yb), as indicated by the green-dashed curves in Fig. \ref{fig:f_evol_heavies}. Our prescription for the r-process yields of massive stars, renders the evolution of these elements  quite similar to that of oxygen: a plateau-like behaviour of [X/Fe] for [Fe/H]$<-1.0$, followed by a decrease of that ratio until its solar value for [Fe/H]$\approx 0.0$. The average observed [X/Fe] trends for these elements are compatible with this scheme. We  note, however, that the "success" of the scheme does not constitute a proof that the main sources of r-elements are indeed CCSN. Various theoretical and observational reasons seem to favor neutron star mergers for that role \citep{wan13,Ishimaru2015} and the recent detection of this event with an inferred presence of r-elements in the ejecta \citep{dro17} reinforces this idea. But even in that case, the product of the (still unknown) rate of neutron star mergers with the corresponding r-yields should result in an evolution not too different from the one depicted here, leaving unchanged our conclusions and inferences below. 

- Having constrained the contribution of the r-component through the quasi-pure r-element evolution, we may try to evaluate its impact on the evolution of the mostly s-elements by observing the behavior of the corresponding green-dashed curves in Fig. \ref{fig:f_evol_heavies}. However, the increasing dispersion of all the observed [X/Fe] ratios with decreasing [Fe/H] makes a quantitative  evaluation of that effect difficult, if not impossible. In any case,  Fig. \ref{fig:f_evol_heavies} shows  that the r-component produces a sizeable fraction of the s-mostly elements at low metallicities [Fe/H]$<-1.0$, and in particular for the second peak s-elements Nd and Sm. However, it is clearly not sufficient to account for the observed trend in the full range of metallicity for any s-element, at least under the assumptions made here (namely those expressed by Eq. \ref{eq:yldmas}).

- The dashed orange curve in Fig. \ref{fig:f_evol_heavies} shows the evolution of the s-component of heavy elements, taking into account the contributions of both LIM stars and rotating massive stars  (but with no r-component). It can be seen that the latter have a larger contribution than LIM stars for all elements and all metallicities [Fe/H]$<-0.6$ ($-1.1$ for Pb).  For elements up to La and Ce, the s-component overwhelms even the r-component down to [Fe/H]$\sim-2$. Its behaviour is primary-like for the lightest s-elements Sr, Y and Zr, down to [Fe/H]$\sim-1.5$ and changes to secondary-like at lower  metallicities. Overall, the s-component resulting from rotating massive stars plays an important role in shaping not only the solar distribution of the heavy elements, but also their evolution during the whole disk phase of the Galaxy, i.e. the last 10 Gyr or so.

- Finally, our baseline model (continuous orange curves in Fig. \ref{fig:f_evol_heavies}) including s-component from LIM stars and rotating massive stars as well as the r-component  improves substantially the situation  in the full range of metallicities for the s-elements, in particular for the lightest ones (Sr, Y, Zr). Indeed, {\it without the weak s-process contribution, the observed (average) abundance evolution of the light s-elements at low metallicities can not be reproduced reasonably well}, their r-component having only a mild impact (compare the two orange curves in Fig. \ref{fig:f_evol_heavies}). Nevertheless, for Zr our baseline model still predicts a small deficit at [Fe/H]$<-2.0$.  This later was also found  by \cite{Tra04} and \cite{Bi17} although in a lesser extent. In addition, for the elements with a significant r-process contribution (Nd, Sm), rotating massive star yields improve considerably the fit to the observed trends at [Fe/H]$<-1.0$. For Pr there is not enough observational data to compare with.

On the other hand, our baseline model produces  a clear decrease  of the second-peak s-element (Ba, La) ratios [X/Fe] at higher than solar metallicities, similar to that recently found in thin disc stars \citep{Bensby2014,Battistini2016,Del17}. The evolution of those elements at  metallicities higher that solar is determined mainly by LIM stars, this result is entirely due to the strong decrease in the yields of second-peak s-elements  in LIM stars of  that metallicitiy range. In contrast, for the light (first-peak) s-elements there is no signature of such a decrease, neither in the observations nor in our model.

Globally, the computed [X/Fe] vs. [Fe/H] evolution for the s-elements agrees with those obtained in previous studies \citep{Tra04,Bi17} for metallicities typical of the disk ([Fe/H]$\geq -1.0$)\footnote{Note that our conclusion about the existence of a solar LEPP is at odd with that from these authors.}. There are, however, significant differences at lower metallicities where the evolution is mainly determined by massive stars and the reason  lies in the different yields that we adopt here. For instance, \citet{Bi17} used rotating massive star yields from \citet{Fri16} and find small differences - and only for [Fe/H]$<-2.0$ - in the evolution of Y and Ba with respect to the case of non-rotating massive star yields.
This is at odd with our main conclusion here (see Fig. \ref{fig:f_evol_heavies}): the weak s-process in rotating massive stars plays a key role in the evolution of the s-elements at low metallicity, in particular for the lightest ones.

It is important to note that 
the observed heavy element ratios [X/Fe] show a huge scatter at [Fe/H]$<-2$ (see Fig. \ref{fig:f_evol_heavies}), in stark contrast to the situation with elements up to the Fe-peak (Fig. \ref{fig:f_elm_evolFepeak}).  This large (2-3 dex) star-to-star scatter in the stellar abundance ratios remains inexplicable by e.g. uncertainties in stellar parameters, NLTE corrections, or sample biases. Obviously, it cannot be reproduced by simple one-zone GCE models as this one. 
It may  suggest, for instance, the need of at least two neutron-capture processes yielding heavy elements at very low metallicities \citep[see e.g.][]{and11,han12}. 

This dispersion is particularly large even for the ratios between different elements (e.g. [Sr,Y/Ba,Eu]) vs [Fe/H], contrary to what would be expected if Sr, Y, Ba, Eu etc., had the same r-process nucleosynthetic origin at these metallicities \citep[e.g.][and references therein]{Fre10,Roe16}. In fact, this observable has been used  as evidence of the existence of a LEPP\footnote{This process is different to the "solar" LEPP discussed in section 3.2.2.} to explain the abundance pattern of the elements in the atomic number range $38 < Z < 47$ found in many halo stars \citep{Tra04,Mo07,Qia08,Am11}. 

It is possible that part of this scatter might be explained as the result of the mix of material ejected in the explosion of massive stars with different rotational velocities, as suggested by \cite{Ces13} and \cite{ces14}. For instance, a range $-0.5\leq$[Sr/Ba]$\leq+0.6$ can be obtained with our GCE model at these metallicities by adding the contribution of the weak s-process from massive stars with v$_{\rm rot}^{\rm ini}=0, 150$ or 300 kms$^{-1}$, respectively, to the dominant r-contribution. Nevertheless, although the abundance pattern found in a particular halo star might be explained by the ejecta from a single massive  star with a specific v$^{\rm ini}_{\rm rot}$,  this is clearly insufficient to understand the full scatter observed in the heavy element abundance ratios, at least with our models.  \cite{ces14}  developed an inhomogeneous GCE model for the galactic halo allowing them to explain the observed scatter by combining the s-process production in  rapidly rotating massive stars ("spinstars") and the r-process contribution, which they assumed to result from massive stars. Here we simply note that the scatter of [X/Fe] at low metallicities concerns also pure r-elements, like Eu. It is not clear whether stellar rotation affects the production of such elements or not, especially if their main source are neutron star mergers. 

We note however, that at such low metallicities, below [Fe/H]$\sim -2$, simple GCE models (inhomogeneous or not) cannot be safely used: indeed, the hierarchical merging scenario for galaxy formation suggests that the early Galaxy was formed from the merging of smaller sub-haloes, each one with its own history and timescale for chemical enrichment (see discussion in the beginning of Sec. \ref{subsec:Element_evolution} and references therein). It appears then that a detailed discussion of the observed scatter of heavy element abundances in halo stars requires a thorough treatment of those factors (imperfect  gas mixing, merging of sub-haloes), which is clearly  beyond the scope of the present study.

\subsubsection{The ratio of heavy to light s-element abundances}
\label{subsub:hsls}

\begin{figure}
\begin{centering}
\includegraphics[width=0.49\textwidth]{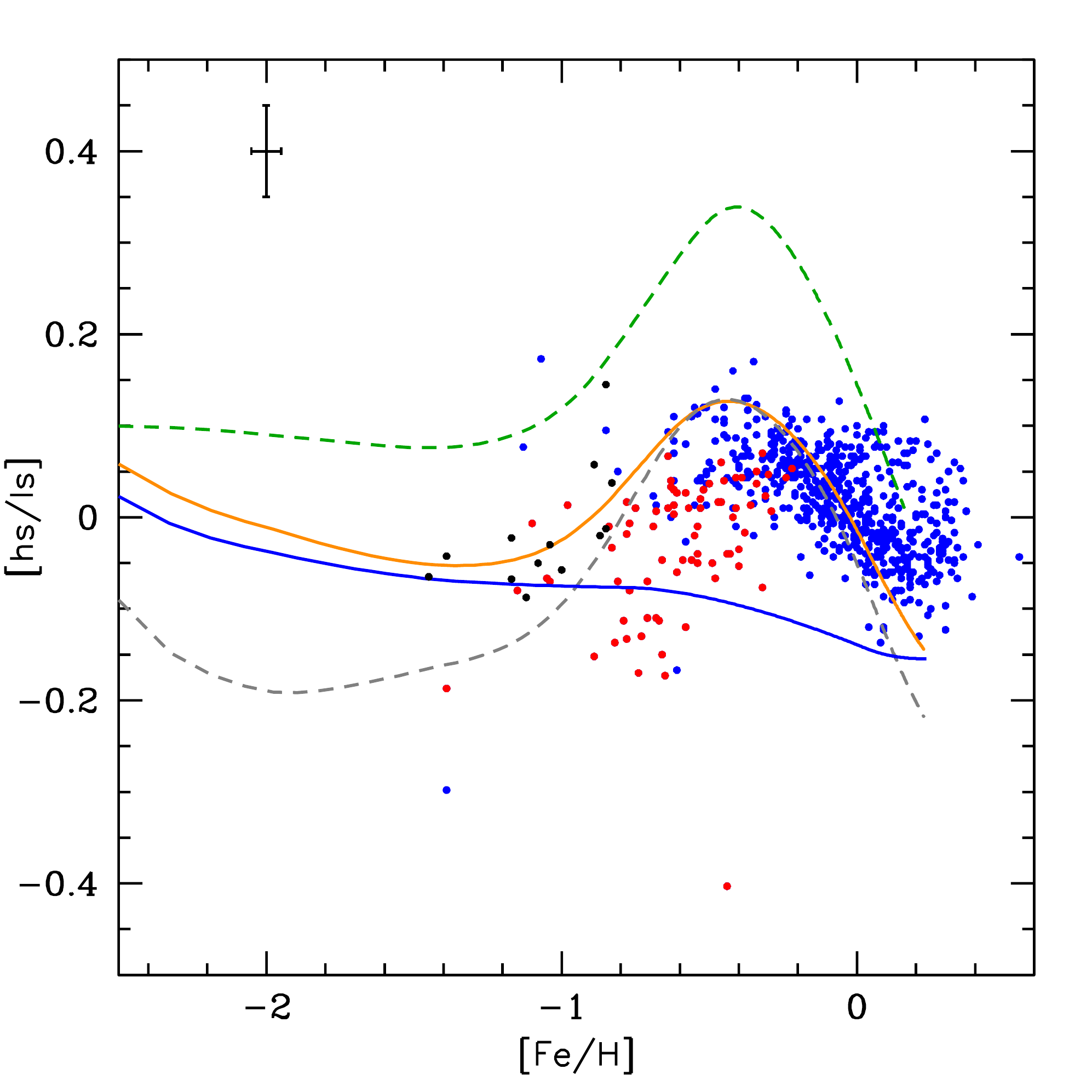} 
\par\end{centering}

\caption{Observed heavy-s (Ba, Ce, Nd) to light-s (Sr, Y, Zr) ratio [hs/ls] vs [Fe/H] compared with our GCE model predictions. Observational data are from
\citet{Del17} for thin (blue dots) and thick (red dots) disk stars. Black dots are the thick disk stars in \citet{fis17}. As in Figs. \ref{fig:f_elm_evolFepeak}    and \ref{fig:f_evol_heavies} solid orange curve shows the prediction from our baseline model, green dashed curve the one for the non-rotating massive stars, and gray dashed curve the non-rotating case where the r-component is not considered. The blue solid curve shows the prediction when the contribution from LIM stars is omitted (see text). 
\label{fig:f_evol_heavies2}}
\end{figure}

A critical test to check the reliability of our s-elements yields is to study the evolution of the abundance ratio between the "heavy s" (hs) and the "light s" (ls) elements. Following \citet{Luc91}, it is common to monitor the s-process efficiency through the relative abundances of the s-elements at the Ba peak (collectively indicated as [hs/Fe]) with respect to those at the Zr peak (indicated as [ls/Fe]). Those nuclei, placed at neutron magic numbers N$ = 50$ and 82, respectively,  are mainly synthesized by the s-process and act as bottlenecks for the s-process path because of their low neutron capture cross sections. As already highlighted in \S \ref{subsec:Lim_Mas}, for relatively low s-process efficiency the neutron flux mainly feeds the nuclei at the Zr peak, while for higher exposures the Ba-peak species are favored. The average [hs/ls] ratio has been extensively used as a measure of the neutron capture efficiency in building up the s-elements and shown to be useful for the interpretation of the evolution of Galactic disk stars. However, until very recently few
relevant observational data were available in unevolved field stars.  

In Figure \ref{fig:f_evol_heavies2} we compare the computed [hs/ls] vs. [Fe/H] evolution with the observed trend derived in thin (blue dots) and thick (red dots) unevolded disk stars by \citet{Del17}.
These authors performed an homogeneous chemical analysis of $\sim 1000$ field stars on the basis of high (S/N$> 100$) quality spectra. We have included also a few thick disk stars (black dots) analyzed by
\citet{fis17}. We obtain the [hs/Fe] ratio using the average
abundance ratio between Ba, Ce and Nd, while for the [ls/Fe] ratio we take the average between Sr, Y and Zr\footnote{Note that the choice of the specific elements to consider in the observational average [hs/Fe ]and [ls/Fe] abundances varies from author to author and depends on the quality of the spectra available and the specific elements analyzed.}. From this figure it is evident that our baseline model (orange solid curve) nicely fits the observed average trend for thin disk stars. From [Fe/H]$\approx 0.0$, the [hs/ls] ratio rises for decreasing metallicities and peaks around [Fe/H]$\sim -0.5$, as observations indicate. Both GCE models and observations  confirm the {\it differential dependence on metallicity of the s-process nucleosynthesis}. To our knowledge, this is the first time that this abundance trend in unevolved stars is reproduced by simple GCE calculations, which reinforces the reliability of the yields used here.

For thick disk stars our predicted trend is slightly above the observed one, although clearly more observations are needed at [Fe/H]$<-1.0$ to extract any definite answer. Note however, that according to  \citet{Del17}, apparently there is not a systematic difference between thick and thin disk stars in the observed [Y,Zr/Fe] ratios at any metallicity. However, this does not hold for the [Ba,Ce/Fe] ratios (with the exception of Nd). For this reason, the observed average [hs/ls] ratio in thick disk stars is below $\sim 0.0$ in Fig. \ref{fig:f_evol_heavies2} . Systematic differences in other elemental ratios (mainly in the $\alpha$-elements [Si+Mg+Ca/Fe]) between thin and thick stars at a given metallicity exist. These differences may be interpreted on the basis of a different SFR history between the thin and thick disk, but also migration of stars and gas across the galactic disk may play a significant role. We remind that this later phenomenon can not be treated in one-zone GCE models. 

 Figure \ref{fig:f_evol_heavies2} clearly shows that the main contributor to the s-elements budget at disk metallicities are LIM stars: their absence, illustrated by the  solid blue curve, does not reproduce the observations. This contribution sets the position of the observed [hs/ls] {\it bump} at [Fe/H]$\sim -0.5$. However, in the absence of rotating massive stars, the computed [hs/ls] maximum value would be much higher than indicated by observations (green dashed curve; note the small observational error). Therefore, according to our yields, rotating massive stars {\it contribute significantly to the light s-elements  at sub-solar metallicities}. This result may change the current view on the role played by the different sources (massive vs LIM stars) on the production of the s-elements in the Galaxy, at least for [Fe/H]$>-1.0$. On the other hand, it appears from this figure that the r-process contribution mainly affects the evolution at very low (halo) metallicities (gray dashed line), with little role for disk metallicities for the s-mostly elements. Unfortunately there are not enough [hs/ls] measurements in unevolved halo stars ([Fe/H]$<-1.0$) to fully test our GCE model predictions. Measurements of the [hs/ls] ratio in very metal-poor stars may be used to evaluate the role of rotating massive stars in the abundance evolution of the s-elements in the  early Galaxy.



\section{Summary and conclusions}
\label{sec:Summary}

In this study we present an analysis of the evolution
of the abundances of elements (H to U) in
the Milky Way halo and the local disk. We use a consistent
GCE model to describe the evolution of those two galactic subsystems (Sec. \ref{subsec:Local_Evol}).

The novelty in the present study is 
threefold: the use of a complete set of isotopes from H to U,  the use of a new grid of stellar yields over the whole stellar mass range and the use of a weighted average of those yields through an empirically calibrated, metallicity-dependent  function of rotation velocities.  

The adopted grid of LIM stars (from the FRUITY database) and
of massive stars (from LC2018)  covers a large  range of masses and metallicities
and, for massive stars, different initial rotational velocities: 0, 150 and 300
kms$^{-1}$. On the basis of recent ideas on massive star explosions, we assume that stars with M$> 25$ \ms \  contribute only through the stellar wind, for all values of initial metallicity and rotational velocity. Our main goal is to test the impact of these new yields in the abundance evolution
of the ensemble of isotopes and elements,
with  particular emphasis on the role of rotating massive stars in the evolution of 
the s-elements. 

Due to our current ignorance on the dependence of the stellar rotation with metallicity, the adopted yields of rotating
massive stars are weighted with a metallicity dependent function. This function is empirically determined, as to obtain both the observed primary behavior of nitrogen versus [Fe/H] (requiring a large average rotational velocity at low metallicities) and to avoid overproduction of s-elements around [Fe/H]$\sim-1$ (requiring lower rotational velocities for disk stars). 

Since we are interested on both isotopic and elemental evolution and
since most heavy elements have a mixed origin, we adopt fiduciary yields for the isotopes of
a r-process origin; namely, we assume that they are produced in CCSNe, their yield being  solar-scaled
to that of $^{16}$O. This permits us to study
the behavior of the other isotopes (of mixed origin), as well as the behavior of the elements
and to constrain the adopted s-element yields.

We find that the resulting elemental and isotopic composition at the epoch of solar
system formation compare remarkably well to the observed proto-solar one. Among the main findings we note:

- The abundances of all major isotopes of the multi-isotopic elements up to Fe ($^{12}$C, $^{14}$N, $^{16}$O, $^{20}$Ne, $^{28}$Si,
$^{32}$S, $^{36}$Ar, $^{40}$Ca, $^{54}$Cr, $^{56}$Fe) are well reproduced,
in most cases to better than 10\%.

- Proto-solar fluorine abundance is well reproduced (at the 85\% level), with no need for $\nu$-induced nucleosynthesis. About 2/3 of the proto-solar F abundance  comes from rotating massive stars, the remaining 1/3 resulting from LIM stars. The isotope $^{15}$N is produced along the same nucleosynthesis path that leads to \nuk{F}{19} and is also well reproduced, with no need for either nova or  $\nu$-induced nucleosynthesis.

- Rotating massive stars are found to have an important impact on  the production of light s-elements (A$<90$), through the increased production of the neutron source \nuk{Ne}{22}. In our model, the proto-solar abundances of  s-only isotopes with A$<90$ are  accounted by rotating massive stars at the 50-85 \% level. In contrast, only a few \% of the s-only isotopes with A$>90$ is made in such stars.  This allows us to obtain an abundance distribution for the s-only isotopes remarkably flat (to better than 10\% for most of them) in the entire mass range $70<$A$<204$. We emphasize that this result is obtained through the combination of  the adopted  SFR and IMF (eliminating the core ejecta of stars with M$>25$ \ms), as well as the adopted IDROV (reducing the role of fast rotating massive stars at metallicities slightly sub-solar), otherwise light s-isotopes would be considerably overproduced. Obviously, other combinations of IMF and IDROV may lead to similar, or even better results. In any case, we have convincingly shown here  that the existence of a solar light element primary process (LEPP) is not necessary, especially considering all the associated uncertainties from stellar and galactic astrophysics.

- Proto-solar elemental abundances
are also found to be well reproduced, especially all the $\alpha-$, Fe-peak
(with the exception of Ni) and heavier than Fe elements. For the latter, we also manage to reproduce satisfactorily their proto-solar s-component (something done for the first time), allowing us to discuss on a firm basis our results for lower metallicities. 

We find that several issues of the proto-solar composition require further studies, the main  ones being the following:

- We underproduce all Mg isotopes  (by $\sim 40\%$ from their solar values) as well as most of the isotopes in the range $38<$A$<50$. The latter is a well known problem, also found with other, widely used, grids of massive-star yields \citep{Woo95,Nomoto2013}. Our rotating massive stars do not help alleviating this problem.

- The overproduction by a factor of two found for
the isotopes $^{54}$Fe and $^{58}$Ni (and for the element Ni as well)   results  from a known problem
of the W7 model for SNIa explosions adopted here.

- The most abundant isotope of Zn ($^{64}$Zn) is underproduced by more than a factor of
two. It is difficult to understand the reason of that since Zn and Cu
have similar production channels and the solar abundance of the latter element is well
reproduced in our model. The explosive component of $^{64}$Zn is perhaps underestimated here and it may be due to high energy supernovae (hypernovae) that are not included in our study.

We also compare  our GCE predictions with a large body of observational
data obtained from a number of recent  large spectroscopic surveys and concerning  [X/Fe] vs [Fe/H]  in halo and disk stars. The main conclusions of this comparison are the following:

- Within the group of light and intermediate mass elements, the evolution of N and F are the two mostly affected by the
rotational massive stars yields, their behavior turning from a secondary (without rotation) to a primary one (with rotation). 
Rotation has already been suggested as the explanation for primary N, but it is the first time that i) this effect is obtained with the use of a metallicity-dependent distribution of rotational velocities and not on the basis of a single velocity, and ii) the concomitant  effect on the evolution of all other elements is carefully studied. We thus find that
our evolution of F vs O is  compatible with the observations at metallicities $\sim\rm{Z_\odot}$ - which suggest a secondary-like behavior - due to the contribution of our LIM stars in this metallicity range. Our prediction for lower metallicities depends critically on the adopted relation between the stellar rotation and metallicity in massive stars, making F an important calibrator of that relation and of the corersponding yields.  Determinations of  [F/Fe] and  and [F/O] ratios in metal-poor unevolved stars are thus urgently needed. 

- We find that the evolution of the isotopic ratio \ciso \ is also affected largely by rotational yields, as suggested previously by \cite{Chiappini2008}. However, we argue here that available observations of this ratio   in red giants cannot help to constrain stellar yields and GCE models, because it is mostly affected by internal stellar processes in those stars. We show quantitatively that, starting with initial \ciso \ provided by our GCE models and introducing internal decrease obtained in standard stellar models, one may understand readily observations of \ciso \ at the red bump, with no need for yields of extremely fast rotating stars.   

- The evolution of the $\alpha$-elements O, Si, S and Ca is barely affected by 
rotational yields. Rotation improves considerably the behavior of K and Sc, but overall, these elements, as well as Mg, Ti and V are  underproduced at all metallicities. Except for Mg (where the yields of \citet{Nomoto2013} are more succesfull), this is a generic problem for all currently used grids of massive star yields.
 
- Among the Fe-peak elements, only the evolution of Cu and Zn is slightly affected by rotational yields, increasing upwards the corresponding [X/Fe] ratios. The evolution of Cu is rather well described by our rotating massive star yields, although recent NLTE abundance determinations of that element in metal-poor
stars casts some doubt on its actual behavior. Despite its similar nucleosynthetic origin, Zn is underproduced at all metallicities by our model. This  problem is shared by other grids of yields and, perhaps, it suggests specific sources at low metallicities, like e.g. hypernovae.  

- Rotating massive stars yields are found to have a dramatic impact in the predicted evolution of the s-elements, in
particular for the lightest ones (Sr,Y,Zr) at low metallicity ([Fe/H]$<-0.5$). The predicted trends are in better agreement with the average observed ones although we find some deficiency for Zr and Mo. For the
heavy s-elements (Ba, La etc) the impact is lower,
but still significant. The evolution for [Fe/H]$\ga -0.5$
of the elements with a predominant s-process origin is dominated by the contribution of LIM stars. Interestingly, we find a decrease
of the [X/Fe] ratios for supersolar metallicities for the heavy s-elements (but not for the lightest ones), resulting from the metallicity dependence of the s-element yields in LIM stars and in agreement with recent observational studies. 

- The combination of yields from rotating massive stars and LIM stars are capable to explain the recently observed trend of heavy to light
 ([hs/ls]) s-elements ratio vs [Fe/H] in unevolved disk stars. This ratio is a measure of the neutron exposure during the
 s-process. Comparison to the observed trend allow us to conclude that rotating massive
 stars have a non-negligible contribution to the evolution of the s-elements also at near solar metallicities. Furthermore, we find that
at [Fe/H]$<-1.0$  the predicted [hs/ls] ratio is found to be extremely sensitive to the role of rotating massive stars.
 Therefore, the [hs/ls] ratio also appears as a calibrator of the yields of rotating massive stars. Further measurements of this ratio in unevolved metal-poor stars of the thin and thick disks will help to improve our understanding of the production of s-elements in that metallicity range.

In summary, we have revisited the chemical evolution of the
halo and the local disk with a consistent GCE model and metallicity-dependent yields from rotating massive stars, as well as LIM stars and SNIa. For the first time, we found that some metallicity-dependent distribution of the initial rotational velocities of massive stars has to be assumed, and we adopted such a distribution on the basis of observed abundances of key elements (nitrogen and s-elements). Under this assumption, we found that the adopted yields can help to improve our understanding of
a large number of observations, particularly regarding the isotopic and elemental
abundances of s-elements at the epoch of solar system formation as well as during Galactic evolution. For some lighter elements, the inclusion of rotation in massive stellar models turns them into primaries (N, F) or improves the situation (Sc), but for others (Mg, K, V and Ti at all metallicities, and some Fe-peak elements at very low metallicities) the situation does not improve and important discrepancies with the observations remain. Finally, we find that
rotating massive star yields may help to explain only partially the large dispersion observed in [X/Fe] at low metallicities for most of the heavy elements. A full explanation probably requires both inhomogeneous chemical evolution of the early ISM and formation of the early Galaxy through hierarchical merging of sub-haloes with different evolutionary histories.

\section*{Acknowledgments}
We are grateful to the referee, G. Meynet, for a careful and thorough reading of the manuscript and many useful comments and suggestions.
C.A. acknowledges to the Spanish grant AYA2015-63588-P
within the European Founds for Regional Development (FEDER).
M.L and A.C. acknowledges PRIN 2014 (P.I.:Pastorello).




\bibliographystyle{mnras}
\bibliography{Reference} 

\begin{thebibliography}{}
\makeatletter
\relax
\def\mn@urlcharsother{\let\do\@makeother \do\$\do\&\do\#\do\^\do\_\do\%\do\~}
\def\mn@doi{\begingroup\mn@urlcharsother \@ifnextchar [ {\mn@doi@}
  {\mn@doi@[]}}
\def\mn@doi@[#1]#2{\def\@tempa{#1}\ifx\@tempa\@empty \href
  {http://dx.doi.org/#2} {doi:#2}\else \href {http://dx.doi.org/#2} {#1}\fi
  \endgroup}
\def\mn@eprint#1#2{\mn@eprint@#1:#2::\@nil}
\def\mn@eprint@arXiv#1{\href {http://arxiv.org/abs/#1} {{\tt arXiv:#1}}}
\def\mn@eprint@dblp#1{\href {http://dblp.uni-trier.de/rec/bibtex/#1.xml}
  {dblp:#1}}
\def\mn@eprint@#1:#2:#3:#4\@nil{\def\@tempa {#1}\def\@tempb {#2}\def\@tempc
  {#3}\ifx \@tempc \@empty \let \@tempc \@tempb \let \@tempb \@tempa \fi \ifx
  \@tempb \@empty \def\@tempb {arXiv}\fi \@ifundefined
  {mn@eprint@\@tempb}{\@tempb:\@tempc}{\expandafter \expandafter \csname
  mn@eprint@\@tempb\endcsname \expandafter{\@tempc}}}

\bibitem[\protect\citeauthoryear{{Abia}}{{Abia}}{2011}]{abi11}
{Abia} C.,  2011, in {Kerschbaum} F.,  {Lebzelter} T.,   {Wing} R.~F.,  eds,
  Astronomical Society of the Pacific Conference Series Vol. 445, Why Galaxies
  Care about AGB Stars II: Shining Examples and Common Inhabitants. p.~13

\bibitem[\protect\citeauthoryear{{Abia}, {Cunha}, {Cristallo}  \& {de
  Laverny}}{{Abia} et~al.}{2015}]{Abi15}
{Abia} C.,  {Cunha} K.,  {Cristallo} S.,   {de Laverny} P.,  2015, \mn@doi
  [\aap] {10.1051/0004-6361/201526586}, \href
  {http://adsabs.harvard.edu/abs/2015A%26A...581A..88A} {581, A88}

\bibitem[\protect\citeauthoryear{{Adibekyan}, {Sousa}, {Santos}, {Delgado
  Mena}, {Gonz{\'a}lez Hern{\'a}ndez}, {Israelian}, {Mayor}  \&
  {Khachatryan}}{{Adibekyan} et~al.}{2012}]{Adi12}
{Adibekyan} V.~Z.,  {Sousa} S.~G.,  {Santos} N.~C.,  {Delgado Mena} E.,
  {Gonz{\'a}lez Hern{\'a}ndez} J.~I.,  {Israelian} G.,  {Mayor} M.,
  {Khachatryan} G.,  2012, \mn@doi [\aap] {10.1051/0004-6361/201219401}, \href
  {http://adsabs.harvard.edu/abs/2012A%26A...545A..32A} {545, A32}

\bibitem[\protect\citeauthoryear{{Andrievsky}, {Spite}, {Korotin}, {Fran{\c
  c}ois}, {Spite}, {Bonifacio}, {Cayrel}  \& {Hill}}{{Andrievsky}
  et~al.}{2011}]{and11}
{Andrievsky} S.~M.,  {Spite} F.,  {Korotin} S.~A.,  {Fran{\c c}ois} P.,
  {Spite} M.,  {Bonifacio} P.,  {Cayrel} R.,   {Hill} V.,  2011, \mn@doi [\aap]
  {10.1051/0004-6361/201116591}, \href
  {http://adsabs.harvard.edu/abs/2011A%26A...530A.105A} {530, A105}

\bibitem[\protect\citeauthoryear{{Andrievsky}, {Bonifacio}, {Caffau},
  {Korotin}, {Spite}, {Spite}, {Sbordone}  \& {Zhukova}}{{Andrievsky}
  et~al.}{2017}]{Andrievsky2017}
{Andrievsky} S.,  {Bonifacio} P.,  {Caffau} E.,  {Korotin} S.,  {Spite} M.,
  {Spite} F.,  {Sbordone} L.,   {Zhukova} A.~V.,  2017, preprint, \href
  {http://adsabs.harvard.edu/abs/2017arXiv170908619A} {} (\mn@eprint {arXiv}
  {1709.08619})

\bibitem[\protect\citeauthoryear{{Arcones} \& {Montes}}{{Arcones} \&
  {Montes}}{2011}]{Am11}
{Arcones} A.,  {Montes} F.,  2011, \mn@doi [\apj] {10.1088/0004-637X/731/1/5},
  \href {http://adsabs.harvard.edu/abs/2011ApJ...731....5A} {731, 5}

\bibitem[\protect\citeauthoryear{{Arlandini}, {K{\"a}ppeler}, {Wisshak},
  {Gallino}, {Lugaro}, {Busso}  \& {Straniero}}{{Arlandini}
  et~al.}{1999}]{Ar99}
{Arlandini} C.,  {K{\"a}ppeler} F.,  {Wisshak} K.,  {Gallino} R.,  {Lugaro} M.,
   {Busso} M.,   {Straniero} O.,  1999, \mn@doi [\apj] {10.1086/307938}, \href
  {http://adsabs.harvard.edu/abs/1999ApJ...525..886A} {525, 886}

\bibitem[\protect\citeauthoryear{{Arnett} \& {Thielemann}}{{Arnett} \&
  {Thielemann}}{1985}]{Arn85}
{Arnett} W.~D.,  {Thielemann} F.-K.,  1985, \mn@doi [\apj] {10.1086/163402},
  \href {http://adsabs.harvard.edu/abs/1985ApJ...295..589A} {295, 589}

\bibitem[\protect\citeauthoryear{{Asplund}, {Grevesse}, {Sauval}  \&
  {Scott}}{{Asplund} et~al.}{2009}]{Asp09}
{Asplund} M.,  {Grevesse} N.,  {Sauval} A.~J.,   {Scott} P.,  2009, \mn@doi
  [\araa] {10.1146/annurev.astro.46.060407.145222}, \href
  {http://adsabs.harvard.edu/abs/2009ARA%26A..47..481A} {47, 481}

\bibitem[\protect\citeauthoryear{{Battino} et~al.,}{{Battino}
  et~al.}{2016}]{Ba16}
{Battino} U.,  et~al., 2016, \mn@doi [\apj] {10.3847/0004-637X/827/1/30}, \href
  {http://adsabs.harvard.edu/abs/2016ApJ...827...30B} {827, 30}

\bibitem[\protect\citeauthoryear{{Battistini} \& {Bensby}}{{Battistini} \&
  {Bensby}}{2016}]{Battistini2016}
{Battistini} C.,  {Bensby} T.,  2016, \mn@doi [\aap]
  {10.1051/0004-6361/201527385}, \href
  {http://adsabs.harvard.edu/abs/2016A%26A...586A..49B} {586, A49}

\bibitem[\protect\citeauthoryear{{Bell} et~al.,}{{Bell}
  et~al.}{2008}]{Bell2008}
{Bell} E.~F.,  et~al., 2008, \mn@doi [\apj] {10.1086/588032}, \href
  {http://adsabs.harvard.edu/abs/2008ApJ...680..295B} {680, 295}

\bibitem[\protect\citeauthoryear{{Bensby}, {Feltzing}  \& {Oey}}{{Bensby}
  et~al.}{2014}]{Bensby2014}
{Bensby} T.,  {Feltzing} S.,   {Oey} M.~S.,  2014, \mn@doi [\aap]
  {10.1051/0004-6361/201322631}, \href
  {http://adsabs.harvard.edu/abs/2014A%26A...562A..71B} {562, A71}

\bibitem[\protect\citeauthoryear{{Bisterzo}, {Travaglio}, {Gallino}, {Wiescher}
   \& {K{\"a}ppeler}}{{Bisterzo} et~al.}{2014}]{Bi14}
{Bisterzo} S.,  {Travaglio} C.,  {Gallino} R.,  {Wiescher} M.,   {K{\"a}ppeler}
  F.,  2014, \mn@doi [\apj] {10.1088/0004-637X/787/1/10}, \href
  {http://adsabs.harvard.edu/abs/2014ApJ...787...10B} {787, 10}

\bibitem[\protect\citeauthoryear{{Bisterzo}, {Travaglio}, {Wiescher},
  {K{\"a}ppeler}  \& {Gallino}}{{Bisterzo} et~al.}{2017}]{Bi17}
{Bisterzo} S.,  {Travaglio} C.,  {Wiescher} M.,  {K{\"a}ppeler} F.,   {Gallino}
  R.,  2017, \mn@doi [\apj] {10.3847/1538-4357/835/1/97}, \href
  {http://adsabs.harvard.edu/abs/2017ApJ...835...97B} {835, 97}

\bibitem[\protect\citeauthoryear{{Brott} et~al.,}{{Brott} et~al.}{2011}]{Bro11}
{Brott} I.,  et~al., 2011, \mn@doi [\aap] {10.1051/0004-6361/201016114}, \href
  {http://adsabs.harvard.edu/abs/2011A%26A...530A.116B} {530, A116}

\bibitem[\protect\citeauthoryear{{Busso}, {Gallino}  \& {Wasserburg}}{{Busso}
  et~al.}{1999}]{Bu99}
{Busso} M.,  {Gallino} R.,   {Wasserburg} G.~J.,  1999, \mn@doi [\araa]
  {10.1146/annurev.astro.37.1.239}, \href
  {http://adsabs.harvard.edu/abs/1999ARA%26A..37..239B} {37, 239}

\bibitem[\protect\citeauthoryear{{Busso}, {Palmerini}, {Maiorca}, {Cristallo},
  {Straniero}, {Abia}, {Gallino}  \& {La Cognata}}{{Busso}
  et~al.}{2010}]{bus10}
{Busso} M.,  {Palmerini} S.,  {Maiorca} E.,  {Cristallo} S.,  {Straniero} O.,
  {Abia} C.,  {Gallino} R.,   {La Cognata} M.,  2010, \mn@doi [\apjl]
  {10.1088/2041-8205/717/1/L47}, \href
  {http://adsabs.harvard.edu/abs/2010ApJ...717L..47B} {717, L47}

\bibitem[\protect\citeauthoryear{{Caffau}, {Bonifacio}, {Faraggiana}, {Fran{\c
  c}ois}, {Gratton}  \& {Barbieri}}{{Caffau} et~al.}{2005}]{Caffau2005}
{Caffau} E.,  {Bonifacio} P.,  {Faraggiana} R.,  {Fran{\c c}ois} P.,  {Gratton}
  R.~G.,   {Barbieri} M.,  2005, \mn@doi [\aap] {10.1051/0004-6361:20052905},
  \href {http://adsabs.harvard.edu/abs/2005A%26A...441..533C} {441, 533}

\bibitem[\protect\citeauthoryear{{Caffau}, {Bonifacio}, {Faraggiana}  \&
  {Steffen}}{{Caffau} et~al.}{2011}]{Caffau2011}
{Caffau} E.,  {Bonifacio} P.,  {Faraggiana} R.,   {Steffen} M.,  2011, \mn@doi
  [\aap] {10.1051/0004-6361/201117313}, \href
  {http://adsabs.harvard.edu/abs/2011A%26A...532A..98C} {532, A98}

\bibitem[\protect\citeauthoryear{{Cartledge}, {Lauroesch}, {Meyer}  \&
  {Sofia}}{{Cartledge} et~al.}{2006}]{Cartledge2006}
{Cartledge} S.~I.~B.,  {Lauroesch} J.~T.,  {Meyer} D.~M.,   {Sofia} U.~J.,
  2006, \mn@doi [\apj] {10.1086/500297}, \href
  {http://cdsads.u-strasbg.fr/abs/2006ApJ...641..327C} {641, 327}

\bibitem[\protect\citeauthoryear{{Casagrande}, {Sch{\"o}nrich}, {Asplund},
  {Cassisi}, {Ram{\'{\i}}rez}, {Mel{\'e}ndez}, {Bensby}  \&
  {Feltzing}}{{Casagrande} et~al.}{2011}]{Cas11}
{Casagrande} L.,  {Sch{\"o}nrich} R.,  {Asplund} M.,  {Cassisi} S.,
  {Ram{\'{\i}}rez} I.,  {Mel{\'e}ndez} J.,  {Bensby} T.,   {Feltzing} S.,
  2011, \mn@doi [\aap] {10.1051/0004-6361/201016276}, \href
  {http://adsabs.harvard.edu/abs/2011A%26A...530A.138C} {530, A138}

\bibitem[\protect\citeauthoryear{{Caughlan} \& {Fowler}}{{Caughlan} \&
  {Fowler}}{1988}]{CF1988}
{Caughlan} G.~R.,  {Fowler} W.~A.,  1988, \mn@doi [Atomic Data and Nuclear Data
  Tables] {10.1016/0092-640X(88)90009-5}, \href
  {http://cdsads.u-strasbg.fr/abs/1988ADNDT..40..283C} {40, 283}

\bibitem[\protect\citeauthoryear{{Cayrel} et~al.,}{{Cayrel}
  et~al.}{2004}]{Cay04}
{Cayrel} R.,  et~al., 2004, \mn@doi [\aap] {10.1051/0004-6361:20034074}, \href
  {http://adsabs.harvard.edu/abs/2004A%26A...416.1117C} {416, 1117}

\bibitem[\protect\citeauthoryear{{Cescutti} \& {Chiappini}}{{Cescutti} \&
  {Chiappini}}{2014}]{ces14}
{Cescutti} G.,  {Chiappini} C.,  2014, \mn@doi [\aap]
  {10.1051/0004-6361/201423432}, \href
  {http://adsabs.harvard.edu/abs/2014A%26A...565A..51C} {565, A51}

\bibitem[\protect\citeauthoryear{{Cescutti}, {Chiappini}, {Hirschi}, {Meynet}
  \& {Frischknecht}}{{Cescutti} et~al.}{2013}]{Ces13}
{Cescutti} G.,  {Chiappini} C.,  {Hirschi} R.,  {Meynet} G.,   {Frischknecht}
  U.,  2013, \mn@doi [\aap] {10.1051/0004-6361/201220809}, \href
  {http://adsabs.harvard.edu/abs/2013A%26A...553A..51C} {553, A51}

\bibitem[\protect\citeauthoryear{{Charbonnel}}{{Charbonnel}}{1995}]{Charbonnel1995}
{Charbonnel} C.,  1995, \mn@doi [\apjl] {10.1086/309744}, \href
  {http://cdsads.u-strasbg.fr/abs/1995ApJ...453L..41C} {453, L41}

\bibitem[\protect\citeauthoryear{{Chen}, {Nissen}, {Zhao}, {Zhang}  \&
  {Benoni}}{{Chen} et~al.}{2000}]{Chen2000}
{Chen} Y.~Q.,  {Nissen} P.~E.,  {Zhao} G.,  {Zhang} H.~W.,   {Benoni} T.,
  2000, \mn@doi [\aaps] {10.1051/aas:2000124}, \href
  {http://adsabs.harvard.edu/abs/2000A%26AS..141..491C} {141, 491}

\bibitem[\protect\citeauthoryear{{Chiappini}, {Matteucci}, {Beers}  \&
  {Nomoto}}{{Chiappini} et~al.}{1999}]{Chiappini1999}
{Chiappini} C.,  {Matteucci} F.,  {Beers} T.~C.,   {Nomoto} K.,  1999, \mn@doi
  [\apj] {10.1086/307006}, \href
  {http://adsabs.harvard.edu/abs/1999ApJ...515..226C} {515, 226}

\bibitem[\protect\citeauthoryear{{Chiappini}, {Hirschi}, {Meynet},
  {Ekstr{\"o}m}, {Maeder}  \& {Matteucci}}{{Chiappini}
  et~al.}{2006}]{Chiappini2006}
{Chiappini} C.,  {Hirschi} R.,  {Meynet} G.,  {Ekstr{\"o}m} S.,  {Maeder} A.,
  {Matteucci} F.,  2006, \mn@doi [\aap] {10.1051/0004-6361:20064866}, \href
  {http://adsabs.harvard.edu/abs/2006A%26A...449L..27C} {449, L27}

\bibitem[\protect\citeauthoryear{{Chiappini}, {Ekstr{\"o}m}, {Meynet},
  {Hirschi}, {Maeder}  \& {Charbonnel}}{{Chiappini}
  et~al.}{2008}]{Chiappini2008}
{Chiappini} C.,  {Ekstr{\"o}m} S.,  {Meynet} G.,  {Hirschi} R.,  {Maeder} A.,
  {Charbonnel} C.,  2008, \mn@doi [\aap] {10.1051/0004-6361:20078698}, \href
  {http://cdsads.u-strasbg.fr/abs/2008A%26A...479L...9C} {479, L9}

\bibitem[\protect\citeauthoryear{{Chieffi} \& {Limongi}}{{Chieffi} \&
  {Limongi}}{2004}]{CL04}
{Chieffi} A.,  {Limongi} M.,  2004, \mn@doi [\apj] {10.1086/392523}, \href
  {http://esoads.eso.org/abs/2004ApJ...608..405C} {608, 405}

\bibitem[\protect\citeauthoryear{{Chieffi} \& {Limongi}}{{Chieffi} \&
  {Limongi}}{2013}]{Chi13}
{Chieffi} A.,  {Limongi} M.,  2013, \mn@doi [\apj]
  {10.1088/0004-637X/764/1/21}, \href
  {http://adsabs.harvard.edu/abs/2013ApJ...764...21C} {764, 21}

\bibitem[\protect\citeauthoryear{{Choplin}, {Hirschi}, {Meynet}  \&
  {Ekstr{\"o}m}}{{Choplin} et~al.}{2017}]{Choplin2017}
{Choplin} A.,  {Hirschi} R.,  {Meynet} G.,   {Ekstr{\"o}m} S.,  2017, \mn@doi
  [\aap] {10.1051/0004-6361/201731948}, \href
  {http://cdsads.u-strasbg.fr/abs/2017A%26A...607L...3C} {607, L3}

\bibitem[\protect\citeauthoryear{{Clayton} \& {Rassbach}}{{Clayton} \&
  {Rassbach}}{1967}]{Cla67}
{Clayton} D.~D.,  {Rassbach} M.~E.,  1967, \mn@doi [\apj] {10.1086/149128},
  \href {http://adsabs.harvard.edu/abs/1967ApJ...148...69C} {148, 69}

\bibitem[\protect\citeauthoryear{{Cohen} et~al.,}{{Cohen} et~al.}{2006}]{coh06}
{Cohen} J.~G.,  et~al., 2006, \mn@doi [\aj] {10.1086/504597}, \href
  {http://adsabs.harvard.edu/abs/2006AJ....132..137C} {132, 137}

\bibitem[\protect\citeauthoryear{{Cristallo}, {Straniero}, {Gallino},
  {Piersanti}, {Dom{\'{\i}}nguez}  \& {Lederer}}{{Cristallo}
  et~al.}{2009}]{Cr09}
{Cristallo} S.,  {Straniero} O.,  {Gallino} R.,  {Piersanti} L.,
  {Dom{\'{\i}}nguez} I.,   {Lederer} M.~T.,  2009, \mn@doi [\apj]
  {10.1088/0004-637X/696/1/797}, \href
  {http://adsabs.harvard.edu/abs/2009ApJ...696..797C} {696, 797}

\bibitem[\protect\citeauthoryear{{Cristallo} et~al.,}{{Cristallo}
  et~al.}{2011}]{Cr11}
{Cristallo} S.,  et~al., 2011, \mn@doi [\apjs] {10.1088/0067-0049/197/2/17},
  \href {http://adsabs.harvard.edu/abs/2011ApJS..197...17C} {197, 17}

\bibitem[\protect\citeauthoryear{{Cristallo}, {Di Leva}, {Imbriani},
  {Piersanti}, {Abia}, {Gialanella}  \& {Straniero}}{{Cristallo}
  et~al.}{2014}]{cr14}
{Cristallo} S.,  {Di Leva} A.,  {Imbriani} G.,  {Piersanti} L.,  {Abia} C.,
  {Gialanella} L.,   {Straniero} O.,  2014, \mn@doi [\aap]
  {10.1051/0004-6361/201424370}, \href
  {http://adsabs.harvard.edu/abs/2014A%26A...570A..46C} {570, A46}

\bibitem[\protect\citeauthoryear{{Cristallo}, {Straniero}, {Piersanti}  \&
  {Gobrecht}}{{Cristallo} et~al.}{2015a}]{Cr15}
{Cristallo} S.,  {Straniero} O.,  {Piersanti} L.,   {Gobrecht} D.,  2015a,
  \mn@doi [\apjs] {10.1088/0067-0049/219/2/40}, \href
  {http://adsabs.harvard.edu/abs/2015ApJS..219...40C} {219, 40}

\bibitem[\protect\citeauthoryear{{Cristallo}, {Abia}, {Straniero}  \&
  {Piersanti}}{{Cristallo} et~al.}{2015b}]{Cr15b}
{Cristallo} S.,  {Abia} C.,  {Straniero} O.,   {Piersanti} L.,  2015b, \mn@doi
  [\apj] {10.1088/0004-637X/801/1/53}, \href
  {http://adsabs.harvard.edu/abs/2015ApJ...801...53C} {801, 53}

\bibitem[\protect\citeauthoryear{{Cristallo}, {Karinkuzhi}, {Goswami},
  {Piersanti}  \& {Gobrecht}}{{Cristallo} et~al.}{2016}]{Cr16}
{Cristallo} S.,  {Karinkuzhi} D.,  {Goswami} A.,  {Piersanti} L.,   {Gobrecht}
  D.,  2016, \mn@doi [\apj] {10.3847/1538-4357/833/2/181}, \href
  {http://adsabs.harvard.edu/abs/2016ApJ...833..181C} {833, 181}

\bibitem[\protect\citeauthoryear{{Cunha}, {Smith}  \& {Gibson}}{{Cunha}
  et~al.}{2008}]{cun08}
{Cunha} K.,  {Smith} V.~V.,   {Gibson} B.~K.,  2008, \mn@doi [\apjl]
  {10.1086/588816}, \href {http://adsabs.harvard.edu/abs/2008ApJ...679L..17C}
  {679, L17}

\bibitem[\protect\citeauthoryear{{Delgado Mena}, {Tsantaki}, {Adibekyan},
  {Sousa}, {Santos}, {Gonz{\'a}lez Hern{\'a}ndez}  \& {Israelian}}{{Delgado
  Mena} et~al.}{2017}]{Del17}
{Delgado Mena} E.,  {Tsantaki} M.,  {Adibekyan} V.~Z.,  {Sousa} S.~G.,
  {Santos} N.~C.,  {Gonz{\'a}lez Hern{\'a}ndez} J.~I.,   {Israelian} G.,  2017,
  preprint, \href {http://adsabs.harvard.edu/abs/2017arXiv170504349D} {}
  (\mn@eprint {arXiv} {1705.04349})

\bibitem[\protect\citeauthoryear{{Denissenkov}}{{Denissenkov}}{2010}]{den10}
{Denissenkov} P.~A.,  2010, \mn@doi [\apj] {10.1088/0004-637X/723/1/563}, \href
  {http://adsabs.harvard.edu/abs/2010ApJ...723..563D} {723, 563}

\bibitem[\protect\citeauthoryear{{Denissenkov} \& {Tout}}{{Denissenkov} \&
  {Tout}}{2003}]{DT03}
{Denissenkov} P.~A.,  {Tout} C.~A.,  2003, \mn@doi [\mnras]
  {10.1046/j.1365-8711.2003.06284.x}, \href
  {http://adsabs.harvard.edu/abs/2003MNRAS.340..722D} {340, 722}

\bibitem[\protect\citeauthoryear{{Doherty}, {Gil-Pons}, {Lau}, {Lattanzio}  \&
  {Siess}}{{Doherty} et~al.}{2014}]{Doherty2014}
{Doherty} C.~L.,  {Gil-Pons} P.,  {Lau} H.~H.~B.,  {Lattanzio} J.~C.,   {Siess}
  L.,  2014, \mn@doi [\mnras] {10.1093/mnras/stt1877}, \href
  {http://adsabs.harvard.edu/abs/2014MNRAS.437..195D} {437, 195}

\bibitem[\protect\citeauthoryear{{Drout} et~al.,}{{Drout} et~al.}{2017}]{dro17}
{Drout} M.~R.,  et~al., 2017, preprint, \href
  {http://adsabs.harvard.edu/abs/2017arXiv171005443D} {} (\mn@eprint {arXiv}
  {1710.05443})

\bibitem[\protect\citeauthoryear{{Dufton} et~al.,}{{Dufton}
  et~al.}{2006}]{duftonetal06}
{Dufton} P.~L.,  et~al., 2006, \mn@doi [\aap] {10.1051/0004-6361:20065392},
  \href {http://adsabs.harvard.edu/abs/2006A%26A...457..265D} {457, 265}

\bibitem[\protect\citeauthoryear{{Eggen}, {Lynden-Bell}  \& {Sandage}}{{Eggen}
  et~al.}{1962}]{Eggen1962}
{Eggen} O.~J.,  {Lynden-Bell} D.,   {Sandage} A.~R.,  1962, \mn@doi [\apj]
  {10.1086/147433}, \href {http://adsabs.harvard.edu/abs/1962ApJ...136..748E}
  {136, 748}

\bibitem[\protect\citeauthoryear{{Ertl}, {Janka}, {Woosley}, {Sukhbold}  \&
  {Ugliano}}{{Ertl} et~al.}{2016}]{ertl16}
{Ertl} T.,  {Janka} H.-T.,  {Woosley} S.~E.,  {Sukhbold} T.,   {Ugliano} M.,
  2016, \mn@doi [\apj] {10.3847/0004-637X/818/2/124}, \href
  {http://esoads.eso.org/abs/2016ApJ...818..124E} {818, 124}

\bibitem[\protect\citeauthoryear{{Esteban}, {Bresolin}, {Peimbert},
  {Garc{\'{\i}}a-Rojas}, {Peimbert}  \& {Mesa-Delgado}}{{Esteban}
  et~al.}{2009}]{est09}
{Esteban} C.,  {Bresolin} F.,  {Peimbert} M.,  {Garc{\'{\i}}a-Rojas} J.,
  {Peimbert} A.,   {Mesa-Delgado} A.,  2009, \mn@doi [\apj]
  {10.1088/0004-637X/700/1/654}, \href
  {http://adsabs.harvard.edu/abs/2009ApJ...700..654E} {700, 654}

\bibitem[\protect\citeauthoryear{{Esteban}, {Garc{\'{\i}}a-Rojas}, {Carigi},
  {Peimbert}, {Bresolin}, {L{\'o}pez-S{\'a}nchez}  \& {Mesa-Delgado}}{{Esteban}
  et~al.}{2014}]{est14}
{Esteban} C.,  {Garc{\'{\i}}a-Rojas} J.,  {Carigi} L.,  {Peimbert} M.,
  {Bresolin} F.,  {L{\'o}pez-S{\'a}nchez} A.~R.,   {Mesa-Delgado} A.,  2014,
  \mn@doi [\mnras] {10.1093/mnras/stu1177}, \href
  {http://adsabs.harvard.edu/abs/2014MNRAS.443..624E} {443, 624}

\bibitem[\protect\citeauthoryear{{Fishlock}, {Yong}, {Karakas}, {Alves-Brito},
  {Mel{\'e}ndez}, {Nissen}, {Kobayashi}  \& {Casey}}{{Fishlock}
  et~al.}{2017}]{fis17}
{Fishlock} C.~K.,  {Yong} D.,  {Karakas} A.~I.,  {Alves-Brito} A.,
  {Mel{\'e}ndez} J.,  {Nissen} P.~E.,  {Kobayashi} C.,   {Casey} A.~R.,  2017,
  \mn@doi [\mnras] {10.1093/mnras/stx047}, \href
  {http://adsabs.harvard.edu/abs/2017MNRAS.466.4672F} {466, 4672}

\bibitem[\protect\citeauthoryear{{Fran{\c c}ois}, {Matteucci}, {Cayrel},
  {Spite}, {Spite}  \& {Chiappini}}{{Fran{\c c}ois}
  et~al.}{2004}]{Francois2004}
{Fran{\c c}ois} P.,  {Matteucci} F.,  {Cayrel} R.,  {Spite} M.,  {Spite} F.,
  {Chiappini} C.,  2004, \mn@doi [\aap] {10.1051/0004-6361:20034140}, \href
  {http://adsabs.harvard.edu/abs/2004A%26A...421..613F} {421, 613}

\bibitem[\protect\citeauthoryear{{Frebel}}{{Frebel}}{2010}]{Fre10}
{Frebel} A.,  2010, \mn@doi [Astronomische Nachrichten]
  {10.1002/asna.201011362}, \href
  {http://adsabs.harvard.edu/abs/2010AN....331..474F} {331, 474}

\bibitem[\protect\citeauthoryear{{Freiburghaus}, {Rosswog}  \&
  {Thielemann}}{{Freiburghaus} et~al.}{1999}]{Fre99}
{Freiburghaus} C.,  {Rosswog} S.,   {Thielemann} F.-K.,  1999, \mn@doi [\apjl]
  {10.1086/312343}, \href {http://adsabs.harvard.edu/abs/1999ApJ...525L.121F}
  {525, L121}

\bibitem[\protect\citeauthoryear{{Frischknecht}, {Hirschi}  \&
  {Thielemann}}{{Frischknecht} et~al.}{2012}]{Frischknecht2012}
{Frischknecht} U.,  {Hirschi} R.,   {Thielemann} F.-K.,  2012, \mn@doi [\aap]
  {10.1051/0004-6361/201117794}, \href
  {http://adsabs.harvard.edu/abs/2012A%26A...538L...2F} {538, L2}

\bibitem[\protect\citeauthoryear{{Frischknecht} et~al.,}{{Frischknecht}
  et~al.}{2016}]{Fri16}
{Frischknecht} U.,  et~al., 2016, \mn@doi [\mnras] {10.1093/mnras/stv2723},
  \href {http://adsabs.harvard.edu/abs/2016MNRAS.456.1803F} {456, 1803}

\bibitem[\protect\citeauthoryear{{Gallino}, {Arlandini}, {Busso}, {Lugaro},
  {Travaglio}, {Straniero}, {Chieffi}  \& {Limongi}}{{Gallino}
  et~al.}{1998}]{Ga98}
{Gallino} R.,  {Arlandini} C.,  {Busso} M.,  {Lugaro} M.,  {Travaglio} C.,
  {Straniero} O.,  {Chieffi} A.,   {Limongi} M.,  1998, \mn@doi [\apj]
  {10.1086/305437}, \href {http://adsabs.harvard.edu/abs/1998ApJ...497..388G}
  {497, 388}

\bibitem[\protect\citeauthoryear{{Gilmore} et~al.,}{{Gilmore}
  et~al.}{2012}]{Gil12}
{Gilmore} G.,  et~al., 2012, The Messenger, \href
  {http://adsabs.harvard.edu/abs/2012Msngr.147...25G} {147, 25}

\bibitem[\protect\citeauthoryear{{Goriely}, {Jorissen}  \& {Arnould}}{{Goriely}
  et~al.}{1990}]{GORI90}
{Goriely} S.,  {Jorissen} A.,   {Arnould} M.,  1990, in {Hillebrandt} W.,
  {M{\"u}ller} E.,  eds, Nuclear Astrophysics, 5th Workshop. p.~60

\bibitem[\protect\citeauthoryear{{Goswami} \& {Prantzos}}{{Goswami} \&
  {Prantzos}}{2000}]{Goswami2000}
{Goswami} A.,  {Prantzos} N.,  2000, \aap, \href
  {http://adsabs.harvard.edu/abs/2000A%26A...359..191G} {359, 191}

\bibitem[\protect\citeauthoryear{{Gratton}, {Sneden}, {Carretta}  \&
  {Bragaglia}}{{Gratton} et~al.}{2000}]{Gratton2000}
{Gratton} R.~G.,  {Sneden} C.,  {Carretta} E.,   {Bragaglia} A.,  2000, \aap,
  \href {http://cdsads.u-strasbg.fr/abs/2000A%26A...354..169G} {354, 169}

\bibitem[\protect\citeauthoryear{{Greggio}}{{Greggio}}{2005}]{Gre05}
{Greggio} L.,  2005, \mn@doi [\aap] {10.1051/0004-6361:20052926}, \href
  {http://adsabs.harvard.edu/abs/2005A%26A...441.1055G} {441, 1055}

\bibitem[\protect\citeauthoryear{{Guerrero} et~al.,}{{Guerrero}
  et~al.}{2013}]{Gu13}
{Guerrero} C.,  et~al., 2013, \mn@doi [European Physical Journal A]
  {10.1140/epja/i2013-13027-6}, \href
  {http://adsabs.harvard.edu/abs/2013EPJA...49...27G} {49, 27}

\bibitem[\protect\citeauthoryear{{Hansen} et~al.,}{{Hansen}
  et~al.}{2012}]{han12}
{Hansen} C.~J.,  et~al., 2012, \mn@doi [\aap] {10.1051/0004-6361/201118643},
  \href {http://adsabs.harvard.edu/abs/2012A%26A...545A..31H} {545, A31}

\bibitem[\protect\citeauthoryear{{Hedrosa}, {Abia}, {Busso}, {Cristallo},
  {Dom{\'{\i}}nguez}, {Palmerini}, {Plez}  \& {Straniero}}{{Hedrosa}
  et~al.}{2013}]{hed13}
{Hedrosa} R.~P.,  {Abia} C.,  {Busso} M.,  {Cristallo} S.,  {Dom{\'{\i}}nguez}
  I.,  {Palmerini} S.,  {Plez} B.,   {Straniero} O.,  2013, \mn@doi [\apjl]
  {10.1088/2041-8205/768/1/L11}, \href
  {http://adsabs.harvard.edu/abs/2013ApJ...768L..11H} {768, L11}

\bibitem[\protect\citeauthoryear{{Heger} \& {Woosley}}{{Heger} \&
  {Woosley}}{2010}]{HW10}
{Heger} A.,  {Woosley} S.~E.,  2010, \mn@doi [\apj]
  {10.1088/0004-637X/724/1/341}, \href
  {http://esoads.eso.org/abs/2010ApJ...724..341H} {724, 341}

\bibitem[\protect\citeauthoryear{{Heger}, {Langer}  \& {Woosley}}{{Heger}
  et~al.}{2000}]{Heg00}
{Heger} A.,  {Langer} N.,   {Woosley} S.~E.,  2000, \mn@doi [\apj]
  {10.1086/308158}, \href {http://adsabs.harvard.edu/abs/2000ApJ...528..368H}
  {528, 368}

\bibitem[\protect\citeauthoryear{{Heijmans} et~al.,}{{Heijmans}
  et~al.}{2012}]{Hei12}
{Heijmans} J.,  et~al., 2012, in Ground-based and Airborne Instrumentation for
  Astronomy IV. p. 84460W, \mn@doi{10.1117/12.925806}

\bibitem[\protect\citeauthoryear{{Herwig}}{{Herwig}}{2005}]{He05}
{Herwig} F.,  2005, \mn@doi [\araa] {10.1146/annurev.astro.43.072103.150600},
  \href {http://adsabs.harvard.edu/abs/2005ARA%26A..43..435H} {43, 435}

\bibitem[\protect\citeauthoryear{{Hirschi}}{{Hirschi}}{2007}]{Hirschi2007}
{Hirschi} R.,  2007, \mn@doi [\aap] {10.1051/0004-6361:20065356}, \href
  {http://adsabs.harvard.edu/abs/2007A%26A...461..571H} {461, 571}

\bibitem[\protect\citeauthoryear{{Hunter}, {Lennon}, {Dufton}, {Trundle},
  {Sim{\'o}n-D{\'{\i}}az}, {Smartt}, {Ryans}  \& {Evans}}{{Hunter}
  et~al.}{2008}]{hunteretal08}
{Hunter} I.,  {Lennon} D.~J.,  {Dufton} P.~L.,  {Trundle} C.,
  {Sim{\'o}n-D{\'{\i}}az} S.,  {Smartt} S.~J.,  {Ryans} R.~S.~I.,   {Evans}
  C.~J.,  2008, \mn@doi [\aap] {10.1051/0004-6361:20078511}, \href
  {http://adsabs.harvard.edu/abs/2008A%26A...479..541H} {479, 541}

\bibitem[\protect\citeauthoryear{{Hunter} et~al.,}{{Hunter}
  et~al.}{2009}]{Hun09}
{Hunter} I.,  et~al., 2009, \mn@doi [\aap] {10.1051/0004-6361/200809925}, \href
  {http://adsabs.harvard.edu/abs/2009A%26A...496..841H} {496, 841}

\bibitem[\protect\citeauthoryear{{Iben} \& {Renzini}}{{Iben} \&
  {Renzini}}{1983}]{IR83}
{Iben} Jr. I.,  {Renzini} A.,  1983, \mn@doi [\araa]
  {10.1146/annurev.aa.21.090183.001415}, \href
  {http://adsabs.harvard.edu/abs/1983ARA%26A..21..271I} {21, 271}

\bibitem[\protect\citeauthoryear{{Indelicato} et~al.,}{{Indelicato}
  et~al.}{2017}]{ind17}
{Indelicato} I.,  et~al., 2017, \mn@doi [\apj] {10.3847/1538-4357/aa7de7},
  \href {http://adsabs.harvard.edu/abs/2017ApJ...845...19I} {845, 19}

\bibitem[\protect\citeauthoryear{{Ishimaru}, {Wanajo}  \&
  {Prantzos}}{{Ishimaru} et~al.}{2015}]{Ishimaru2015}
{Ishimaru} Y.,  {Wanajo} S.,   {Prantzos} N.,  2015, \mn@doi [\apjl]
  {10.1088/2041-8205/804/2/L35}, \href
  {http://adsabs.harvard.edu/abs/2015ApJ...804L..35I} {804, L35}

\bibitem[\protect\citeauthoryear{{Iwamoto}, {Brachwitz}, {Nomoto}, {Kishimoto},
  {Umeda}, {Hix}  \& {Thielemann}}{{Iwamoto} et~al.}{1999}]{Iwa99}
{Iwamoto} K.,  {Brachwitz} F.,  {Nomoto} K.,  {Kishimoto} N.,  {Umeda} H.,
  {Hix} W.~R.,   {Thielemann} F.-K.,  1999, \mn@doi [\apjs] {10.1086/313278},
  \href {http://adsabs.harvard.edu/abs/1999ApJS..125..439I} {125, 439}

\bibitem[\protect\citeauthoryear{{J{\"o}nsson} et~al.,}{{J{\"o}nsson}
  et~al.}{2014}]{Jon14}
{J{\"o}nsson} H.,  et~al., 2014, \mn@doi [\aap] {10.1051/0004-6361/201423597},
  \href {http://adsabs.harvard.edu/abs/2014A%26A...564A.122J} {564, A122}

\bibitem[\protect\citeauthoryear{{J{\"o}nsson}, {Ryde}, {Spitoni}, {Matteucci},
  {Cunha}, {Smith}, {Hinkle}  \& {Schultheis}}{{J{\"o}nsson}
  et~al.}{2017}]{Jon17}
{J{\"o}nsson} H.,  {Ryde} N.,  {Spitoni} E.,  {Matteucci} F.,  {Cunha} K.,
  {Smith} V.,  {Hinkle} K.,   {Schultheis} M.,  2017, \mn@doi [\apj]
  {10.3847/1538-4357/835/1/50}, \href
  {http://adsabs.harvard.edu/abs/2017ApJ...835...50J} {835, 50}

\bibitem[\protect\citeauthoryear{{Jos\'e}}{{Jos\'e}}{2016}]{jos16}
{Jos\'e} J.,  2016, {Stellar Explosions: Hydrodynamics and Nucleosynthesis},
  \mn@doi{10.1201/b19165.
}

\bibitem[\protect\citeauthoryear{{K{\"a}ppeler}, {Gallino}, {Bisterzo}  \&
  {Aoki}}{{K{\"a}ppeler} et~al.}{2011}]{Kap11}
{K{\"a}ppeler} F.,  {Gallino} R.,  {Bisterzo} S.,   {Aoki} W.,  2011, \mn@doi
  [Reviews of Modern Physics] {10.1103/RevModPhys.83.157}, \href
  {http://adsabs.harvard.edu/abs/2011RvMP...83..157K} {83, 157}

\bibitem[\protect\citeauthoryear{{Karakas} \& {Lattanzio}}{{Karakas} \&
  {Lattanzio}}{2014}]{KL14}
{Karakas} A.~I.,  {Lattanzio} J.~C.,  2014, \mn@doi [\pasa]
  {10.1017/pasa.2014.21}, \href
  {http://adsabs.harvard.edu/abs/2014PASA...31...30K} {31, 30}

\bibitem[\protect\citeauthoryear{{Karakas} \& {Lugaro}}{{Karakas} \&
  {Lugaro}}{2016}]{KL16}
{Karakas} A.~I.,  {Lugaro} M.,  2016, \mn@doi [\apj]
  {10.3847/0004-637X/825/1/26}, \href
  {http://adsabs.harvard.edu/abs/2016ApJ...825...26K} {825, 26}

\bibitem[\protect\citeauthoryear{{Kasen}, {Badnell}  \& {Barnes}}{{Kasen}
  et~al.}{2013}]{Kasen2013}
{Kasen} D.,  {Badnell} N.~R.,   {Barnes} J.,  2013, \mn@doi [\apj]
  {10.1088/0004-637X/774/1/25}, \href
  {http://adsabs.harvard.edu/abs/2013ApJ...774...25K} {774, 25}

\bibitem[\protect\citeauthoryear{{Kobayashi}, {Karakas}  \&
  {Umeda}}{{Kobayashi} et~al.}{2011a}]{Kobayashi2011}
{Kobayashi} C.,  {Karakas} A.~I.,   {Umeda} H.,  2011a, \mn@doi [\mnras]
  {10.1111/j.1365-2966.2011.18621.x}, \href
  {http://adsabs.harvard.edu/abs/2011MNRAS.414.3231K} {414, 3231}

\bibitem[\protect\citeauthoryear{{Kobayashi}, {Izutani}, {Karakas}, {Yoshida},
  {Yong}  \& {Umeda}}{{Kobayashi} et~al.}{2011b}]{Kobayashi2011b}
{Kobayashi} C.,  {Izutani} N.,  {Karakas} A.~I.,  {Yoshida} T.,  {Yong} D.,
  {Umeda} H.,  2011b, \mn@doi [\apjl] {10.1088/2041-8205/739/2/L57}, \href
  {http://adsabs.harvard.edu/abs/2011ApJ...739L..57K} {739, L57}

\bibitem[\protect\citeauthoryear{{Komiya}, {Yamada}, {Suda}  \&
  {Fujimoto}}{{Komiya} et~al.}{2014}]{Komiya2014}
{Komiya} Y.,  {Yamada} S.,  {Suda} T.,   {Fujimoto} M.~Y.,  2014, \mn@doi
  [\apj] {10.1088/0004-637X/783/2/132}, \href
  {http://adsabs.harvard.edu/abs/2014ApJ...783..132K} {783, 132}

\bibitem[\protect\citeauthoryear{{Kroupa}}{{Kroupa}}{2002}]{Kro02}
{Kroupa} P.,  2002, in {Grebel} E.~K.,  {Brandner} W.,  eds,  Astronomical
  Society of the Pacific Conference Series Vol. 285, Modes of Star Formation
  and the Origin of Field Populations. p.~86 (\mn@eprint {} {astro-ph/0102155})

\bibitem[\protect\citeauthoryear{{Kubryk}, {Prantzos}  \&
  {Athanassoula}}{{Kubryk} et~al.}{2015}]{Kubryk2015a}
{Kubryk} M.,  {Prantzos} N.,   {Athanassoula} E.,  2015, \mn@doi [\aap]
  {10.1051/0004-6361/201424171}, \href
  {http://adsabs.harvard.edu/abs/2015A%26A...580A.126K} {580, A126}

\bibitem[\protect\citeauthoryear{{Kunz}, {Fey}, {Jaeger}, {Mayer}, {Hammer},
  {Staudt}, {Harissopulos}  \& {Paradellis}}{{Kunz} et~al.}{2002}]{kun02}
{Kunz} R.,  {Fey} M.,  {Jaeger} M.,  {Mayer} A.,  {Hammer} J.~W.,  {Staudt} G.,
   {Harissopulos} S.,   {Paradellis} T.,  2002, \mn@doi [\apj]
  {10.1086/338384}, \href {http://adsabs.harvard.edu/abs/2002ApJ...567..643K}
  {567, 643}

\bibitem[\protect\citeauthoryear{{Lai}, {Bolte}, {Johnson}, {Lucatello},
  {Heger}  \& {Woosley}}{{Lai} et~al.}{2008}]{Lai2008}
{Lai} D.~K.,  {Bolte} M.,  {Johnson} J.~A.,  {Lucatello} S.,  {Heger} A.,
  {Woosley} S.~E.,  2008, \mn@doi [\apj] {10.1086/588811}, \href
  {http://adsabs.harvard.edu/abs/2008ApJ...681.1524L} {681, 1524}

\bibitem[\protect\citeauthoryear{{Lattimer}, {Schramm}  \&
  {Grossman}}{{Lattimer} et~al.}{1977}]{Lat77}
{Lattimer} J.~M.,  {Schramm} D.~N.,   {Grossman} L.,  1977, \mn@doi [\nat]
  {10.1038/269116a0}, \href {http://adsabs.harvard.edu/abs/1977Natur.269..116L}
  {269, 116}

\bibitem[\protect\citeauthoryear{{Levshakov}, {Centuri{\'o}n}, {Molaro}  \&
  {Kostina}}{{Levshakov} et~al.}{2006}]{les06}
{Levshakov} S.~A.,  {Centuri{\'o}n} M.,  {Molaro} P.,   {Kostina} M.~V.,  2006,
  \mn@doi [\aap] {10.1051/0004-6361:200600001}, \href
  {http://cdsads.u-strasbg.fr/abs/2006A%26A...447L..21L} {447, L21}

\bibitem[\protect\citeauthoryear{{Li}, {Ludwig}, {Caffau}, {Christlieb}  \&
  {Zhao}}{{Li} et~al.}{2013}]{li13}
{Li} H.~N.,  {Ludwig} H.-G.,  {Caffau} E.,  {Christlieb} N.,   {Zhao} G.,
  2013, \mn@doi [\apj] {10.1088/0004-637X/765/1/51}, \href
  {http://adsabs.harvard.edu/abs/2013ApJ...765...51L} {765, 51}

\bibitem[\protect\citeauthoryear{{Limongi} \& {Chieffi}}{{Limongi} \&
  {Chieffi}}{2003}]{LC03}
{Limongi} M.,  {Chieffi} A.,  2003, \mn@doi [\apj] {10.1086/375703}, \href
  {http://cdsads.u-strasbg.fr/abs/2003ApJ...592..404L} {592, 404}

\bibitem[\protect\citeauthoryear{{Limongi} \& {Chieffi}}{{Limongi} \&
  {Chieffi}}{2012}]{LC12}
{Limongi} M.,  {Chieffi} A.,  2012, \mn@doi [\apjs]
  {10.1088/0067-0049/199/2/38}, \href
  {http://esoads.eso.org/abs/2012ApJS..199...38L} {199, 38}

\bibitem[\protect\citeauthoryear{{Limongi} \& {Chieffi}}{{Limongi} \&
  {Chieffi}}{2018}]{LC18}
{Limongi} M.,  {Chieffi} A.,  2018, in preparation

\bibitem[\protect\citeauthoryear{{Limongi}, {Straniero}  \&
  {Chieffi}}{{Limongi} et~al.}{2000}]{LSC00}
{Limongi} M.,  {Straniero} O.,   {Chieffi} A.,  2000, \mn@doi [\apjs]
  {10.1086/313424}, \href {http://esoads.eso.org/abs/2000ApJS..129..625L} {129,
  625}

\bibitem[\protect\citeauthoryear{{Lodders}, {Palme}  \& {Gail}}{{Lodders}
  et~al.}{2009}]{Lod09}
{Lodders} K.,  {Palme} H.,   {Gail} H.-P.,  2009, \mn@doi [Landolt
  B{\"o}rnstein] {10.1007/978-3-540-88055-4_34}, \href
  {http://adsabs.harvard.edu/abs/2009LanB...4B...44L} {}

\bibitem[\protect\citeauthoryear{{Lucatello}, {Gratton}, {Cohen}, {Beers},
  {Christlieb}, {Carretta}  \& {Ram{\'{\i}}rez}}{{Lucatello}
  et~al.}{2003}]{luc03}
{Lucatello} S.,  {Gratton} R.,  {Cohen} J.~G.,  {Beers} T.~C.,  {Christlieb}
  N.,  {Carretta} E.,   {Ram{\'{\i}}rez} S.,  2003, \mn@doi [\aj]
  {10.1086/345886}, \href {http://adsabs.harvard.edu/abs/2003AJ....125..875L}
  {125, 875}

\bibitem[\protect\citeauthoryear{{Luck} \& {Bond}}{{Luck} \&
  {Bond}}{1991}]{Luc91}
{Luck} R.~E.,  {Bond} H.~E.,  1991, \mn@doi [\apjs] {10.1086/191615}, \href
  {http://adsabs.harvard.edu/abs/1991ApJS...77..515L} {77, 515}

\bibitem[\protect\citeauthoryear{{Maas}, {Pilachowski}  \& {Hinkle}}{{Maas}
  et~al.}{2016}]{Maas2016}
{Maas} Z.~G.,  {Pilachowski} C.~A.,   {Hinkle} K.,  2016, \mn@doi [\aj]
  {10.3847/0004-6256/152/6/196}, \href
  {http://adsabs.harvard.edu/abs/2016AJ....152..196M} {152, 196}

\bibitem[\protect\citeauthoryear{{Maas}, {Pilachowski}  \& {Cescutti}}{{Maas}
  et~al.}{2017}]{Maas2017}
{Maas} Z.~G.,  {Pilachowski} C.~A.,   {Cescutti} G.,  2017, \mn@doi [\apj]
  {10.3847/1538-4357/aa7050}, \href
  {http://adsabs.harvard.edu/abs/2017ApJ...841..108M} {841, 108}

\bibitem[\protect\citeauthoryear{{Maeder}}{{Maeder}}{1983}]{Maeder1983}
{Maeder} A.,  1983, \aap, \href
  {http://adsabs.harvard.edu/abs/1983A%26A...120..113M} {120, 113}

\bibitem[\protect\citeauthoryear{{Maeder}}{{Maeder}}{1992}]{Maeder1992}
{Maeder} A.,  1992, \aap, \href
  {http://cdsads.u-strasbg.fr/abs/1992A%26A...264..105M} {264, 105}

\bibitem[\protect\citeauthoryear{{Maeder} \& {Meynet}}{{Maeder} \&
  {Meynet}}{2012}]{Maeder2012}
{Maeder} A.,  {Meynet} G.,  2012, \mn@doi [Reviews of Modern Physics]
  {10.1103/RevModPhys.84.25}, \href
  {http://adsabs.harvard.edu/abs/2012RvMP...84...25M} {84, 25}

\bibitem[\protect\citeauthoryear{{Maeder}, {Meynet}  \& {Chiappini}}{{Maeder}
  et~al.}{2015}]{Maeder2015}
{Maeder} A.,  {Meynet} G.,   {Chiappini} C.,  2015, \mn@doi [\aap]
  {10.1051/0004-6361/201424153}, \href
  {http://cdsads.u-strasbg.fr/abs/2015A%26A...576A..56M} {576, A56}

\bibitem[\protect\citeauthoryear{{Maiorca}, {Uitenbroek}, {Uttenthaler},
  {Randich}, {Busso}  \& {Magrini}}{{Maiorca} et~al.}{2014}]{mai14}
{Maiorca} E.,  {Uitenbroek} H.,  {Uttenthaler} S.,  {Randich} S.,  {Busso} M.,
   {Magrini} L.,  2014, \mn@doi [\apj] {10.1088/0004-637X/788/2/149}, \href
  {http://adsabs.harvard.edu/abs/2014ApJ...788..149M} {788, 149}

\bibitem[\protect\citeauthoryear{{Maoz} \& {Mannucci}}{{Maoz} \&
  {Mannucci}}{2012}]{Mao12}
{Maoz} D.,  {Mannucci} F.,  2012, \mn@doi [\pasa] {10.1071/AS11052}, \href
  {http://adsabs.harvard.edu/abs/2012PASA...29..447M} {29, 447}

\bibitem[\protect\citeauthoryear{{Marty}, {Chaussidon}, {Wiens}, {Jurewicz}  \&
  {Burnett}}{{Marty} et~al.}{2011}]{mar11}
{Marty} B.,  {Chaussidon} M.,  {Wiens} R.~C.,  {Jurewicz} A.~J.~G.,   {Burnett}
  D.~S.,  2011, \mn@doi [Science] {10.1126/science.1204656}, \href
  {http://adsabs.harvard.edu/abs/2011Sci...332.1533M} {332, 1533}

\bibitem[\protect\citeauthoryear{{Masseron}, {Johnson}, {Lucatello}, {Karakas},
  {Plez}, {Beers}  \& {Christlieb}}{{Masseron} et~al.}{2012}]{mas12}
{Masseron} T.,  {Johnson} J.~A.,  {Lucatello} S.,  {Karakas} A.,  {Plez} B.,
  {Beers} T.~C.,   {Christlieb} N.,  2012, \mn@doi [\apj]
  {10.1088/0004-637X/751/1/14}, \href
  {http://adsabs.harvard.edu/abs/2012ApJ...751...14M} {751, 14}

\bibitem[\protect\citeauthoryear{{Matteucci}}{{Matteucci}}{1996}]{Matteucci1996}
{Matteucci} F.,  1996, \fcp, \href
  {http://adsabs.harvard.edu/abs/1996FCPh...17..283M} {17, 283}

\bibitem[\protect\citeauthoryear{{Metzger} et~al.,}{{Metzger}
  et~al.}{2010}]{Metzger2010}
{Metzger} B.~D.,  et~al., 2010, \mn@doi [\mnras]
  {10.1111/j.1365-2966.2010.16864.x}, \href
  {http://adsabs.harvard.edu/abs/2010MNRAS.406.2650M} {406, 2650}

\bibitem[\protect\citeauthoryear{{Meynet} \& {Maeder}}{{Meynet} \&
  {Maeder}}{1997}]{Mey97}
{Meynet} G.,  {Maeder} A.,  1997, \aap, \href
  {http://adsabs.harvard.edu/abs/1997A%26A...321..465M} {321, 465}

\bibitem[\protect\citeauthoryear{{Meynet} \& {Maeder}}{{Meynet} \&
  {Maeder}}{2002a}]{MM02}
{Meynet} G.,  {Maeder} A.,  2002a, \mn@doi [\aap] {10.1051/0004-6361:20011554},
  \href {http://cdsads.u-strasbg.fr/abs/2002A%26A...381L..25M} {381, L25}

\bibitem[\protect\citeauthoryear{{Meynet} \& {Maeder}}{{Meynet} \&
  {Maeder}}{2002b}]{MM2002b}
{Meynet} G.,  {Maeder} A.,  2002b, \mn@doi [\aap] {10.1051/0004-6361:20020755},
  \href {http://adsabs.harvard.edu/abs/2002A%26A...390..561M} {390, 561}

\bibitem[\protect\citeauthoryear{{Mishenina}, {Pignatari}, {Korotin},
  {Soubiran}, {Charbonnel}, {Thielemann}, {Gorbaneva}  \& {Basak}}{{Mishenina}
  et~al.}{2013}]{Mishenina2013}
{Mishenina} T.~V.,  {Pignatari} M.,  {Korotin} S.~A.,  {Soubiran} C.,
  {Charbonnel} C.,  {Thielemann} F.-K.,  {Gorbaneva} T.~I.,   {Basak} N.~Y.,
  2013, \mn@doi [\aap] {10.1051/0004-6361/201220687}, \href
  {http://adsabs.harvard.edu/abs/2013A%26A...552A.128M} {552, A128}

\bibitem[\protect\citeauthoryear{{Montes} et~al.,}{{Montes}
  et~al.}{2007}]{Mo07}
{Montes} F.,  et~al., 2007, \mn@doi [\apj] {10.1086/523084}, \href
  {http://adsabs.harvard.edu/abs/2007ApJ...671.1685M} {671, 1685}

\bibitem[\protect\citeauthoryear{{Nakajima} \& {Sorahana}}{{Nakajima} \&
  {Sorahana}}{2016}]{nak16}
{Nakajima} T.,  {Sorahana} S.,  2016, \mn@doi [\apj]
  {10.3847/0004-637X/830/2/159}, \href
  {http://adsabs.harvard.edu/abs/2016ApJ...830..159N} {830, 159}

\bibitem[\protect\citeauthoryear{{Nault} \& {Pilachowski}}{{Nault} \&
  {Pilachowski}}{2013}]{nau13}
{Nault} K.~A.,  {Pilachowski} C.~A.,  2013, \mn@doi [\aj]
  {10.1088/0004-6256/146/6/153}, \href
  {http://adsabs.harvard.edu/abs/2013AJ....146..153N} {146, 153}

\bibitem[\protect\citeauthoryear{{Nissen}, {Chen}, {Carigi}, {Schuster}  \&
  {Zhao}}{{Nissen} et~al.}{2014}]{nis14}
{Nissen} P.~E.,  {Chen} Y.~Q.,  {Carigi} L.,  {Schuster} W.~J.,   {Zhao} G.,
  2014, \mn@doi [\aap] {10.1051/0004-6361/201424184}, \href
  {http://adsabs.harvard.edu/abs/2014A%26A...568A..25N} {568, A25}

\bibitem[\protect\citeauthoryear{{Nomoto}, {Tominaga}, {Umeda}, {Kobayashi}  \&
  {Maeda}}{{Nomoto} et~al.}{2006}]{NTUKM06}
{Nomoto} K.,  {Tominaga} N.,  {Umeda} H.,  {Kobayashi} C.,   {Maeda} K.,  2006,
  \mn@doi [Nuclear Physics A] {10.1016/j.nuclphysa.2006.05.008}, \href
  {http://esoads.eso.org/abs/2006NuPhA.777..424N} {777, 424}

\bibitem[\protect\citeauthoryear{{Nomoto}, {Kobayashi}  \& {Tominaga}}{{Nomoto}
  et~al.}{2013}]{Nomoto2013}
{Nomoto} K.,  {Kobayashi} C.,   {Tominaga} N.,  2013, \mn@doi [\araa]
  {10.1146/annurev-astro-082812-140956}, \href
  {http://cdsads.u-strasbg.fr/abs/2013ARA%26A..51..457N} {51, 457}

\bibitem[\protect\citeauthoryear{{O'Connor} \& {Ott}}{{O'Connor} \&
  {Ott}}{2011}]{oo11}
{O'Connor} E.,  {Ott} C.~D.,  2011, \mn@doi [\apj]
  {10.1088/0004-637X/730/2/70}, \href
  {http://esoads.eso.org/abs/2011ApJ...730...70O} {730, 70}

\bibitem[\protect\citeauthoryear{{Palacios}, {Charbonnel}, {Talon}  \&
  {Siess}}{{Palacios} et~al.}{2006}]{Palacios2006}
{Palacios} A.,  {Charbonnel} C.,  {Talon} S.,   {Siess} L.,  2006, \mn@doi
  [\aap] {10.1051/0004-6361:20053065}, \href
  {http://cdsads.u-strasbg.fr/abs/2006A%26A...453..261P} {453, 261}

\bibitem[\protect\citeauthoryear{{Pejcha} \& {Prieto}}{{Pejcha} \&
  {Prieto}}{2015}]{pp15}
{Pejcha} O.,  {Prieto} J.~L.,  2015, \mn@doi [\apj]
  {10.1088/0004-637X/806/2/225}, \href
  {http://esoads.eso.org/abs/2015ApJ...806..225P} {806, 225}

\bibitem[\protect\citeauthoryear{{Pian} et~al.,}{{Pian}
  et~al.}{2017}]{Pian2017}
{Pian} E.,  et~al., 2017, \mn@doi [\nat] {10.1038/nature24298}, \href
  {http://adsabs.harvard.edu/abs/2017Natur.551...67P} {551, 67}

\bibitem[\protect\citeauthoryear{{Piersanti}, {Cristallo}  \&
  {Straniero}}{{Piersanti} et~al.}{2013}]{Pi13}
{Piersanti} L.,  {Cristallo} S.,   {Straniero} O.,  2013, \mn@doi [\apj]
  {10.1088/0004-637X/774/2/98}, \href
  {http://adsabs.harvard.edu/abs/2013ApJ...774...98P} {774, 98}

\bibitem[\protect\citeauthoryear{{Pignatari}, {Gallino}, {Heil}, {Wiescher},
  {K{\"a}ppeler}, {Herwig}  \& {Bisterzo}}{{Pignatari}
  et~al.}{2010}]{Pignatari2010}
{Pignatari} M.,  {Gallino} R.,  {Heil} M.,  {Wiescher} M.,  {K{\"a}ppeler} F.,
  {Herwig} F.,   {Bisterzo} S.,  2010, \mn@doi [\apj]
  {10.1088/0004-637X/710/2/1557}, \href
  {http://adsabs.harvard.edu/abs/2010ApJ...710.1557P} {710, 1557}

\bibitem[\protect\citeauthoryear{{Pilachowski} \& {Pace}}{{Pilachowski} \&
  {Pace}}{2015}]{Pil16}
{Pilachowski} C.~A.,  {Pace} C.,  2015, \mn@doi [\aj]
  {10.1088/0004-6256/150/3/66}, \href
  {http://adsabs.harvard.edu/abs/2015AJ....150...66P} {150, 66}

\bibitem[\protect\citeauthoryear{{Pilyugin}, {V{\'{\i}}lchez}  \&
  {Thuan}}{{Pilyugin} et~al.}{2010}]{pil10}
{Pilyugin} L.~S.,  {V{\'{\i}}lchez} J.~M.,   {Thuan} T.~X.,  2010, \mn@doi
  [\apj] {10.1088/0004-637X/720/2/1738}, \href
  {http://adsabs.harvard.edu/abs/2010ApJ...720.1738P} {720, 1738}

\bibitem[\protect\citeauthoryear{{Pizzone} et~al.,}{{Pizzone}
  et~al.}{2017}]{piz17}
{Pizzone} R.~G.,  et~al., 2017, \mn@doi [\apj] {10.3847/1538-4357/836/1/57},
  \href {http://adsabs.harvard.edu/abs/2017ApJ...836...57P} {836, 57}

\bibitem[\protect\citeauthoryear{{Prantzos}}{{Prantzos}}{2003}]{Prantzos2003}
{Prantzos} N.,  2003, preprint, \href
  {http://adsabs.harvard.edu/abs/2003astro.ph..1043P} {} (\mn@eprint {arXiv}
  {astro-ph/0301043})

\bibitem[\protect\citeauthoryear{{Prantzos}}{{Prantzos}}{2006}]{Prantzos2006}
{Prantzos} N.,  2006, preprint, \href
  {http://adsabs.harvard.edu/abs/2006astro.ph.12633P} {} (\mn@eprint {arXiv}
  {astro-ph/0612633})

\bibitem[\protect\citeauthoryear{{Prantzos}}{{Prantzos}}{2012a}]{Prantzos2012a}
{Prantzos} N.,  2012a, \mn@doi [\aap] {10.1051/0004-6361/201117448}, \href
  {http://adsabs.harvard.edu/abs/2012A%26A...538A..80P} {538, A80}

\bibitem[\protect\citeauthoryear{{Prantzos}}{{Prantzos}}{2012b}]{Prantzos2012b}
{Prantzos} N.,  2012b, \mn@doi [\aap] {10.1051/0004-6361/201219043}, \href
  {http://adsabs.harvard.edu/abs/2012A%26A...542A..67P} {542, A67}

\bibitem[\protect\citeauthoryear{{Prantzos}, {Arnould}  \&
  {Arcoragi}}{{Prantzos} et~al.}{1987}]{Prantzos1987}
{Prantzos} N.,  {Arnould} M.,   {Arcoragi} J.-P.,  1987, \mn@doi [\apj]
  {10.1086/165125}, \href {http://adsabs.harvard.edu/abs/1987ApJ...315..209P}
  {315, 209}

\bibitem[\protect\citeauthoryear{{Prantzos}, {Hashimoto}  \&
  {Nomoto}}{{Prantzos} et~al.}{1990}]{Pra90}
{Prantzos} N.,  {Hashimoto} M.,   {Nomoto} K.,  1990, \aap, \href
  {http://adsabs.harvard.edu/abs/1990A%26A...234..211P} {234, 211}

\bibitem[\protect\citeauthoryear{{Prantzos}, {Casse}  \&
  {Vangioni-Flam}}{{Prantzos} et~al.}{1993}]{Prantzos1993}
{Prantzos} N.,  {Casse} M.,   {Vangioni-Flam} E.,  1993, \mn@doi [\apj]
  {10.1086/172233}, \href {http://adsabs.harvard.edu/abs/1993ApJ...403..630P}
  {403, 630}

\bibitem[\protect\citeauthoryear{{Prantzos}, {Vangioni-Flam}  \&
  {Chauveau}}{{Prantzos} et~al.}{1994}]{Prantzos1994}
{Prantzos} N.,  {Vangioni-Flam} E.,   {Chauveau} S.,  1994, \aap, \href
  {http://adsabs.harvard.edu/abs/1994A%26A...285..132P} {285, 132}

\bibitem[\protect\citeauthoryear{{Prantzos}, {Aubert}  \& {Audouze}}{{Prantzos}
  et~al.}{1996}]{Prantzos1996}
{Prantzos} N.,  {Aubert} O.,   {Audouze} J.,  1996, \aap, \href
  {http://cdsads.u-strasbg.fr/abs/1996A%26A...309..760P} {309, 760}

\bibitem[\protect\citeauthoryear{{Qian} \& {Wasserburg}}{{Qian} \&
  {Wasserburg}}{2008}]{Qia08}
{Qian} Y.-Z.,  {Wasserburg} G.~J.,  2008, \mn@doi [\apj] {10.1086/591545},
  \href {http://adsabs.harvard.edu/abs/2008ApJ...687..272Q} {687, 272}

\bibitem[\protect\citeauthoryear{{Raiteri}, {Busso}, {Picchio}, {Gallino}  \&
  {Pulone}}{{Raiteri} et~al.}{1991}]{Rai91}
{Raiteri} C.~M.,  {Busso} M.,  {Picchio} G.,  {Gallino} R.,   {Pulone} L.,
  1991, \mn@doi [\apj] {10.1086/169622}, \href
  {http://adsabs.harvard.edu/abs/1991ApJ...367..228R} {367, 228}

\bibitem[\protect\citeauthoryear{{Ram{\'{\i}}rez-Agudelo}
  et~al.,}{{Ram{\'{\i}}rez-Agudelo} et~al.}{2017}]{ram17}
{Ram{\'{\i}}rez-Agudelo} O.~H.,  et~al., 2017, \mn@doi [\aap]
  {10.1051/0004-6361/201628914}, \href
  {http://adsabs.harvard.edu/abs/2017A%26A...600A..81R} {600, A81}

\bibitem[\protect\citeauthoryear{{Recio-Blanco}, {de Laverny}, {Worley},
  {Santos}, {Melo}  \& {Israelian}}{{Recio-Blanco} et~al.}{2012a}]{Ale12}
{Recio-Blanco} A.,  {de Laverny} P.,  {Worley} C.,  {Santos} N.~C.,  {Melo} C.,
    {Israelian} G.,  2012a, \mn@doi [\aap] {10.1051/0004-6361/201118261}, \href
  {http://adsabs.harvard.edu/abs/2012A%26A...538A.117R} {538, A117}

\bibitem[\protect\citeauthoryear{{Recio-Blanco}, {de Laverny}, {Worley},
  {Santos}, {Melo}  \& {Israelian}}{{Recio-Blanco} et~al.}{2012b}]{rec12}
{Recio-Blanco} A.,  {de Laverny} P.,  {Worley} C.,  {Santos} N.~C.,  {Melo} C.,
    {Israelian} G.,  2012b, \mn@doi [\aap] {10.1051/0004-6361/201118261}, \href
  {http://adsabs.harvard.edu/abs/2012A%26A...538A.117R} {538, A117}

\bibitem[\protect\citeauthoryear{{Renda} et~al.,}{{Renda} et~al.}{2004}]{ren04}
{Renda} A.,  et~al., 2004, \mn@doi [\mnras] {10.1111/j.1365-2966.2004.08215.x},
  \href {http://adsabs.harvard.edu/abs/2004MNRAS.354..575R} {354, 575}

\bibitem[\protect\citeauthoryear{{Roederer}, {Preston}, {Thompson}, {Shectman},
  {Sneden}, {Burley}  \& {Kelson}}{{Roederer} et~al.}{2014}]{Roederer2014}
{Roederer} I.~U.,  {Preston} G.~W.,  {Thompson} I.~B.,  {Shectman} S.~A.,
  {Sneden} C.,  {Burley} G.~S.,   {Kelson} D.~D.,  2014, \mn@doi [\aj]
  {10.1088/0004-6256/147/6/136}, \href
  {http://adsabs.harvard.edu/abs/2014AJ....147..136R} {147, 136}

\bibitem[\protect\citeauthoryear{{Roederer}, {Karakas}, {Pignatari}  \&
  {Herwig}}{{Roederer} et~al.}{2016}]{Roe16}
{Roederer} I.~U.,  {Karakas} A.~I.,  {Pignatari} M.,   {Herwig} F.,  2016,
  \mn@doi [\apj] {10.3847/0004-637X/821/1/37}, \href
  {http://adsabs.harvard.edu/abs/2016ApJ...821...37R} {821, 37}

\bibitem[\protect\citeauthoryear{{Romano} \& {Matteucci}}{{Romano} \&
  {Matteucci}}{2007}]{Romano2007}
{Romano} D.,  {Matteucci} F.,  2007, \mn@doi [\mnras]
  {10.1111/j.1745-3933.2007.00320.x}, \href
  {http://adsabs.harvard.edu/abs/2007MNRAS.378L..59R} {378, L59}

\bibitem[\protect\citeauthoryear{{Romano}, {Karakas}, {Tosi}  \&
  {Matteucci}}{{Romano} et~al.}{2010}]{Romano2010}
{Romano} D.,  {Karakas} A.~I.,  {Tosi} M.,   {Matteucci} F.,  2010, \mn@doi
  [\aap] {10.1051/0004-6361/201014483}, \href
  {http://adsabs.harvard.edu/abs/2010A%26A...522A..32R} {522, A32}

\bibitem[\protect\citeauthoryear{{Rosswog}, {Korobkin}, {Arcones}, {Thielemann}
   \& {Piran}}{{Rosswog} et~al.}{2014}]{Ros14}
{Rosswog} S.,  {Korobkin} O.,  {Arcones} A.,  {Thielemann} F.-K.,   {Piran} T.,
   2014, \mn@doi [\mnras] {10.1093/mnras/stt2502}, \href
  {http://adsabs.harvard.edu/abs/2014MNRAS.439..744R} {439, 744}

\bibitem[\protect\citeauthoryear{{Salvadori}, {Schneider}  \&
  {Ferrara}}{{Salvadori} et~al.}{2007}]{Salvadori2007}
{Salvadori} S.,  {Schneider} R.,   {Ferrara} A.,  2007, \mn@doi [\mnras]
  {10.1111/j.1365-2966.2007.12133.x}, \href
  {http://adsabs.harvard.edu/abs/2007MNRAS.381..647S} {381, 647}

\bibitem[\protect\citeauthoryear{{Sch{\"o}nrich} \& {Binney}}{{Sch{\"o}nrich}
  \& {Binney}}{2009}]{sch09}
{Sch{\"o}nrich} R.,  {Binney} J.,  2009, \mn@doi [\mnras]
  {10.1111/j.1365-2966.2009.14750.x}, \href
  {http://adsabs.harvard.edu/abs/2009MNRAS.396..203S} {396, 203}

\bibitem[\protect\citeauthoryear{{Sellwood} \& {Binney}}{{Sellwood} \&
  {Binney}}{2002}]{sel02}
{Sellwood} J.~A.,  {Binney} J.~J.,  2002, \mn@doi [\mnras]
  {10.1046/j.1365-8711.2002.05806.x}, \href
  {http://adsabs.harvard.edu/abs/2002MNRAS.336..785S} {336, 785}

\bibitem[\protect\citeauthoryear{{Shen}, {Cooke}, {Ramirez-Ruiz}, {Madau},
  {Mayer}  \& {Guedes}}{{Shen} et~al.}{2015}]{she15}
{Shen} S.,  {Cooke} R.~J.,  {Ramirez-Ruiz} E.,  {Madau} P.,  {Mayer} L.,
  {Guedes} J.,  2015, \mn@doi [\apj] {10.1088/0004-637X/807/2/115}, \href
  {http://adsabs.harvard.edu/abs/2015ApJ...807..115S} {807, 115}

\bibitem[\protect\citeauthoryear{{Sneden}, {Cowan}  \& {Gallino}}{{Sneden}
  et~al.}{2008}]{Sne08}
{Sneden} C.,  {Cowan} J.~J.,   {Gallino} R.,  2008, \mn@doi [\araa]
  {10.1146/annurev.astro.46.060407.145207}, \href
  {http://adsabs.harvard.edu/abs/2008ARA%26A..46..241S} {46, 241}

\bibitem[\protect\citeauthoryear{{Spite} et~al.,}{{Spite} et~al.}{2005}]{Spi05}
{Spite} M.,  et~al., 2005, \mn@doi [\aap] {10.1051/0004-6361:20041274}, \href
  {http://adsabs.harvard.edu/abs/2005A%26A...430..655S} {430, 655}

\bibitem[\protect\citeauthoryear{{Spite} et~al.,}{{Spite}
  et~al.}{2006}]{Spite2006}
{Spite} M.,  et~al., 2006, \mn@doi [\aap] {10.1051/0004-6361:20065209}, \href
  {http://cdsads.u-strasbg.fr/abs/2006A%26A...455..291S} {455, 291}

\bibitem[\protect\citeauthoryear{{Stancliffe}, {Tout}  \& {Pols}}{{Stancliffe}
  et~al.}{2004}]{St04}
{Stancliffe} R.~J.,  {Tout} C.~A.,   {Pols} O.~R.,  2004, \mn@doi [\mnras]
  {10.1111/j.1365-2966.2004.07987.x}, \href
  {http://adsabs.harvard.edu/abs/2004MNRAS.352..984S} {352, 984}

\bibitem[\protect\citeauthoryear{{Steinmetz}}{{Steinmetz}}{2003}]{Ste03}
{Steinmetz} M.,  2003, in {Munari} U.,  ed.,  Astronomical Society of the
  Pacific Conference Series Vol. 298, GAIA Spectroscopy: Science and
  Technology. p.~381 (\mn@eprint {} {astro-ph/0211417})

\bibitem[\protect\citeauthoryear{{Straniero}, {Gallino}, {Busso}, {Chiefei},
  {Raiteri}, {Limongi}  \& {Salaris}}{{Straniero} et~al.}{1995}]{St95}
{Straniero} O.,  {Gallino} R.,  {Busso} M.,  {Chiefei} A.,  {Raiteri} C.~M.,
  {Limongi} M.,   {Salaris} M.,  1995, \mn@doi [\apjl] {10.1086/187767}, \href
  {http://adsabs.harvard.edu/abs/1995ApJ...440L..85S} {440, L85}

\bibitem[\protect\citeauthoryear{{Straniero}, {Gallino}  \&
  {Cristallo}}{{Straniero} et~al.}{2006}]{St06}
{Straniero} O.,  {Gallino} R.,   {Cristallo} S.,  2006, \mn@doi [Nuclear
  Physics A] {10.1016/j.nuclphysa.2005.01.011}, \href
  {http://adsabs.harvard.edu/abs/2006NuPhA.777..311S} {777, 311}

\bibitem[\protect\citeauthoryear{{Sukhbold}, {Ertl}, {Woosley}, {Brown}  \&
  {Janka}}{{Sukhbold} et~al.}{2016}]{suck16}
{Sukhbold} T.,  {Ertl} T.,  {Woosley} S.~E.,  {Brown} J.~M.,   {Janka} H.-T.,
  2016, \mn@doi [\apj] {10.3847/0004-637X/821/1/38}, \href
  {http://esoads.eso.org/abs/2016ApJ...821...38S} {821, 38}

\bibitem[\protect\citeauthoryear{{Takahashi} \& {Yokoi}}{{Takahashi} \&
  {Yokoi}}{1987}]{TY87}
{Takahashi} K.,  {Yokoi} K.,  1987, \mn@doi [Atomic Data and Nuclear Data
  Tables] {10.1016/0092-640X(87)90010-6}, \href
  {http://adsabs.harvard.edu/abs/1987ADNDT..36..375T} {36, 375}

\bibitem[\protect\citeauthoryear{{Tautvai{\v s}ien{\.e}}, {Drazdauskas},
  {Bragaglia}, {Randich}  \& {{\v Z}enovien{\.e}}}{{Tautvai{\v s}ien{\.e}}
  et~al.}{2016}]{Taut2016}
{Tautvai{\v s}ien{\.e}} G.,  {Drazdauskas} A.,  {Bragaglia} A.,  {Randich} S.,
   {{\v Z}enovien{\.e}} R.,  2016, \mn@doi [\aap]
  {10.1051/0004-6361/201629273}, \href
  {http://cdsads.u-strasbg.fr/abs/2016A%26A...595A..16T} {595, A16}

\bibitem[\protect\citeauthoryear{{Thielemann}, {Nomoto}  \&
  {Yokoi}}{{Thielemann} et~al.}{1986}]{Thielemann1986}
{Thielemann} F.-K.,  {Nomoto} K.,   {Yokoi} K.,  1986, \aap, \href
  {http://cdsads.u-strasbg.fr/abs/1986A%26A...158...17T} {158, 17}

\bibitem[\protect\citeauthoryear{{Timmes}, {Woosley}  \& {Weaver}}{{Timmes}
  et~al.}{1995}]{Timmes1995}
{Timmes} F.~X.,  {Woosley} S.~E.,   {Weaver} T.~A.,  1995, \mn@doi [\apjs]
  {10.1086/192172}, \href {http://adsabs.harvard.edu/abs/1995ApJS...98..617T}
  {98, 617}

\bibitem[\protect\citeauthoryear{{Travaglio}, {Gallino}, {Arnone}, {Cowan},
  {Jordan}  \& {Sneden}}{{Travaglio} et~al.}{2004}]{Tra04}
{Travaglio} C.,  {Gallino} R.,  {Arnone} E.,  {Cowan} J.,  {Jordan} F.,
  {Sneden} C.,  2004, \mn@doi [\apj] {10.1086/380507}, \href
  {http://adsabs.harvard.edu/abs/2004ApJ...601..864T} {601, 864}

\bibitem[\protect\citeauthoryear{{Trippella}, {Busso}, {Palmerini}, {Maiorca}
  \& {Nucci}}{{Trippella} et~al.}{2016}]{Tr16}
{Trippella} O.,  {Busso} M.,  {Palmerini} S.,  {Maiorca} E.,   {Nucci} M.~C.,
  2016, \mn@doi [\apj] {10.3847/0004-637X/818/2/125}, \href
  {http://adsabs.harvard.edu/abs/2016ApJ...818..125T} {818, 125}

\bibitem[\protect\citeauthoryear{{Umeda} \& {Nomoto}}{{Umeda} \&
  {Nomoto}}{2002}]{UN02}
{Umeda} H.,  {Nomoto} K.,  2002, \mn@doi [\apj] {10.1086/323946}, \href
  {http://esoads.eso.org/abs/2002ApJ...565..385U} {565, 385}

\bibitem[\protect\citeauthoryear{{Vassiliadis} \& {Wood}}{{Vassiliadis} \&
  {Wood}}{1993}]{vw93}
{Vassiliadis} E.,  {Wood} P.~R.,  1993, \mn@doi [\apj] {10.1086/173033}, \href
  {http://adsabs.harvard.edu/abs/1993ApJ...413..641V} {413, 641}

\bibitem[\protect\citeauthoryear{{Ventura}, {Di Criscienzo}, {Carini}  \&
  {D'Antona}}{{Ventura} et~al.}{2013}]{Ve13}
{Ventura} P.,  {Di Criscienzo} M.,  {Carini} R.,   {D'Antona} F.,  2013,
  \mn@doi [\mnras] {10.1093/mnras/stt444}, \href
  {http://adsabs.harvard.edu/abs/2013MNRAS.431.3642V} {431, 3642}

\bibitem[\protect\citeauthoryear{{Wanajo}}{{Wanajo}}{2013}]{wan13}
{Wanajo} S.,  2013, \mn@doi [\apjl] {10.1088/2041-8205/770/2/L22}, \href
  {http://adsabs.harvard.edu/abs/2013ApJ...770L..22W} {770, L22}

\bibitem[\protect\citeauthoryear{{Weiss} \& {Ferguson}}{{Weiss} \&
  {Ferguson}}{2009}]{WF09}
{Weiss} A.,  {Ferguson} J.~W.,  2009, \mn@doi [\aap]
  {10.1051/0004-6361/200912043}, \href
  {http://adsabs.harvard.edu/abs/2009A%26A...508.1343W} {508, 1343}

\bibitem[\protect\citeauthoryear{{Wilson} et~al.,}{{Wilson}
  et~al.}{2010}]{Wil10}
{Wilson} J.~C.,  et~al., 2010, in Ground-based and Airborne Instrumentation for
  Astronomy III. p. 77351C, \mn@doi{10.1117/12.856708}

\bibitem[\protect\citeauthoryear{{Woosley} \& {Weaver}}{{Woosley} \&
  {Weaver}}{1995}]{Woo95}
{Woosley} S.~E.,  {Weaver} T.~A.,  1995, \mn@doi [\apjs] {10.1086/192237},
  \href {http://adsabs.harvard.edu/abs/1995ApJS..101..181W} {101, 181}

\bibitem[\protect\citeauthoryear{{Woosley}, {Hartmann}, {Hoffman}  \&
  {Haxton}}{{Woosley} et~al.}{1990}]{Woosley1990}
{Woosley} S.~E.,  {Hartmann} D.~H.,  {Hoffman} R.~D.,   {Haxton} W.~C.,  1990,
  \mn@doi [\apj] {10.1086/168839}, \href
  {http://cdsads.u-strasbg.fr/abs/1990ApJ...356..272W} {356, 272}

\bibitem[\protect\citeauthoryear{{Yamaguchi} et~al.,}{{Yamaguchi}
  et~al.}{2015}]{yam15}
{Yamaguchi} H.,  et~al., 2015, \mn@doi [\apjl] {10.1088/2041-8205/801/2/L31},
  \href {http://adsabs.harvard.edu/abs/2015ApJ...801L..31Y} {801, L31}

\bibitem[\protect\citeauthoryear{{Yan}, {Shi}  \& {Zhao}}{{Yan}
  et~al.}{2015}]{Yan2015}
{Yan} H.~L.,  {Shi} J.~R.,   {Zhao} G.,  2015, \mn@doi [\apj]
  {10.1088/0004-637X/802/1/36}, \href
  {http://adsabs.harvard.edu/abs/2015ApJ...802...36Y} {802, 36}

\bibitem[\protect\citeauthoryear{{Yanny} et~al.,}{{Yanny} et~al.}{2009}]{Yan09}
{Yanny} B.,  et~al., 2009, \mn@doi [\aj] {10.1088/0004-6256/137/5/4377}, \href
  {http://adsabs.harvard.edu/abs/2009AJ....137.4377Y} {137, 4377}

\bibitem[\protect\citeauthoryear{{Yong} et~al.,}{{Yong}
  et~al.}{2013}]{Yong2013}
{Yong} D.,  et~al., 2013, \mn@doi [\apj] {10.1088/0004-637X/762/1/26}, \href
  {http://adsabs.harvard.edu/abs/2013ApJ...762...26Y} {762, 26}

\makeatother
\end{thebibliography}








\bsp	
\label{lastpage}
\end{document}